%% file: non-abelian-twists.tex
\documentclass[10pt,a4paper]{article}

\pdfoutput=1        
\usepackage[utf8]{inputenc}
\usepackage{amsmath}
\usepackage{amsfonts}
\usepackage{amssymb}
\usepackage{dsfont}
\usepackage{upgreek}
\usepackage{physics}
\usepackage{import}     
\usepackage{graphicx}
\usepackage{cite}
\usepackage{fancyhdr}   
\usepackage{caption}    

\usepackage{bookmark}   

\fancyheadoffset{0pt}

\newcommand{\iD}[1]{(#1)}       
\newcommand{\iOm}[1]{\vb{ #1}}  
\newcommand{\UL}{U_{L}}         
\newcommand{\UR}{U_{R}}         
\newcommand{\myperp}{\perp}

\newcommand{\COMMENTOOK}[1]{}

\author{
  Riccardo Finotello\thanks{E-mail: riccardo.finotello@to.infn.it} \and
  Igor Pesando\thanks{E-mail: ipesando@to.infn.it} \\[0.75cm]
  Dipartimento di Fisica, Universit\`{a} di Torino \\
  and I.N.F.N. - sezione di Torino \\[0.3cm]
  Via P. Giuria 1, I-10125 Torino, Italy \\
}
\title{
  The Classical Solution for the Bosonic String in the Presence of Three
  D-branes Rotated by Arbitrary $\mathrm{SO}(4)$ Elements
}
\date{}

\hypersetup
{
  pdftitle={The Classical Solution for the Bosonic String in the Presence of
    Three D-branes Rotated by Arbitrary SO(4) Elements},
  pdfsubject={String Theory},
  pdfauthor={Riccardo Finotello, Igor Pesando},
  pdfkeywords={string theory, conformal field theory, high energy physics,
    d-brane, rotations, hypergeometric},
  pdfstartview={FitH},
  pdfcreator={pdfTeX Tex Live 2016/Debian},
  pdfproducer={pdfLaTeX},
  pdflang={en-GB},
  pdfpagemode={UseOutlines},
  bookmarksopen={true},
  bookmarksnumbered={true},
  hidelinks
}

\begin{document}

  \maketitle

  \begin{abstract}
    We consider the classical instantonic contribution to the open string
    configuration associated with three D-branes with relative rotation
    matrices in $\mathrm{SO}(4)$ which corresponds to the computation of the
    classical part of the correlator of three non Abelian twist fields. We
    write the classical solution as a sum of a product of two hypergeometric
    functions. Differently from all the previous cases with three D-branes, the
    solution is not holomorphic and suggests that the classical bosonic string
    knows when the configuration may be supersymmetric. We show how this
    configuration reduces to the standard Abelian twist field computation. From
    the phenomenological point of view, the Yukawa couplings between  chiral
    matter at the intersection in this configuration are more suppressed with
    respect to the factorized case in the literature.
  \end{abstract}

  \thispagestyle{empty}

  \clearpage
  \tableofcontents

  \numberwithin{equation}{section}

  \thispagestyle{empty}

  \clearpage
  \section{Introduction and Conclusions}

  The study of viable phenomenological models in the framework of String
  Theory often involves the analysis of the properties of systems of
  D-branes. Clearly the inclusion of the physical requirements needed for a
  consistent theory deeply constrains the possible scenarios. In particular the
  chiral spectrum of the Standard Model acts as a strong restriction on the
  possible D-brane setup. Intersecting branes
  represent a relevant class of such models with interacting chiral matter.

  In this paper we focus on the development of technical tools for the
  computation of Yukawa interactions for D-branes at angles
  \cite{Chamoun:2003pf, Cremades:2003qj, Cvetic:2009mt, Abel:2006yk,
  Chen:2008rx, Abel:2004ue}. These couplings, as well as the study of flavour
  changing neutral currents \cite{Abel:2003fk}, are crucial in determining the
  validity of the different models. Moreover, several similar computations
  heavily require the ability to compute correlation functions of twist fields
  and excited twist fields and Green functions in the presence of twists.

  The computation of the correlation functions of Abelian twist fields is the
  subject of a vast and solid literature
  and play a prominent role in many scenarios, such as magnetic branes with
  commuting magnetic fluxes \cite{Angelantonj:2000hi, Bertolini:2005qh,
  Bianchi:2005yz, Pesando:2009tt, Forste:2018kub}, strings propagating in a
  gravitational wave background \cite{Kiritsis:1993jk, DAppollonio:2003zow,
  Berkooz:2004yy, DAppollonio:2004ppq}, D-brane bound states \cite{Gava:1997jt,
  Duo:2007he, David:2000um} and tachyon condensation in Superstring Field
  Theory \cite{David:2000um, David:2000yn, David:2001vm, Hashimoto:2003xz}. A
  similar investigation can be extended to the properties of excited twist
  fields even though they are slightly more subtle to treat and hide many more
  delicate aspects \cite{Burwick:1990tu, Stieberger:1992bj, Erler:1992gt,
  Anastasopoulos:2011gn, Anastasopoulos:2011hj, Anastasopoulos:2013sta}.
  Nonetheless, many results were found starting from the old dual models up to
  more modern interpretations of String Theory \cite{Sciuto:1969vz,
  DellaSelva:1970bj}. The generalization of the correlation function of pure
  twists fields to an arbitrary number of plain and excited twist fields (in
  combination with the usual vertex operators) is however more recent
  \cite{Pesando:2014owa, Pesando:2012cx, Pesando:2011ce} and blends the CFT
  techniques with the path integral approach and the study of the Reggeon
  vertex \cite{Pesando:2003ww, DiVecchia:2007dh, Pesando:2011yd,
  DiVecchia:2011mf, Pesando:2013qda}. The same result has also been recovered
  in the framework of the canonical quantization \cite{Pesando:2014sca} and
  shows a global picture behind the computation of the correlators instead of a
  case-by-case dependence.

  In the framework of intersecting D6-branes at angles we study the case of the
  D-branes whose relative rotations are non Abelian and, as a consequence,
  present non Abelian twist fields at the intersections. We try to understand
  the  subtleties and technical issues arising from such scenario which has
  been studied only in few cases: in older days in the formulation of non
  Abelian orbifolds \cite{Inoue:1987ak, Inoue:1990ci, Gato:1990mx,
  Frampton:2000mq} and more recently for a D-branes system whose relative
  rotations are in $\mathrm{SU}(2)$ \cite{Pesando:2015fpj}.

  The configuration for which we develop the technical tools needed to study
  the Yukawa couplings is three D6-branes inside $\mathds{R}^{1,9}$ with an
  internal space of the form $\mathds{R}^4 \times \mathds{R}^2$, prior to
  compactification to a torus, where the branes are embedded as lines in
  $\mathds{R}^2$ and as bi-dimensional surfaces inside $\mathds{R}^4$. In
  particular we focus on the relative rotations which characterize each brane
  in $\mathds{R}^4$ with respect to the others. They will generally be non
  commuting $\mathrm{SO}(4)$ matrices.

  In this paper we study the classical solution of the bosonic string which
  governs the behavior of the correlator of twist field and consequently the
  Yukawa couplings. In fact, once we separate the classical contribution of the
  string from the quantum fluctuations, using the path integral approach, we
  can write the correlator of $N_B$ twist fields as
  \begin{equation*}
    \left\langle \prod\limits_{t = 1}^{N_B} \upsigma_{M_{\iD{t}}}(x_t)
    \right\rangle
    =
    \mathcal{N}\left( \left\lbrace x_t, M_{\iD{t}} \right\rbrace_{1 \le t \le
    N_B} \right) e^{-S_E\left( \left\lbrace x_t, M_{\iD{t}} \right\rbrace_{1
    \le t \le N_B} \right)},
  \end{equation*}
  where $M_{\iD{t}}$ ($1 \le t \le N_B$) are the monodromies induced by the
  twist fields, $N_B$ is the number of D-branes (and their intersections) and
  $x_t$ are the interaction points on the string worldsheet. Even though the
  quantum corrections in $\mathcal{N}\left( \left\lbrace x_t, M_{\iD{t}}
  \right\rbrace_{1 \le t \le N_B} \right)$ are crucial to the complete
  determination of the correlator, the classical contribution to the Euclidean
  action represents the leading term of the Yukawa couplings. In this paper we
  address only this point in order to better understand the differences
  from the usual factorized case and generalize the results of the previous
  analysis on non Abelian rotations of the branes. We will not consider the
  quantum corrections since they cannot be computed with the actual techniques
  and their determination requires the computation of the 4 twists correlator
  which requires knowledge of the connection formula for Heun functions which
  is not known.

  In the second section of this paper we study the boundary conditions for the
  open string describing the D-branes embedded in $\mathds{R}^4$. We first
  define the embedding of a brane locally in a well adapted frame of reference
  where all branes have the same embedding conditions, then we connect all
  these local descriptions using a global coordinate system. In this reference
  frame each brane is rotated with respect to the others and this gives raise
  to monodromies of the doubled string coordinate fields.

  In the third section we choose the monodromies in $\mathrm{SO}(4)$ and we
  rewrite the boundary conditions problem in spinor representation by means of
  the local isomorphism $\mathrm{SO}(4) \cong \mathrm{SU}(2) \times
  \mathrm{SU}(2)$. In doing so we recast the issue of finding the solution
  intended as $4$ real vector in the search of two solutions in the fundamental
  of $\mathrm{SU}(2)$, one for each $\mathrm{SU}(2)$.

  In the fourth section we solve the previous problem of finding two functions
  transforming as a vector of $\mathrm{SU}(2)$ by means of a basis of
  hypergeometric functions. In particular we show how to relate the parameters
  of the rotations and the parameters of the hypergeometric equation and the
  fact that a rescaling factor is needed with respect to the conventionally
  normalized basis of solutions of the hypergeometric equation. Given the
  infinite number of solutions representing the same rotations and labeled by
  the choice of integer factors, we isolate the correct and finite number of
  solutions, actually two, by looking for independent hypergeometric functions
  and a finite Euclidean action.

  The fifth section is dedicated to recovering the previous results from the
  general case. We compute the Abelian limit of the monodromies and we connect
  the parameters of the $\mathrm{SU}(2) \times \mathrm{SU}(2)$ to the usual
  parameters used in the geometrical construction. We then show how the known
  result follows naturally and we encounter the same analyticity properties of
  the field which have been shown in the past. We check also that the case of
  $\mathrm{SU}(2)$ monodromies is smoothly recovered.

  Eventually, in the last section we give a natural interpretation of the
  result highlighting the key differences between the case of Abelian twist
  fields and the general setup and showing the physical consequences on the
  Yukawa couplings. The final result shows a substantial difference between the
  Abelian and the non Abelian case and even between the $\mathrm{SU}(2) \times
  \mathrm{SU}(2)$ and $\mathrm{SU}(2)$ cases. In the Abelian formulation the
  contribution of the Euclidean action is exactly the area of the triangle
  formed by the intersecting branes in $\mathds{R}^2$, that is the string
  worldsheet is completely contained inside the polygon and the action is
  indeed proportional to its area. In the non Abelian case, even though the
  three intersection points still define a 2-dimensional plane in
  $\mathds{R}^4$, the string worldsheet is no longer flat and spans a larger
  area with respect to the previous case. Intuitively, because of the non
  Abelian nature of the D-brane rotation, the string has to bend in order to
  stretch between the branes and cannot entirely reside on a flat surface. The
  difference between the $\mathrm{SO}(4)$ and $\mathrm{SU}(2)$ cases is more
  subtle: in the $\mathrm{SU}(2)$ case there exist complex coordinates for
  $\mathds{R}^{4}$ for which the classical string solution is holomorphic in
  the upper half plane while in $\mathrm{SO}(4)$ case this does not happen. The
  reason of this can probably be traced back to supersymmetry, even if we are
  dealing with the bosonic part only. In fact, for branes rotated by
  $\mathrm{SU}(2)$ elements, part of the spacetime supersymmetry is preserved.
  The further suppression with respect to the Abelian case of the Yukawa
  interactions represents the  physical interpretation of the result.

  \section{D-brane Configuration and Boundary Conditions}

  Even though we are ultimately interested to the framework of superstrings and
  D6-branes intersecting at angles in the internal space, we will focus on the
  bosonic string embedded in $\mathds{R}^{1,d+4}$. The branes are seen as
  2-dimensional Euclidean planes in $\mathds{R}^{4}$ times possible further
  dimensions in $\mathds{R}^{1,d}$. We then specifically concentrate on the
  Euclidean explicit solution for the classical bosonic string in this
  scenario.

  The mathematical analysis is however more general and can be applied to any
  $Dp$-brane embedded in a generic Euclidean space $\mathds{R}^q$. The full
  classical solution can in principle be written also in this case provided one
  can find the explicit form of the basis of functions with the proper boundary
  and monodromy conditions. This is possible in the case of three intersecting
  branes but in general it is an open mathematical issue. In fact, in the case
  of three branes with generic embedding we can usually connect a local basis
  around one intersection point to a local basis around a second intersection
  point, the third depending on the first two intersections, by means of
  Mellin-Barnes integrals. This way the solution can be explicitly and globally
  constructed. However, with more than three D-branes (consequently,
  intersection points) the situation is by far more difficult since the
  explicit form of the connection formulas is not known and therefore we cannot
  write any local basis with respect to the others. Hence the global solution
  cannot be fully specified.

  \subsection{Intersecting D-branes at Angles}

  First of all we describe more precisely the embedding of the D-branes in
  $\mathds{R}^{1,d+4}$ associated to the Euclidean space $\mathds{R}^{4}$
  which is the main focus of this paper. Let $N_B$ be the total number of
  D-branes and $t = 1,\, 2,\, \dots,\, N_B$ be an index defined modulo $N_B$
  to label them, then we can describe one of those D-branes in a well adapted
  system of coordinates $X_{\iD{t}}^I$, where $I = 1,\, 2,\, 3,\, 4$, as:
  \begin{equation}
    X_{\iD{t}}^3 = X_{\iD{t}}^4 = 0.
    \label{eq:well-adapt-embed}
  \end{equation}
  That is, we choose $X_{\iD{t}}^1$ and $X_{\iD{t}}^2$ to be the coordinates
  parallel to the brane labeled with $D_{\iD{t}}$ while $X_{\iD{t}}^3$ and
  $X_{\iD{t}}^4$ are the coordinates orthogonal to it.

  \begin{figure}[t]
    \centering
    \def\svgwidth{0.6\textwidth}
    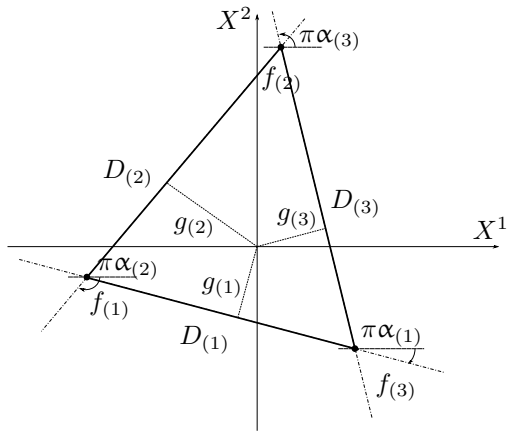
    \caption{Geometry of branes at angles in the factorized case: branes are
      embedded in the plane $\mathds{R}^2$ as lines. Here, the geometrical
      interpretation of $g_{\iD{t}}$ is straightforward and such that
      $g_{\iD{t}}^{\myperp} \ge 0 $ and we describe the planar rotation
      $R_{\iD{t}}( \upalpha_{\iD{t}} ) \in \mathrm{SO}(2)
      \mathrm{U}(1)$ by means of a single parameter $-1 \le
      \upalpha_{\iD{t}} < 1$.}
    \label{fig:branes_at_angles}
  \end{figure}

  This well adapted reference coordinates system is connected to the global
  $\mathds{R}^{4}$ coordinates $X^I$, which we use to study the entire set of
  D-branes, as:
  \begin{equation}
    X^I_{\iD{t}} = \left( R_{\iD{t}} \right)^I_{\,J} X^J - g^I_{\iD{t}} \qfor
    I,J = 1,\,2,\,3,\,4,
    \label{eq:brane_rotation}
  \end{equation}
  where $R_{\iD{t}}$ represents the rotation of the D-brane $D_{\iD{t}}$ and
  $g_{\iD{t}}$ its translation with respect to the origin of the global set of
  coordinates (see Figure~\ref{fig:branes_at_angles} for a 2-dimensional
  example). While we could naively consider $R_{\iD{t}} \in \mathrm{SO}(4)$,
  rotating separately the subset of coordinates parallel and orthogonal to the
  D-brane does not affect the embedding and it just amounts to a trivial
  redefinition of the initial well adapted coordinates. Therefore $R_{\iD{t}}$
  is actually defined in the Grassmannian:
  \begin{equation}
    R_{\iD{t}} \in \frac{\mathrm{SO}(4)}{\mathrm{S}\left( \mathrm{O}(2) \times
    \mathrm{O}(2) \right)},
  \end{equation}
  that is we need only consider the left coset where $R_{\iD{t}}$ is a
  representative of an equivalence relation of the form
  \begin{equation*}
    R_{\iD{t}} \sim \mathcal{O}_{\iD{t}} R_{\iD{t}},
  \end{equation*}
  where the ${\mathrm{S}\left( \mathrm{O}(2) \times \mathrm{O}(2) \right)}$
  element $\mathcal{O}_{\iD{t}}$ is defined as
  \begin{equation*}
    \mathcal{O}_{\iD{t}} = \mqty(
      \mathrm{O}^{\parallel}_{\iD{t}} &
      \\
      & \mathrm{O}^{\myperp}_{\iD{t}}
    )
  \end{equation*}
  with $\mathrm{O}^{\parallel}_{\iD{t}} \in \mathrm{O}(2)$,
  $\mathrm{O}^{\myperp}_{\iD{t}} \in \mathrm{O}(2)$ and
  $\det(\mathcal{O}_{\iD{t}}) = 1$. Here we adopted a shorthand notation which
  we will use again later: the superscript $\parallel$ represents any of the
  coordinates parallel to the brane, while $\myperp$ any of the orthogonal
  directions. Notice that we write $\mathrm{S}\left( \mathrm{O}(2) \times
  \mathrm{O}(2) \right)$ and not $\mathrm{SO}(2) \times \mathrm{SO}(2)$ since
  the additional $\mathds{Z}_2$ group can be used to set $g_{\iD{t}}^{\myperp}
  \ge 0$.

  \subsection{Boundary Conditions for Branes at Angles}

\begin{figure}[t]
    \centering
    \def\svgwidth{0.6\textwidth}
    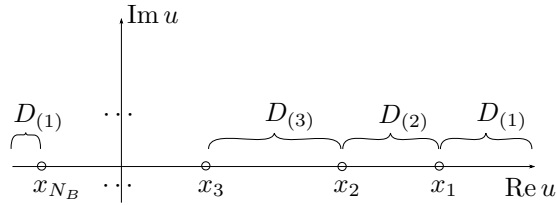
    \caption{Each interval $\left[ x_t, x_{t-1} \right]$ for $t = 2,\, 3,\,
      \dots,\, N_B$ defines the brane $D_{\iD{t}}$. The brane $D_{\iD{1}}$ is
      actually defined on the union of the intervals $\left( -\infty, x_{N_B}
      \right]$ and $\left[ x_1, +\infty \right)$.}
    \label{fig:branes-conformal-gauge}
  \end{figure}

  We now consider the implications of the embedding of the branes on the
  boundary conditions of the open strings. Let $\uptau_E = i \uptau$ be the
  Euclidean time direction, then we define the usual upper plane coordinates:
  \begin{eqnarray*}
    u = x + i y = e^{\uptau_E + i \upsigma} & \in & \mathrm{H} \cup
    \left\lbrace z \in \mathds{C} \mid \Im z = 0 \right\rbrace,
    \\
    \overline{u} = x - i y = e^{\uptau_E - i \upsigma} & \in &
    \overline{\mathrm{H}} \cup \left\lbrace z \in \mathds{C} \mid \Im z = 0
    \right\rbrace,
  \end{eqnarray*}
  where $\mathrm{H} = \left\lbrace z \in \mathds{C} \mid \Im z > 0
  \right\rbrace$ is the upper complex plane, $\overline{\mathrm{H}} =
  \left\lbrace z \in \mathds{C} \mid \Im z < 0 \right\rbrace$ is the lower
  complex plane and $\overline{u} = u^*$ by definition.
  In the conformal coordinates $u$ and $\overline{u}$, D-branes are mapped to
  the real axis $\Im z = 0$ and we use the symbol $D_{\iD{t}}$ to specify both
  the brane and the interval representing it on the real axis of the upper half
  plane:
  \begin{equation*}
    D_{\iD{t}} = \left[ x_t, x_{t-1} \right],
  \end{equation*}
  where $t = 2,\, 3,\, \dots,\, N_B$ and $x_t < x_{t-1}$. The points $x_t$ and
  $x_{t-1}$ represent the worldsheet intersection points of the brane
  $D_{\iD{t}}$ with the branes $D_{\iD{t+1}}$ and $D_{\iD{t-1}}$ respectively.
  With this choice we have to consider carefully the interval $\left[ x_1,
  x_{N_B} \right]$ representing the brane $D_{\iD{1}}$: since the branes are
  defined modulo $N_B$, as shown in Figure~\ref{fig:branes-conformal-gauge}, it
  actually is:
  \begin{equation*}
    D_{\iD{1}} = \left[ x_1, +\infty \right) \cup \left( -\infty, x_{N_B}
    \right].
  \end{equation*}

  In the global coordinates system $X^I$ ($I = 1,\, 2,\, 3,\, 4$), associated
  to the subspace $\mathds{R}^{4} \subset \mathds{R}^{1,d+4}$ where branes are
  generically rotated by a non Abelian rotation, the relevant part of the
  string action in conformal gauge is:
  \begin{equation}
    \begin{split}
      S_{\mathds{R}^{4}} & =
      \frac{1}{2\uppi \upalpha'} \iint\limits_{\mathrm{H}} \dd[2]{u} \partial
      X^I \overline{\partial} X^I =
      \\
      & = \frac{1}{4\uppi \upalpha'} \iint\limits_{\mathds{R} \times
      \mathds{R}^+} \dd{x}\dd{y} \left( \left( \pdv{X^I}{x} \right)^2 + \left(
      \pdv{X^I}{y} \right)^2 \right),
    \end{split}
    \label{eq:string_action}
  \end{equation}
  where $\dd[2]{u} = \dd{u}\dd{\overline{u}} = 2 \dd{x}\dd{y}$ and
  \begin{eqnarray*}
    \partial = \pdv{u} = \frac{1}{2} \left( \pdv{x} - i\pdv{y} \right),
    \\
    \overline{\partial} = \pdv{\overline{u}} = \frac{1}{2} \left( \pdv{x} +
    i\pdv{y} \right).
  \end{eqnarray*}
  Clearly, the equations of motion in these coordinates are:
  \begin{equation}
    \partial \overline{\partial} X^I( u, \overline{u} ) = \frac{1}{4} \left(
    \pdv[2]{x} + \pdv[2]{y} \right) X^I( x+iy, x-iy ) = 0,
    \label{eq:string_equation_of_motion}
  \end{equation}
  and their solution factorizes as usual in left and right moving parts:
  \begin{equation*}
    X^I(u,\overline{u}) = X^I_L(u) + X^I_R(\overline{u}).
  \end{equation*}
  The information on the D-branes is in the boundary conditions which we now
  discuss.

  In the well adapted coordinates, where the embedding is given by
  \eqref{eq:well-adapt-embed}, we describe an open string with one of the
  endpoints on the brane $D_{\iD{t}}$ through the relations:
  \begin{eqnarray}
    \eval{\partial_{\upsigma} X^i_{\iD{t}}( \uptau, \upsigma )}_{\upsigma = 0}
    = & \eval{\partial_{y} X^i_{\iD{t}}(u, \overline{u} )}_{y = 0} & =  0 \qfor
    i = 1,\, 2,
    \label{eq:neumann_bc}
    \\
    X^m_{\iD{t}}( \uptau, 0 ) = & X^m_{\iD{t}}(x,x ) & =  0 \qfor m = 3,\, 4,
    \label{eq:dirichlet_bc}
  \end{eqnarray}
  where $x \in D_{\iD{t}} = \left[ x_t, x_{t-1} \right]$ and the index $i$
  labels the Neumann boundary conditions associated with the parallel
  directions while $m$ labels the Dirichlet coordinates associated to the
  normal ones. As argued in the previous section, this well adapted set of
  coordinates is connected to the global coordinates $X^I$ as in
  \eqref{eq:brane_rotation}.

  In order to deal with the presence of $g_{\iD{t}}^m$ in
  \eqref{eq:brane_rotation} and \eqref{eq:dirichlet_bc} and to get simpler
  boundary conditions, we consider the derivative along the boundary direction
  of \eqref{eq:dirichlet_bc} in such a way to remove the dependence on the
  translation $g_{\iD{t}}^m$. This procedure produces simpler boundary
  conditions which are nevertheless not equivalent to the original ones: they
  will be recovered later by adding further constraints. The simpler boundary
  conditions for the global coordinates are:
  \begin{eqnarray*}
    \left( R_{\iD{t}} \right)^i_{\,J} \eval{\partial_{\upsigma} X^J( \uptau,
    \upsigma )}_{\upsigma = 0} & = & 0 \qfor i = 1,\, 2,
    \\
    \left( R_{\iD{t}} \right)^m_{\,J} \eval{\partial_{\uptau} X^J( \uptau,
    \upsigma )}_{\upsigma = 0} & = & 0 \qfor m = 3,\, 4.
  \end{eqnarray*}
  In the upper half plane coordinates and using the solution of the equations
  of motions they become:
  \begin{eqnarray*}
    \left( R_{\iD{t}} \right)^i_{\,J} \left( \partial X^J_L(x+i0^+) -
    \overline{\partial} X^J_R(x-i0^+) \right) & = & 0 \qfor i = 1,\, 2,
    \\
    \left( R_{\iD{t}} \right)^m_{\,J} \left( \partial X^J_L(x+i0^+) +
    \overline{\partial} X^J_R(x-i0^+) \right) & = & 0 \qfor m = 3,\, 4,
  \end{eqnarray*}
  where $x \in D_{\iD{t}}$.

  Introducing the matrix
  \begin{equation}
    \mathcal{S} = \mqty( \dmat{ 1, 1, -1, -1 } ),
\label{eq:reflection_S}
  \end{equation}
  we can write the full boundary conditions (not just the simplified version we
  have just discussed) in terms of discontinuities along the branes and space
  time interactions points as:
  \begin{equation}
    \begin{cases}
      \partial X^I_L(x+i0^+) = \left( U_{\iD{t}} \right)^I_{\,J}
      \overline{\partial} X^J_R(x-i0^+) \qfor x_t \le x < x_{t-1}
      \\
      X^I(x_t,x_t) = f^I_{\iD{t}}
    \end{cases},
    \label{eq:discontinuity_bc}
  \end{equation}
  where
  \begin{equation}
    U_{\iD{t}} = R^{-1}_{\iD{t}} \mathcal{S} R_{\iD{t}} \in
    \frac{\mathrm{SO}(4)}{\mathrm{S}( \mathrm{O}(2) \times \mathrm{O}(2) )}
\label{eq:U_brane_t}
  \end{equation}
  and $f_{\iD{t}}$ is the target space embedding of the worldsheet interaction
  point between the brane $D_{\iD{t}}$ and $D_{\iD{t+1}}$. Given its
  definition, it is trivial but nonetheless critical to show that $U_{\iD{t}}$
  satisfies
  \begin{equation*}
    U_{\iD{t}} = U^{-1}_{\iD{t}} = U^T_{\iD{t}}.
  \end{equation*}
  On the other hand, another key point is the fact that $f_{\iD{t}}$ recovers
  the apparent loss of information on the translation $g_{\iD{t}}$. Consider
  for instance the embedding equations \eqref{eq:dirichlet_bc} for any two
  intersecting branes $D_{\iD{t}}$ and $D_{\iD{t+1}}$, then introducing the
  auxiliary quantities
  \begin{eqnarray*}
    \mathcal{R}_{\iD{t,t+1}} = \mqty( R_{\iD{t}}^m \\ R_{\iD{t+1}}^n ) & \in &
    \mathrm{GL}_4(\mathds{R}) \qfor m, n = 3, 4,
    \\
    \mathcal{G}_{\iD{t,t+1}} = \mqty( g_{\iD{t}}^m \\ g_{\iD{t+1}}^n ) & \in &
    \mathds{R}^4 \qfor m, n = 3, 4,
  \end{eqnarray*}
  we compute the intersection point as:
  \begin{equation*}
    f_{\iD{t}} = \left( \mathcal{R}_{\iD{t,t+1}} \right)^{-1}
    \mathcal{G}_{\iD{t,t+1}}.
  \end{equation*}
  The result shows that information on the parameter $g_{\iD{t}}$ is recovered
  through the global boundary conditions in the second equation of
  \eqref{eq:discontinuity_bc}.

  \subsection{Doubling Trick and Branch Cut Structure}

\begin{figure}[t]
    \centering
    \def\svgwidth{0.8\textwidth}
    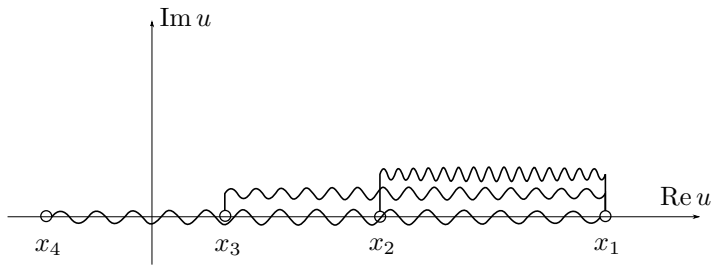
    \caption{The appearance of non trivial discontinuities or monodromies on
      the branes shows that the complex plane has a branch cut structure. The
      particular nature of the rotations of the D-branes is such that the
      branch cuts run along the boundary between finite points. Here we show
      the case of four D-branes, i.e. $N_B = 4$.}
    \label{fig:finite_cuts}
  \end{figure}

  In going from the boundary conditions \eqref{eq:neumann_bc} and
  \eqref{eq:dirichlet_bc} to \eqref{eq:discontinuity_bc} we introduced
  discontinuities across each D-brane thus defining a non trivial cut structure
  on the complex plane. We introduce the doubling trick to deal with fields
  which take values on the whole complex plane by gluing the relations along an
  arbitrary but fixed D-brane $D_{\iD{\overline{t}}}$:
  \begin{equation}
    \partial \mathcal{X}(z) =
    \begin{cases}
      \partial X_L(u) & \qif z = u \qand \Im z > 0 \qor z \in
      D_{\iD{\overline{t}}}
      \\
      U_{\iD{\overline{t}}} \overline{\partial} X_R(\overline{u}) & \qif z =
      \overline{u} \qand \Im z < 0 \qor z \in D_{\iD{\overline{t}}}
    \end{cases}.
    \label{eq:real_doubling_trick}
  \end{equation}
  Let
  \begin{eqnarray*}
    \mathcal{U}_{\iD{t,t+1}} & = & U_{\iD{t+1}} U_{\iD{t}},
    \\
    \widetilde{\mathcal{U}}_{\iD{t,t+1}} & = & U_{\iD{\overline{t}}} U_{\iD{t}}
    U_{\iD{t+1}} U_{\iD{\overline{t}}},
  \end{eqnarray*}
  then we restate the boundary conditions in terms of the doubling field:
  \begin{eqnarray}
    \partial \mathcal{X}( x_t + e^{2\uppi i}( \upeta + i0^+ ) ) & = &
    \mathcal{U}_{\iD{t,t+1}} \partial \mathcal{X}( x_t + \upeta + i0^+ ),
    \label{eq:top_monodromy}
    \\
    \partial \mathcal{X}( x_t + e^{2\uppi i}( \upeta - i0^+ ) ) & = &
    \widetilde{\mathcal{U}}_{\iD{t,t+1}} \partial \mathcal{X}(x_t + \upeta -
    i0^+),
    \label{eq:bottom_monodromy}
  \end{eqnarray}
  for $0 < \upeta < \min\left( \abs{x_{t-1}-x_t}, \abs{x_t-x_{t+1}} \right)$ in
  order to consider only the two intersecting D-branes $D_{\iD{t}}$ and
  $D_{\iD{t+1}}$. The matrices $\mathcal{U}_{\iD{t,t+1}}$ and
  $\widetilde{\mathcal{U}}_{\iD{t,t+1}}$ represent the non trivial monodromies,
  consequence of the rotation of the branes and their boundary conditions.
  Notice however that they are somewhat special $\mathrm{SO}(4)$ matrices and
  in section \ref{sect:special_SO4} we give the general parametrization
  in term of $\mathrm{SU}_L(2) \times\mathrm{SU}_R(2)$ parameters.
  Given the non Abelian characteristic of the rotations, there are two
  different monodromies depending on the base point: one for paths starting in
  the upper plane $\mathrm{H}$ and one for paths starting in
  $\overline{\mathrm{H}}$. As a consequence of the nature of the rotations of
  the D-branes, a path on the complex plane enclosing all the branes
  simultaneously does not show any monodromy:
  \begin{equation*}
    \prod\limits_{t = 1}^{N_B}
    \mathcal{U}_{\iD{\overline{t} - t, \overline{t} + 1 - t}}
    = \prod\limits_{t = 1}^{N_B}
    \widetilde{\mathcal{U}}_{\iD{\overline{t} +t, \overline{t} + 1 +
        t}}
    = \mathds{1}_4,
  \end{equation*}
  where $t$ is, as always, defined modulo $N_B$. These relations reveal that
  the complex plane has branch cuts running between the branes, at finite, as
  shown in Figure~\ref{fig:finite_cuts}. We therefore translated the rotations
  of the D-branes in terms of $\mathcal{U}_{\iD{t,t+1}}$ and
  $\widetilde{\mathcal{U}}_{\iD{t,t+1}}$ which are the matrix representation of
  the homotopy group of the complex plane with the described branch cut
  structure.

  As a consistency check of the procedure, the string action
  \eqref{eq:string_action} can be computed in terms of the new doubling field:
  the map
  \begin{equation*}
    x_t + \upeta \pm i0^+ \mapsto x_t + e^{2\uppi i}( \upeta \pm i0^+)
  \end{equation*}
  must leave the action invariant since it does not depend on the branch cut
  structure in the first place. In fact, it is easy to show that
  \begin{equation*}
    S = \frac{1}{4\uppi \upalpha'} \iint\limits_{\mathds{C}} \dd{z}
    \dd{\overline{z}} \partial \mathcal{X}^T(z) U_{\iD{\overline{t}}}
    \overline{\partial} \mathcal{X}(\overline{z}),
  \end{equation*}
  where $\dd[2] z = \dd{z} \dd{\overline{z}} = 2 \dd{x}\dd{y}$, is left
  untouched by the map.

  \section{D-branes at Angles in Spinor Representation}

  In the previous section we showed how to encode the rotations of the D-branes
  in matrices representing the non trivial monodromies of the doubling field.
  In order to find a solution to the equations of motion with the boundary
  conditions determined by the brane rotations, we should now find an explicit
  solution $\partial \mathcal{X}(z)$ such the non trivial monodromies in
  \eqref{eq:top_monodromy} and \eqref{eq:bottom_monodromy} can be reproduced.

  At first analysis $\partial \mathcal{X}(z)$ is a 4-dimensional \emph{real}
  vector which has $N_B$ non trivial monodromies factors represented by $4
  \times 4$ real matrices, one for each interaction point $x_t$. The solution
  to the string equations of motion is therefore represented by four linearly
  independent functions with $N_B$ branch points. We can try to look for them
  among the solutions to fourth order differential equations with $N_B$ finite
  Fuchsian points. This is however an open mathematical problem in its general
  statement: the basis of such functions around each branch point are usually
  complicated and defined up to several free parameters. Moreover, and more
  importantly, the connection between any two of these basis is an unsolved
  mathematical problem. Using contour integrals and representing the functions
  as Mellin-Barnes integrals it might be possible to solve the issue in the
  very special case $N_B = 3$ but it is certainly not the best course of
  action.

  On the other hand our main interest is to find a solution precisely for $N_B
  = 3$. We then use the isomorphism
  \begin{equation*}
    \mathrm{SO}(4) \cong \frac{\mathrm{SU}(2) \times
    \mathrm{SU}(2)}{\mathds{Z}_2}
  \end{equation*}
  in order to restate the problem of finding a 4-dimensional \emph{real}
  solution to the equations of motion to a quest for a $2 \times 2$ complex
  matrix. This matrix can be seen as a linear superposition of tensor products
  of two (complex) vectors in the fundamental representation of two different
  $\mathrm{SU}(2)$. We can think of these vectors as a solution to a second
  order differential equation with three Fuchsian points, possibly the
  hypergeometric equation. Our task is then to map the original
  $\mathrm{SO}(4)$ monodromies into two sets of $\mathrm{SU}(2)$ monodromies
  and then to find the corresponding parameters of the hypergeometric
  functions.

  \subsection{Review of the Isomorphism}

  In order to carry out the computations we consider the isomorphism between
  $\mathrm{SO}(4)$ and two different copies of $\mathrm{SU}(2)$. Here we sketch
  how operatively the isomorphism works in order to fix our notations while in
  Appendix~\ref{sec:isomorphism} we review it in more details.

  We first consider a basis
  \begin{equation*}
    \uptau = \left( i \mathds{1}_2, \vb{\upsigma} \right)
  \end{equation*}
  where $\vb{\upsigma}$ is a vector containing the usual Pauli matrices. We
  then choose to parameterize any matrix of $\mathrm{SU}(2)$ with a
  3-dimensional vector
  \begin{equation}
    \small
    \vb{n} \in \left\lbrace \left( n^1, n^2, n^3 \right) \in \mathds{R}^3 \mid
    0 \le n \le \frac{1}{2} \qand \vb{n} \equiv \vb{n}' \qq{when} n = n' =
    \frac{1}{2} \right\rbrace,
    \label{eq:n_definition_set}
  \end{equation}
  such that:
  \begin{equation}
    U(\vb{n}) = \cos(2\uppi n) \mathds{1}_2 + i \frac{\vb{n} \cdot
    \vb{\upsigma}}{n} \sin(2\uppi n) \in \mathrm{SU}(2),
    \label{eq:explicit_expr_generic_SU2}
  \end{equation}
  where $n = \norm{\vb{n}}$ so that the following properties hold:
  \begin{eqnarray}
    \left( U(\vb{n}) \right)^* = & \upsigma^2 U(\vb{n}) \upsigma^2 & =
    U(\widetilde{\vb{n}}),
    \label{eq:complex_conjugate_SU2}
    \\
    \left( U(\vb{n}) \right)^{\dagger} = & \left( U(\widetilde{\vb{n}})
    \right)^T & = U(-\vb{n}),
    \label{eq:dagger_SU2}
    \\
    -U(\vb{n}) = & U(\widehat{\vb{n}}), &
    \label{eq:minus_SU2}
  \end{eqnarray}
  where $\widetilde{\vb{n}} = \left( -n^1, +n^2, -n^3 \right)$ and
  $\widehat{\vb{n}} = - \left( \frac{1}{2} - n   \right) \vb{n}/n$.

  We then define a new set of coordinates $X_{(s)}$ in this representation:
  \begin{equation}
    X_{(s)}(u, \overline{u}) = X^I(u, \overline{u}) \uptau_I,
    \label{eq:isomorphism_coordinates}
  \end{equation}
  where a rotation of $\mathrm{SU}_L(2) \times \mathrm{SU}_R(2)$ acts
  as\footnote{
    In the following we write $\UL(\vb{n})$ and $\UR(\vb{m})$ even if
    it is not necessary to specify whether the group element is in the
    left or right $\mathrm{SU}(2)$ since the parameters are explicitly given.
  }
  \begin{equation*}
    X_{(s)}'(u, \overline{u}) = \UL(\vb{n}) X_{(s)}(u, \overline{u})
    \UR^{\dagger}(\vb{m})
  \end{equation*}
  and it is equivalent to a 4-dimensional rotation
  \begin{equation*}
    \left( X'(u, \overline{u}) \right)^I = R^I_{\,J}(\vb{n}, \vb{m}) X^J(u,
    \overline{u}),
  \end{equation*}
  where
  \begin{equation*}
    R_{IJ}(\vb{n}, \vb{m}) = \frac{1}{2} \tr(\uptau_I^{\dagger} \UL(\vb{n})
    \uptau_J \UR^{\dagger}(\vb{m})) \in \mathrm{SO}(4).
  \end{equation*}

  We can therefore work directly on a representation of $\mathrm{SU}_L(2)
  \times \mathrm{SU}_R(2)$ since we have built an isomorphism which maps the
  two sets of three real parameters of $\mathrm{SU}(2)$ matrix (encoded in the
  vectors $\vb{n}$ and $\vb{m}$) to a $\mathrm{SO}(4)$ matrix which is in fact
  described by six real parameters. Notice that there is a residual
  $\mathds{Z}_2$ symmetry acting as $\left\lbrace \UL, \UR \right\rbrace
  \leftrightarrow \left\lbrace -\UL, -\UR \right\rbrace$ so that we can fix the
  first non vanishing component of $\vb{n}$ to be positive.
  The correct isomorphism is therefore:
  \begin{equation*}
    \mathrm{SO}(4) \cong \frac{\mathrm{SU}_L(2) \times
    \mathrm{SU}_R(2)}{\mathds{Z}_2}.
  \end{equation*}

  \subsection{Doubling Trick and Rotations in Spinor Representation}

  In the light of the possibility to use the spinor representation of the
  rotations to find the solutions to the equations of motion of the classical
  bosonic string, we need to reproduce \eqref{eq:real_doubling_trick} as two
  separate $\mathrm{SU}(2)$ rotations. Consider then:
  \begin{equation}
    \partial \mathcal{X}_{(s)}(z) =
    \begin{cases}
      \partial X_{(s),\, L}(u) & \qif z \in \mathrm{H} \qor z \in
      D_{\iD{\overline{t}}}
      \\
      \UL(\vb{n}_{\iD{\overline{t}}}) \overline{\partial} X_{(s),\,
      R}(\overline{u}) \UR^{\dagger}(\vb{m}_{\iD{\overline{t}}}) & \qif z \in
      \overline{\mathrm{H}} \qor z \in D_{\iD{\overline{t}}}
    \end{cases},
    \label{eq:spinor_doubling_trick}
  \end{equation}
  where $\partial \mathcal{X}_{(s)}(z)$, $\partial X_{(s),\,L}(u)$ and
  $\overline{\partial} X_{(s),\,R}(\overline{u})$ are 2-dimensional square
  matrices in the sense of \eqref{eq:isomorphism_coordinates}.

  As in the real representation, we read the discontinuities on the branes with
  respect to the brane $D_{\iD{\overline{t}}}$ in terms of monodromies of
  $\partial \mathcal{X}(z)$, leaving the branch cut structure and the homotopy
  group considerations unchanged as long as we consider both left and right
  sectors of $\mathrm{SU}_L(2) \times \mathrm{SU}_R(2)$ at the same time. In
  particular, let $0 < \upeta < \min\left( \abs{x_t-x_{t-1}}, \abs{x_{t+1}-x_t}
  \right)$, then we find:
  \begin{eqnarray}
    \partial \mathcal{X}_{(s)}( x_t + e^{2\uppi i}( \upeta + i0^+) )
    & = &
    \mathcal{L}_{\iD{t, t+1}} \partial \mathcal{X}_{(s)}( x_t + \upeta + i0^+ )
    \mathcal{R}_{\iD{t, t+1}}^{\dagger},
    \label{eq:top_spinor_monodromy}
    \\
    \partial \mathcal{X}_{(s)}( x_t + e^{2\uppi i}( \upeta - i0^+) )
    & = &
          \widetilde{\mathcal{L}}_{\iD{t, t+1}}
          \partial \mathcal{X}_{(s)}( x_t + \upeta
    - i0^+ ) \widetilde{\mathcal{R}}_{\iD{t, t+1}}^{\dagger},
    \label{eq:bottom_spinor_monodromy}
  \end{eqnarray}
  where:
  \begin{eqnarray*}
    \mathcal{L}_{\iD{t,t+1}}
    & = &
          \UL(\vb{n}_{\iD{t+1}}) \UL^{\dagger}(\vb{n}_{\iD{t}}),
    \\
    \widetilde{\mathcal{L}}_{\iD{t,t+1}}
    & = &
          \UL(\vb{n}_{\iD{\overline{t}}})
    \UL^{\dagger}(\vb{n}_{\iD{t}}) \UL(\vb{n}_{\iD{t+1}})
    \UL^{\dagger}(\vb{n}_{\iD{\overline{t}}}),
    \\
    \mathcal{R}_{\iD{t,t+1}}
    & = & \UR(\vb{m}_{\iD{t+1}}) \UR^{\dagger}(\vb{m}_{\iD{t}}),
    \\
    \widetilde{\mathcal{R}}_{\iD{t,t+1}}
    & = & \UR(\vb{m}_{\iD{\overline{t}}})
    \UR^{\dagger}(\vb{m}_{\iD{t}}) \UR(\vb{m}_{\iD{t+1}})
    \UR^{\dagger}(\vb{m}_{\iD{\overline{t}}}).
  \end{eqnarray*}

  In the spinor representation the action \eqref{eq:string_action} becomes
  \begin{equation*}
    S = \frac{1}{4 \uppi \upalpha'} \iint\limits_{\mathrm{H}} \dd[2]{u}
    \tr(\partial X_{(s)}(u) \cdot \overline{\partial}
    X^{\dagger}_{(s)}(\overline{u}))
  \end{equation*}
  or, in terms of the doubling fields:
  \begin{equation}
    S = \frac{1}{8 \uppi \upalpha'} \iint\limits_{\mathds{C}} \dd[2]{z}
    \tr(\UL(\vb{n}_{\iD{\overline{t}}}) \partial \mathcal{X}_{(s)}(z)
    \UR^{\dagger}(\vb{m}_{\iD{\overline{t}}}) \overline{\partial}
    \mathcal{X}_{(s)}^{\dagger}(\overline{z})).
    \label{eq:action_doubling_fields_spinor_representation}
  \end{equation}
  Even in this case, the map $x_t + \upeta \pm i0^+ \mapsto x_t + e^{2\uppi i}(
  \upeta \pm i0^+ )$ does not generate additional contributions, leaving the
  action unchanged.

  \subsection{Special Form of \texorpdfstring{$\mathrm{SU}(2)$}{SU(2)} Matrices
  for Branes at Angles} \label{sect:special_SO4}

  We now show that the $\mathrm{SU}(2)$ involved in the branes at angles are of
  a very special form. For the left $\mathrm{SU}(2)$ sector we have:
  \begin{equation*}
    \mathcal{L}_{\iD{t,t+1}}
    =
    \UL(\vb{n}_{\iD{t,t+1}})
    =
    -\vb{v}_{\iD{t+1}}\cdot \vb{v}_{\iD{t}}
    +
    i (\vb{v}_{\iD{t+1}}\times \vb{v}_{\iD{t}})\cdot \vb{\upsigma} ,
  \end{equation*}
  with $\vb{v}_{\iD{t}}^2=1$, and similarly for the right sector.
  This follows from the fact that the $\mathrm{SO}(4)$ matrix $U_{\iD{t}}$
  defined in \eqref{eq:U_brane_t} has also  special properties
  and hence the corresponding the $\mathrm{SU}_L(2)\times \mathrm{SU}_R(2)$
  element $(\UL(\vb{n}_{\iD{t}}), \UL(\vb{m}_{\iD{t}}) )$ is special. In
  particular for the left part we have
  \begin{equation}
    \UL(\vb{n}_{\iD{t}})
    =
    i\vb{v}_{\iD{t}} \cdot \vb{\upsigma} \qc
    \vb{v}_{\iD{t}}^2=1,
    \label{eq:special_UL_brane_t}
  \end{equation}
  similarly for the right part.
  In fact the matrix $\mathcal{S}$ in \eqref{eq:reflection_S} can be
  represented as $\UL=\UR= i \upsigma_1$, then any matrix
  $\UL(\vb{n}_{\iD{t}})$ is of the form $\UL(\vb{n}_{\iD{t}})=
  U(\vb{r}_{\iD{t}}) \cdot \left(i \upsigma_1\right) \cdot
  U^\dagger(\vb{r}_{\iD{t}})$, for some $\vb{r}_{\iD{t}}$ as follows from
  \eqref{eq:U_brane_t}. Now this matrix has vanishing trace and squares to
  $-\mathds{1}_2$ hence the term  proportional to $\mathds{1}_2$ in the
  expression of the generic $\mathrm{SU}(2)$ element given in
  \eqref{eq:explicit_expr_generic_SU2} vanishes and therefore
  $n_{\iD{t}}=\frac{1}{4}$ so that \eqref{eq:special_UL_brane_t} follows.

  \section{The Classical Solution}

  \subsection{The Choice of Hypergeometric Functions}

\begin{figure}[t]
    \centering
    \def\svgwidth{0.8\textwidth}
    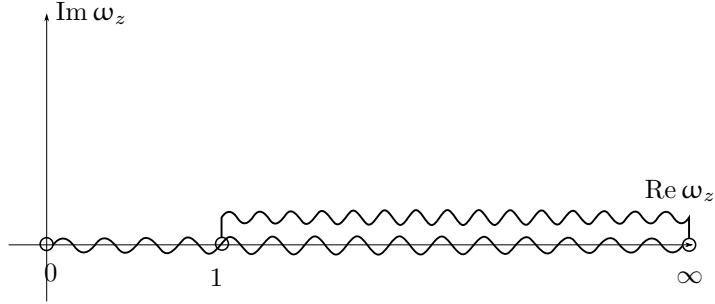
    \caption{Fixing the $\mathrm{SL}_2(\mathds{R})$ invariance for $N_B = 3$
      and $\overline{t} = 1$ leads to a cut structure with all the cuts defined
      on the real axis towards $\upomega_{\overline{t}} = \infty$.}
    \label{fig:hypergeometric_cuts}
  \end{figure}

  As anticipated in the previous section, the spinorial representation entails
  using $\mathrm{SU}(2)$ matrices: it greatly simplifies the search for the
  basis of functions satisfying the desired boundary conditions. Fixing the
  $\mathrm{SL}_2(\mathds{R})$ invariance naturally leads to consider a basis of
  hypergeometric functions in order to reproduce the monodromy matrices in
  \eqref{eq:top_spinor_monodromy} and \eqref{eq:bottom_spinor_monodromy}.
  Specifically, since we are interested in a solution with $N_B = 3$, we fix
  the three intersection points $x_{\overline{t}-1}$, $x_{\overline{t}+1}$ and
  $x_{\overline{t}}$ to $\upomega_{\overline{t}-1} = 0$,
  $\upomega_{\overline{t}+1} = 1$ and $\upomega_{\overline{t}} = \infty$
  respectively through the map:
  \begin{equation}
    \upomega_u = \frac{u - x_{\overline{t}-1}}{u - x_{\overline{t}}}
    \cdot \frac{x_{\overline{t}+1} - x_{\overline{t}-1}} {x_{\overline{t}+1} -
    x_{\overline{t}}}
    \label{eq:def_omega}
  \end{equation}
  The new cut structure for this choice is presented in
  Figure~\ref{fig:hypergeometric_cuts} and fixes $\arg(\upomega_t - \upomega_z)
  \in \left[ 0, 2\uppi \right)$ for $t = \overline{t}-1,\, \overline{t}+1$. We
  then choose for example $\overline{t} = 1$.

  The natural choice of the functions to reproduce the given monodromies is a
  basis of hypergeometric functions. In particular we define:
  \begin{equation*}
    F(a,b;c;z)
    = \sum\limits_{k = 0}^{+\infty} \frac{(a)_k  (b)_k}{\Upgamma(c+k)} \cdot
    \frac{z^k}{k!}
    = \frac{1}{\Upgamma(c)} {}_2 F_1(a,b;c;z),
  \end{equation*}
  where ${}_2 F_1(a,b;c;z)$ is the usual Gauss hypergeometric function and
  $\Upgamma(s)$ is the Euler Gamma function. With this choice, $F(a,b;c;z)$ is
  well defined for any value of $a$, $b$ and $c$ (not simply when $c$ is a
  strictly positive integer, as in the definition of the Gauss hypergeometric
  function). We then choose:
  \begin{equation}
    B_{\iOm{0}}(z) = \mqty(
      F(a,b;c;z)
      \\
      (-z)^{1-c} F(a+1-c,b+1-c;2-c;z)
    )
    \label{eq:basis_0}
  \end{equation}
  as a basis of hypergeometric functions around $z = 0$, with a branch cut on
  the interval $\left[ 0, +\infty \right)$. The choice of the branch cuts
  follows from the cut on $\left[ 1, +\infty \right)$ coming from $F(a,b;c;z)$
  which has a singularity at $z=1$ and the cut on $\left[ 0, +\infty \right)$
  descending from $(-z)^{1-c}$, given the usual cut for $z^\upalpha$ on the
  negative axis.

  As we previously observed, the homotopy group of the branch cut plane, a
  sphere with three marked points, is such that a path enclosing all the
  singularities is homotopically trivial. Therefore the corresponding product
  of the monodromy matrices (in the inverse order with respect to the products
  of paths) is the unit matrix. That is, let
  $\mathcal{M}_{\upomega_t}^{\pm}$ be the  monodromy matrix which represents
  the homotopy loop around $\upomega_t$ ($\mathcal{M}^+$ represents a path
  starting in $\mathrm{H}$ and $\mathcal{M}^-$ a path with base point in
  $\overline{\mathrm{H}}$). Then they satisfy:
  \begin{equation}
    \mathcal{M}_{\iOm{0}}^+ \mathcal{M}_{\iOm{1}}^+
    \mathcal{M}_{\iOm{\infty}}^+
    =
    \mathcal{M}_{\iOm{\infty}}^- \mathcal{M}_{\iOm{1}}^-
    \mathcal{M}_{\iOm{0}}^-
    =
    \mathds{1}_2
    \label{eq:monodromy_relations}
  \end{equation}
  which shows that we can recover the matrix of the monodromy factors in
  $\upomega_{\overline{t}+1} = 1$ as a product of monodromies around $0$ and
  $\infty$ given the properties
  \begin{eqnarray*}
    \mathcal{M}_{\iOm{0}}^+ & = \mathcal{M}_{\iOm{0}}^- & =
    \mathcal{M}_{\iOm{0}},
    \\
    \mathcal{M}_{\iOm{\infty}}^+ & = \mathcal{M}_{\iOm{\infty}}^- & =
    \mathcal{M}_{\iOm{\infty}}.
  \end{eqnarray*}
  These matrices are an abstract representation of the monodromy group since
  they are in an arbitrary basis: to have an explicit representation we need to
  fix an explicit basis.

  Using the basis around $z = 0$ in \eqref{eq:basis_0}, it is straightforward
  to find the explicit representation ${M}_{\iOm{0}}$ of the abstract monodromy
  $\mathcal{M}_{\iOm{0}}$:
  \begin{eqnarray}
    {M}_{\iOm{0}} & = & \mqty( 1 & \\ & e^{-2\uppi i c} ).
    \label{eq:monodromy_zero}
  \end{eqnarray}
  In order to compute the monodromy matrix ${M}_{\iOm{\infty}}$ at $\infty$ in
  the basis \eqref{eq:basis_0}, it is best to first compute the explicit
  monodromy representation $\widetilde{M}_{\iOm{\infty}}$ of the abstract
  monodromy $\mathcal{M}_{\iOm{\infty}}$ in the canonical basis of
  hypergeometric functions around $z = \infty$:
  \begin{equation*}
    B_{\iOm{\infty}}(z) = \mqty(
      (-z)^{-a} F(a,a+1-c;a+1-b;z^{-1})
      \\
      (-z)^{-b} F(b,b+1-c;b+1-a;z^{-1})
    ).
  \end{equation*}
  Then we use how this basis is connected by the matrix
  \begin{equation}
    \mathcal{C} = \frac{\uppi}{\sin(\uppi(a-b))}
    \mqty(
      \frac{1}{\Upgamma(b)\Upgamma(c-a)}
      &
      -\frac{1}{\Upgamma(a) \Upgamma(c-b)}
      \\
      \frac{1}{\Upgamma(1-a)\Upgamma(b+1-c)}
      &
      -\frac{}{\Upgamma(1-b)\Upgamma(a+1-c)}
    ),
    \label{eq:transition_matrix}
  \end{equation}
  to $B_{\iOm{0}}(z) = \mathcal{C} B_{\iOm{\infty}}(z)$ in order to compute the
  explicit monodromy representation ${M}_{\iOm{\infty}}$. In fact, performing
  the loop around the infinity as $z \rightarrow z e^{-i 2 \uppi}$, we find
  \begin{equation*}
    \widetilde{M}_{\iOm{\infty}} = \mqty(
      e^{2\uppi i a} &
      \\
      & e^{2\uppi i b}
    ),
  \end{equation*}
  and finally
  \begin{eqnarray}
    M_{\iOm{\infty}} & = & \mathcal{C} \widetilde{M}_{\iOm{\infty}}
    \mathcal{C}^{-1}.
    \label{eq:monodromy_infty}
  \end{eqnarray}

  \subsection{The Monodromy Factors}

  The stage being set, our task is now to reproduce the monodromies of the
  doubling field in spinor representation \eqref{eq:top_spinor_monodromy} (we
  do not need to consider \eqref{eq:bottom_spinor_monodromy} since they are the
  same monodromies) by taking tensor products of two basis of hypergeometric
  functions: the first basis reproduces the monodromies defined as
  $\mathcal{L}$ and the second one those defined as $\mathcal{R}$.

  In principle there can be several combinations of parameters of the
  hypergeometric function yielding the same monodromies, therefore we consider
  the full solution\footnote{In the following we use only the spinor
  representation and for simplicity we write $\partial \mathcal{X}(z)$ instead
  of $ \partial \mathcal{X}_{(s)}(z)$.} to be a linear superposition of all
  possible contributions:
  \begin{equation}
    \partial \mathcal{X}(z) = \pdv{\upomega_z}{z} \sum\limits_{l,r} c_{lr}
    \partial \mathcal{X}_{l,r}(\upomega_z).
    \label{eq:formal_solution}
  \end{equation}
  Explicitly we write any possible solution in a factorized form as
  \begin{equation}
    \partial \mathcal{X}_{l,r}(\upomega_z) = (-\upomega_z)^{A_{lr}}
    (1-\upomega_z)^{B_{lr}} \mathcal{B}_{\iOm{0},\, l}^{(L)}(\upomega_z)
    \mathcal{B}_{\iOm{0},\, r}^{(R)}(\upomega_z) ^T,
    \label{eq:formal_solution_lr}
  \end{equation}
  where  $l$ and $r$ label the possible parameters associate with the left and
  right hypergeometric. We have also introduced the left basis element
  \begin{equation}
    \begin{split}
      \mathcal{B}_{\iOm{0},\, l}^{(L)}(\upomega_z)
      & = D^{(L)}_l B_{\iOm{0},\,l}^{(L)}(\upomega_z) =
      \\
      & = \mqty(
        F(a_l, b_l; c_l; \upomega_z)
        \\
        K_l^{(L)} (-z)^{(1-c_l)} F(a_l+1-c_l, b_l+1-c_l; 2-c_l; \upomega_z)
      )
    \end{split}
  \end{equation}
  where
  \begin{equation}
    D^{(L)}_l = \mqty(
      1 &
      \\
      & K^{(L)}_l
    ) \in \mathrm{GL}_2(\mathds{C})
  \end{equation}
  is a relative normalization of the two components of each basis. The right
  sector follows in a similar way. These may be different for each solution.
  Notice that the matrices $D^{(L)}_l$ do not fix an absolute factor which is
  contained in $c_{l r}$ but only the normalization of one component of the
  basis with respect to the other.

  After the determination of the possible solutions, we need to select the
  truly independent ones and among them those with a finite action. It will
  turn out actually to be easier to determine the solutions with finite action
  and then verify that they are independent.

  \subsubsection{Fixing the Parameters in the Most Obvious Case}

  We now determine the possible $\partial \mathcal{X}_{l,r}(\upomega_z)$ which
  have the right monodromies. We will do this for the most general
  $\mathrm{SU}(2)$ matrices despite the fact the the ones involved in our
  problem are of a very special form.

  In order to reproduce the monodromies, consider the matrices in
  \eqref{eq:monodromy_zero} and \eqref{eq:monodromy_infty}. We impose:
  {\footnotesize
  \begin{eqnarray}
    &&\begin{cases}
      D^{(L)} M_{\iOm{0}}^{(L)} \left( D^{(L)} \right)^{-1} = e^{-2\uppi i
      \updelta_{\iOm{0}}^{(L)}} \mathcal{L}(\vb{n}_{\iOm{0}})
      \\
      D^{(R)} M_{\iOm{0}}^{(R)} \left( D^{(R)} \right)^{-1}
      =
      e^{-2\uppi i \updelta_{\iOm{0}}^{(R)}} \mathcal{R}^*(\vb{m}_{\iOm{0}})
      =
      e^{-2\uppi i \updelta_{\iOm{0}}^{(R)}}
      \mathcal{R}(\widetilde{\vb{m}}_{\iOm{0}})
      \\
      e^{2\uppi i ( A_{lr} - \updelta_{\iOm{0}}^{(L)} -
      \updelta_{\iOm{0}}^{(R)} )} = 1
      \end{cases},
    \label{eq:parameters_equality_zero}
    \\
    &&\begin{cases}
      D^{(L)} M_{\iOm{\infty}}^{(L)} \left( D^{(L)} \right)^{-1} = e^{-2\uppi i
      \updelta_{\iOm{\infty}}^{(L)}} \mathcal{L}(\vb{n}_{\iOm{\infty}})
      \\
      D^{(R)} M_{\iOm{\infty}}^{(R)} \left( D^{(R)} \right)^{-1}
      =
      e^{-2\uppi i \updelta_{\iOm{\infty}}^{(R)}}
      \mathcal{R}^*(\vb{m}_{\iOm{\infty}})
      =
      e^{-2\uppi i \updelta_{\iOm{\infty}}^{(R)}}
      \mathcal{R}(\widetilde{\vb{m}}_{\iOm{\infty}})
      \\
      e^{2\uppi i ( A_{lr} + B_{lr} - \updelta_{\iOm{\infty}}^{(L)} -
      \updelta_{\iOm{\infty}}^{(R)} )} = 1
    \end{cases},
    \label{eq:parameters_equality_infty}
  \end{eqnarray}
  }
  where we defined
  \begin{eqnarray*}
    \mathcal{L}(\vb{n}_{\iOm{0}})
    & = &
    \mathcal{L}_{\iD{\overline{t}-1,\overline{t}}}
    =
    \UL(\vb{n}_{\iD{\overline{t}}})
    \UL^{\dagger}(\vb{n}_{\iD{\overline{t}-1}}),
    \\
    \mathcal{L}(\vb{n}_{\iOm{\infty}})
    & = &
    \mathcal{L}_{\iD{\overline{t},\overline{t}+1}}
    =
    \UL(\vb{n}_{\iD{\overline{t}+1}})
    \UL^{\dagger}(\vb{n}_{\iD{\overline{t}}}),
    \\
    \mathcal{R}(\vb{m}_{\iOm{0}})
    & = &
    \mathcal{R}_{\iD{\overline{t}-1,\overline{t}}}
    =
    \UR(\vb{n}_{\iD{\overline{t}}})
    \UR^{\dagger}(\vb{n}_{\iD{\overline{t}-1}}),
    \\
    \mathcal{R}(\vb{m}_{\iOm{\infty}})
    & = &
    \mathcal{R}_{\iD{\overline{t},\overline{t}+1}}
    =
    \UR(\vb{n}_{\iD{\overline{t}+1}})
    \UR^{\dagger}(\vb{n}_{\iD{\overline{t}}}).
  \end{eqnarray*}
  That is we highlighted the dependence on the parameters of the
  $\mathrm{SU}(2)$ matrices on the $\upomega_t$ interaction point instead of
  the branes between which the interaction develops.

  Notice that the range of definition of $\updelta_{\iOm{0}}^{(L)}$ is
  \begin{equation*}
    \upalpha \le \updelta_{\iOm{0}}^{(L)} \le  \upalpha +\frac{1}{2},
  \end{equation*}
  i.e. the width of the range is only $\frac{1}{2}$ and not $1$ as one would
  naively expect since $e^{i 4 \uppi \updelta_{\iOm{0}}^{(L)}}$ is the
  determinant of the right hand side of the first equation in
  \eqref{eq:parameters_equality_zero}. We will choose $\upalpha = 0$ for
  simplicity. The same is true for all the other additional parameters
  $\updelta_{\iOm{0}}^{(R)}$ and $\updelta_{\iOm{\infty}}^{(L,\,R)}$.

  As we are interested in relative rotations of the branes, we can fix the
  rotation in $\upomega_{\overline{t}-1} = 0$ to be in the maximal torus of
  $\mathrm{SU}_L(2) \times \mathrm{SU}_R(2)$ without loss of generality. Stated
  otherwise, since we have two independent groups we can choose the orientation
  of both the vectors $\vb{n}_{\iOm{0}}$ and $\vb{m}_{\iOm{0}}$. In particular
  we set:
  \begin{eqnarray}
    \vb{n}_{\iOm{0}} & = & ( 0, 0, n_{\iOm{0}}^3 ) \in \mathds{R}^3
    \qq{where} 0 < n_{\iOm{0}}^3 < \frac{1}{2},
    \label{eq:maximal_torus_left}
    \\
    \widetilde{\vb{m}}_{\iOm{0}} & = & ( 0, 0, -m_{\iOm{0}}^3 ) \in
    \mathds{R}^3 \qq{where} 0 < m_{\iOm{0}}^3 < \frac{1}{2},
    \label{eq:maximal_torus_right}
  \end{eqnarray}
  where the case $n_{\iOm{0}}^3=0$ is excluded because we consider a non
  trivial rotation. We take the parameters of the rotation in $\upomega =
  \infty$ to be the most general
  \begin{eqnarray*}
    \vb{n}_{\iOm{\infty}} & = & ( n_{\iOm{\infty}}^1, n_{\iOm{\infty}}^2,
    n_{\iOm{\infty}}^3 ),
    \\
    \widetilde{\vb{m}}_{\iOm{\infty}} & = & ( -m_{\iOm{\infty}}^1,
    m_{\iOm{\infty}}^2, -m_{\iOm{\infty}}^3 ),
  \end{eqnarray*}
  even if we can set $n_{\iOm{\infty}}^2=0$ (and also $m_{\iOm{\infty}}^2=0$)
  because, after fixing $\vb{n}_{\iOm{0}}$, we can still perform $U(1)$
  rotations which leave it invariant but mix $n_{\iOm{\infty}}^1,
  n_{\iOm{\infty}}^2$. We nevertheless keep the general expression in order to
  check our computations.

  As we show in Appendix~\ref{sec:parameters}, solving
  \eqref{eq:parameters_equality_zero} and \eqref{eq:parameters_equality_infty}
  links the parameters of the hypergeometric function to the parameter of the
  rotations, thus reproducing the boundary conditions of the intersecting
  D-branes through the non trivial monodromies of the basis. We find:
  \begin{eqnarray*}
    a_l^{(L)} & = & n_{\iOm{0}} + (-1)^{f^{(L)}} n_{\iOm{1}} + n_{\iOm{\infty}}
    + \mathfrak{a}^{(L)}_l \qq{where} \mathfrak{a}^{(L)}_l \in \mathds{Z},
    \\
    b_l^{(L)} & = & n_{\iOm{0}} + (-1)^{f^{(L)}} n_{\iOm{1}} - n_{\iOm{\infty}}
    + \mathfrak{b}^{(L)}_l \qq{where} \mathfrak{b}^{(L)}_l \in \mathds{Z},
    \\
    c_l^{(L)} & = & 2 n_{\iOm{0}} + \mathfrak{c}^{(L)}_l \qq{where}
    \mathfrak{c}^{(L)}_l \in \mathds{Z},
    \\
    \updelta_{\iOm{0}}^{(L)} & = & n_{\iOm{0}},
    \\
    \updelta_{\iOm{\infty}}^{(L)} & = & -n_{\iOm{0}} - (-1)^{f^{(L)}}
    n_{\iOm{1}},
    \\
    K^{(L)}_l & = & -\frac{1}{2 \uppi^2}
                    \Upgamma(1-a_l^{(L)}) \Upgamma(1-b_l^{(L)})
                    \Upgamma(a_l^{(L)}+1-c_l^{(L)})
                    \Upgamma(b_l^{(L)}+1-c_l^{(L)}) \times
    \\
    & \times & \sin(\uppi c_l^{(L)}) \sin(\uppi (a_l^{(L)}-b_l^{(L)}) )
    \frac{n^1_{\iOm{\infty}}+ i~n^2_{\iOm{\infty}}}{n_{\iOm{\infty}}},
  \label{eq:K_factor_value}
  \end{eqnarray*}
  where $f^{(L)} \in \left\lbrace 0,\, 1 \right\rbrace$ and we also introduced
  the norm $n_{\iOm{1}}$ of parameters of the rotation around
  $\upomega_{\overline{t}+1} = 1$, that is $\vb{n}_{\iOm{1}}$, which depends on
  the other parameters through:
  \begin{equation}
    \cos(2\uppi n_{\iOm{1}}) = \cos(2\uppi n_{\iOm{0}}) \cos(2\uppi
    n_{\iOm{\infty}}) - \sin(2\uppi n_{\iOm{0}}) \sin(2\uppi n_{\iOm{\infty}})
    \frac{n_{\iOm{\infty}}^3}{n_{\iOm{\infty}}}.
    \label{eq:dependent_monodromy_main_text}
  \end{equation}
  This relation follows from \eqref{eq:monodromy_relations} for the monodromy
  $M_{\iOm{1}}^+ = M_{\iOm{0}}^{-1} M_{\iOm{\infty}}^{-1}$ and the standard
  composition rule for the $\mathrm{SU}(2)$ parameters given in
  \eqref{eq:product_in_SU2}. The same relations for the right sector follow
  under the exchange $(L) \leftrightarrow (R)$ and $\vb{n} \leftrightarrow
  \widetilde{\vb{m}}$.

  The other parameters $A_{lr}$ and $B_{lr}$ are then a consequence of the
  previous results and the equations \eqref{eq:parameters_equality_zero} and
  \eqref{eq:parameters_equality_infty}:
  \begin{eqnarray*}
    A_{lr} & = & n_{\iOm{0}} + m_{\iOm{0}} + \mathfrak{A}_{lr},
    \\
    B_{lr} & = & (-1)^{f^{(L)}} n_{\iOm{1}} + (-1)^{f^{(R)}} m_{\iOm{1}} +
    \mathfrak{B}_{lr}
  \end{eqnarray*}
  where
  $\mathfrak{A}_{lr} \in \mathds{Z}$ and $\mathfrak{B}_{lr} \in \mathds{Z}$.

  \subsubsection{Solutions with Different \texorpdfstring{$f^{(L)}$}{fL} and
  \texorpdfstring{$f^{(R)}$}{fR} Are the Same}

  As we see from the equations above the parameters of the hypergeometric
  function are still affected by some ambiguities: the choice of $f^{(L)}$ and
  $f^{(R)}$ seems an arbitrary decision leading to an undefined solution.
  However we can use the properties of the hypergeometric functions to show
  that any choice of their values does not affect the final result.
  Specifically, we could choose to start with certain values but we can recover
  the others through:
  \begin{equation*}
    \mathrm{P}\left\lbrace \mqty{
      0   & 1     & \infty & \\
      0   & 0     & a      & z \\
      1-c & c-a-b & b      &
    } \right\rbrace
    = (1-z)^{c-a-b} \mathrm{P}\left\lbrace \mqty{
      0   & 1     & \infty & \\
      0   & 0     & c-b    & z \\
      1-c & a+b-c & c-a    &
    } \right\rbrace,
  \end{equation*}
  where $\mathrm{P}$ is the Papperitz-Riemann symbol for the hypergeometric
  functions. This way we can assign any of the possible values to $f^{(L)}$ and
  $f^{(R)}$ and then recover the other identifying:
  \begin{eqnarray*}
    f^{(L)}{}' & = & 1 + f^{(L)} \qq{mod} 2
    \\
    \mathfrak{a}_l' & = & \mathfrak{c}_l - \mathfrak{b}_l,
    \\
    \mathfrak{b}_l' & = & \mathfrak{c}_l - \mathfrak{a}_l,
    \\
    \mathfrak{c}_l' & = &\mathfrak{c}_l,
    \\
    \mathfrak{A}_{lr}' & = & \mathfrak{A}_{lr},
    \\
    \mathfrak{B}_{lr}' & = & \mathfrak{B}_{lr} - \mathfrak{a}^{(L)}_l -
    \mathfrak{a}^{(R)}_r - \mathfrak{b}^{(L)}_l - \mathfrak{b}^{(R)}_r +
    \mathfrak{c}^{(L)}_l + \mathfrak{c}^{(R)}_r,
  \end{eqnarray*}
  and similarly the parameters of the right sector. This means that the
  choice of $f^{(L,R)}$ is simply a convenient relabeling of parameters. In
  what follows we choose $f^{(L)} = f^{(R)} = 0$ for simplicity.

  As previously stated, in order to get a well defined solution we must impose
  some constraints on the hypergeometric parameters. Specifically we require:
  \begin{eqnarray*}
    c_l^{(L)} & \not\in & \mathds{Z},
    \\
    a_l^{(L)} + b_l^{(L)} & \not\in & \mathds{Z} + \frac{1}{2}.
  \end{eqnarray*}

  All these relations link the parameters of the hypergeometric function to the
  monodromies associated to the boundary conditions of the intersecting
  D-branes. They are however more general than actually needed: the number of
  parameters necessary to fix our configuration is $6$ (i.e. $n_{\iOm{0}}^3,
  n_{\iOm{\infty}}^1, n_{\iOm{\infty}}^3$ and $m_{\iOm{0}}^3,
  m_{\iOm{\infty}}^1, m_{\iOm{\infty}}^3$), since as noticed before we can
  always fix $n_{\iOm{\infty}}^2 = m_{\iOm{\infty}}^2 = 0$. This is a
  consequence of the fact that all parameters depend on the moduli, exception
  made for $K^{(L)}$ and $K^{(R)}$ which depend on $n_{\iOm{\infty}}^1 + i
  n_{\iOm{\infty}}^2$ and $m_{\iOm{\infty}}^1 + i m_{\iOm{\infty}}^2$.
  Performing a $SU_L(2)$ and $SU_R(2)$ rotation around the third axis and a
  shift of the parameters $\updelta_{\iOm{\infty}}$, the phases of
  $K$ can then be made to vanish.

  \subsubsection{The Importance of the Normalization Factors
  \texorpdfstring{$K$}{K}}
  Using the Papperitz-Riemann symbol the solutions found can be symbolically
  written as
  {\small
  \begin{equation}
    \begin{split}
    & (-\upomega)^\mathfrak{A} (1-\upomega)^\mathfrak{B} \times
    \\
    & \times \mathrm{P}\left\lbrace
      \mqty{
          0  & 1 & \infty &
          \\
          n_{\iOm{0}} & n_{\iOm{1}} & n_{\iOm{\infty}}+\mathfrak{a}^{(L)}
          & \upomega
          \\
          -n_{\iOm{0}}+1 -\mathfrak{c}^{(L)} &
          -n_{\iOm{1}}-\mathfrak{a}^{(L)}-\mathfrak{b}^{(L)}+\mathfrak{c}^{(L)}
          & -n_{\iOm{\infty}}+\mathfrak{b}^{(L)} &
      }
    \right\rbrace \times
    \\
    & \times \mathrm{P}\left\lbrace
      \mqty{
          0 & 1 & \infty &
          \\
          m_{\iOm{0}}   & m_{\iOm{1}} & m_{\iOm{\infty}}+\mathfrak{a}^{(R)}
          & \upomega
          \\
          -m_{\iOm{0}}+1 -\mathfrak{c}^{(R)} &
          -m_{\iOm{1}}-\mathfrak{a}^{(R)}-\mathfrak{b}^{(R)}+\mathfrak{c}^{(R)}
          & -m_{\iOm{\infty}}+\mathfrak{b}^{(R)} &
      }
      \right\rbrace.
    \end{split}
  \label{eq:symbolic_solutions_using_P}
  \end{equation}
  }
  This is exactly what one would have expected but the parameters $K$ cannot be
  guessed from the $\mathrm{P}$ symbol. They nevertheless play a very important
  role for the consistency of the solution.

  Hypergeometric functions can be connected by relations between contiguous
  functions. It is indeed possible to show that any hypergeometric function
  $F(a + \mathfrak{a}, b + \mathfrak{b}; c + \mathfrak{c}; z)$ can be written
  as a combination of $F(a,b;c;z)$ and any of its contiguous functions
  \cite{NIST:DLMF}. For example we could consider:
  \begin{equation}
    \small
    F(a + \mathfrak{a}, b + \mathfrak{b}; c + \mathfrak{c}; z) = h_1(a,b,c;z)
    F(a+1,b;c;z) + h_2(a,b,c;z) F(a,b;c;z),
    \label{eq:reduction_F_F+}
  \end{equation}
  where $h_1(a,b,c;z)$ and $h_2(a,b,c;z)$ are in $C[\frac{1}{1-z}, \frac{1}{z},
  z]$, i.e. they are finite sums of integer (both positive and negative) powers
  of $z$ and negative powers of $1-z$. For simplicity let:
  \begin{eqnarray*}
    F & = &  F(a,b;c;z),
    \\
    F(a+k) & = &  F(a+k,b;c;z),
    \\
    F(b+k) & = &  F(a,b+k;c;z),
    \\
      & \dots &
  \end{eqnarray*}
  Similarly we use the shorthand notation\footnote{Here we are a sloppy in
  writing $K_{a,b,c}$ since it depends on a phase which is not a function of
  $a,b,c$. See \eqref{eq:K_factor_value}  and \eqref{eq: n12+n22}.} for the
  vector:
  \begin{equation}
    \mathcal{B}_{\iOm{0}}(a, b, c; z) =
    \mqty(
      F(a, b; c; z)
      \\
      K_{a,b,c} (-z)^{(1-c)} F(a+1-c, b+1-c; 2-c; z)
    ).
  \end{equation}
  Then we can algorithmically use relations such as
  \begin{equation}
    \begin{split}
      (c-a) F(a-1) + (2a-c+(b-a)z) F - a(1-z) F(a+1) & = 0.
      \\
      (b-a) F + a F(a+1) - b F(b+1) & = 0,
      \\
      (c-a-b) F + a (1-z) F(a+1) - (c-b) F(b-1) & = 0,
      \\
       (a+(b-c)z)F - a (1-z) F(a+1) + (c-a)(c-b)z F(c+1) & = 0,
      \\
      (c-a-1) F + a F(a+1) -  F(c-1) & = 0,
    \end{split}
    \label{eq:contiguous_functions}
  \end{equation}
  in order to eliminate unwanted integer factors from each parameter and
  to keep only $F$ and any of its contiguous functions.

  Now $\mathcal{B}_{\iOm{0}}$, considered as whole and made of two independent
  hypergeometric functions, is a basis element for the possible solutions of
  the classical and quantum string e.o.m.. Using any relation in
  \eqref{eq:contiguous_functions} we can change $a, b$ or $c$ by $\pm 1$ in a
  coherent way in both hypergeometric functions, then the result must still be
  a linear combination of solutions $\mathcal{B}_{\iOm{0}}$. For example from
  the first equation in \eqref{eq:contiguous_functions} we expect:
  \begin{equation}
    (c-a) \mathcal{B}_{\iOm{0}}(a-1) + (2a-c+(b-a)z) \mathcal{B}_{\iOm{0}} -
    a(1-z)\mathcal{B}_{\iOm{0}}(a+1) = 0,
  \end{equation}
  which can be used to lower and rise $a$. This relation holds only because of
  the presence of $K$. In fact the coefficients in this equation are
  exactly equal to those in the relation for $F$ for the first component of
  $\mathcal{B}_{\iOm{0}}$ but this is a non trivial fact for the second
  component where the factor $K$ plays a fundamental role.

  In a similar way, even if more complicated to prove, the relation which is
  needed to lower $c$ which reads:
  \begin{equation}
    (a-c)(b-c) \mathcal{B}_{\iOm{0}}(c+1) + (a+(b-c)z) \mathcal{B}_{\iOm{0}} -
    a(1-z) \mathcal{B}_{\iOm{0}}(a+1) = 0.
  \end{equation}

  \subsection{Constraints from the Finite Euclidean Action}

  In the previous section we found all the most obvious possible
  solutions to the classical string e.o.m. However we should look for a
  solution with finite action, thus restricting our attention to such property.

In principle it would be obvious to use
\eqref{eq:contiguous_functions} to restrict the possible arbitrary
integers entering the solution to
  \begin{eqnarray*}
    \mathfrak{a}^{(L)} & \in & \left\lbrace -1, 0 \right\rbrace,
    \\
    \mathfrak{b}^{(L)} & = & 0,
    \\
    \mathfrak{c}^{(L)} & = & 0,
  \end{eqnarray*}
  and analogously for the right counterparts and then to use
  \eqref{eq:reduction_F_F+} to write the possible solution as
  \begin{equation}
    \begin{split}
      \partial \mathcal{X}(z) &=
      \pdv{\upomega_z}{z} (-\upomega_z)^{n_{\iOm{0}}+m_{\iOm{0}}}
      (1-\upomega_z)^{n_{\iOm{1}}+m_{\iOm{1}}} \times
      \\
      &\times \sum\limits_{\mathfrak{a}^{(L,R)} \in \left\lbrace -1, 0
      \right\rbrace} h(\upomega_z, \mathfrak{a}^{(L,R)}) \times
      \\
      &\times
      \mathcal{B}_{\iOm{0}}^{(L)}(a^{(L)}+\mathfrak{a}^{(L)},b,c;\upomega_z)
      \left(
      \mathcal{B}_{\iOm{0}}^{(R)}(a^{(R)}+\mathfrak{a}^{(R)},b,c;\upomega_z)
      \right)^T.
    \end{split}
    \label{eq:doubling_field_expansion}
  \end{equation}
  We should finally find an explicit form for $h(\upomega_z,
  \mathfrak{a}^{(L,R)})$ which yield a finite action.

  It turns however out to be by far simpler to use the symbolic
  solution \eqref{eq:symbolic_solutions_using_P} to find the possible
  basis elements with finite action.
  Actually finding the possible solutions with finite action can be recast to
  the issue of finding finite solution, i.e. such that the field $\partial
  \mathcal{X}(z)$ is finite by itself. The latter formulation is by far simpler
  than the former since it is
  linear while the former is quadratic.
  From \eqref{eq:action_doubling_fields_spinor_representation}
  it is clear that the action can be expressed as the sum of the
  product of any possible couple of elements of the
  over-complete expansion of the solution \eqref{eq:formal_solution}.
  Therefore we must exam all the possible behaviors of
  any couple
  $\partial \mathcal{X}_{l_1 r_1}(z) \overline{\partial}
  \mathcal{X}_{l_2 r_2}(\overline{z})$.
  In proximity of any singular point the behavior of any element of solution
  \eqref{eq:formal_solution} can be easily read from its symbolic
  representation given by \eqref{eq:symbolic_solutions_using_P} and it
  is of the form:
  \begin{equation*}
    \partial \mathcal{X}(z) \sim \upomega_t^{C_t} \mqty(
    \upomega_t^{k_{t 1}} \\
    \upomega_t^{k_{t 2}} ) \mqty( \upomega_t^{h_{t 1}} &
    \upomega_t^{h_{t 2}} ) \qfor
    \upomega_z \to \upomega_t.
  \end{equation*}
  It is then easy to verify that imposing the convergence of the
  action both at finite and infinite
  intersection points yields the same constraints as imposing the
  convergence at any point of the classical solution
  (in spinor representation as follows from \eqref{eq:spinor_doubling_trick})
  \begin{equation}
    \small
    X_{(s)}(u,\overline{u}) = f_{\iD{\overline{t}-1}\,(s)}
    + \int\limits_{x_{\overline{t}-1}}^u \dd{u'} \partial \mathcal{X}(u')
    +
    \UL^{\dagger}(\vb{n}_{\iD{\overline{t}}})
    ~
    \int\limits_{x_{\overline{t}-1}}^{\bar u} \dd{\bar u'} \partial
    \mathcal{X}(\bar u')
    ~
    \UR(\vb{m}_{\iD{\overline{t}}}),
  \label{eq:classical_solution}
  \end{equation}
  where $f_{\overline{t}-1\,(s)}=f^I_{\overline{t}-1} \uptau_I$. They are:
  \begin{equation}
    \begin{split}
      C_t+ k_{t i}+h_{t j}>-1, ~~~~i,j\in\{1,2\},&~~\upomega_t\in\{0, 1\},
      \\
      C_t+ k_{t i}+h_{t j}<-1, ~~~~i,j\in\{1,2\},&~~\upomega_t=\infty.
    \end{split}
  \label{eq:constraints_finite_X}
  \end{equation}
  To explain the approach in the easiest setup let us consider the case where
  the right rotation is trivial. In this case
  \eqref{eq:symbolic_solutions_using_P} becomes
  \begin{equation}
    \begin{split}
      & (-\upomega)^\mathfrak{A} (1-\upomega)^\mathfrak{B} \times
      \\
      &\times \mathrm{P}\left\lbrace
      \mqty{
        0 & 1 & \infty &
        \\
        n_{\iOm{0}}   & n_{\iOm{1}} & n_{\iOm{\infty}}+\mathfrak{a}^{(L)} &
        \upomega
        \\
        -n_{\iOm{0}}+1 -\mathfrak{c}^{(L)} &
        -n_{\iOm{1}}-\mathfrak{a}^{(L)}-\mathfrak{b}^{(L)}+\mathfrak{c}^{(L)}
        & -n_{\iOm{\infty}}+\mathfrak{b}^{(L)}&}
        \right\rbrace.
    \end{split}
  \end{equation}
  Then it is easy to see that the only possible solution compatible with
  \eqref{eq:constraints_finite_X} is
   \begin{equation}
     \mathrm{P}\left\lbrace
         \mqty{ 0        & 1       & \infty      & \\
           n_{\iOm{0}}-1   & n_{\iOm{1}}-1 & n_{\iOm{\infty}}+1 & \upomega \\
           -n_{\iOm{0}}
            &
            -n_{\iOm{1}}
            & -n_{\iOm{\infty}}+2
            &}
        \right\rbrace
              ,
      \label{eq:X_solution_pure_L}
   \end{equation}
   i.e. $\mathfrak{a}^{(L)}=-1$, $\mathfrak{b}^{(L)}=0$,
   $\mathfrak{c}^{(L)}=0$, $\mathfrak{A}=-1$ and  $\mathfrak{B}=-1$. In the
   general case the situation is more complicated one could think that  taking
   the product \eqref{eq:X_solution_pure_L} and the corresponding solution for
   the right sector would yield the answer. Unfortunately it is not the case
   since for $\upomega=0$ we get $C_{0}+h_{0 1}+k_{0
   1}=n_{\iOm{0}}+m_{\iOm{0}}-2<-1$. To find the solution however we start from
   such  product and we try to move integer factors between indices and between
   the left and right solutions. For each possible case the solution is unique
   and it is given by
\begin{enumerate}
\item $n_{\iOm{0}}>m_{\iOm{0}}$ and $n_{\iOm{1}}>m_{\iOm{1}}$
   \begin{align}
       \mathrm{P}\left\lbrace
         \mqty{
           0        & 1         & \infty      & \\
           n_{\iOm{0}}-1 & n_{\iOm{1}}-1 & n_{\iOm{\infty}}+1
           & \upomega \\
           -n_{\iOm{0}}  & -n_{\iOm{1}}  & -n_{\iOm{\infty}}+2
            &}
              \right\rbrace
              ~
       \mathrm{P}\left\lbrace
         \mqty{
           0         & 1         & \infty      & \\
           m_{\iOm{0}}    & m_{\iOm{1}} & m_{\iOm{\infty}}
           & \upomega \\
           -m_{\iOm{0}}+1 & -m_{\iOm{1}} & -m_{\iOm{\infty}}+1
            &}
        \right\rbrace
              .
              \label{eq:X_solution>>}
   \end{align}
 \item $n_{\iOm{0}}>m_{\iOm{0}}$, $n_{\iOm{1}}<m_{\iOm{1}}$ and $n_{\iOm{\infty}}>m_{\iOm{\infty}}$
   \begin{align}
       \mathrm{P}\left\lbrace
         \mqty{
           0        & 1         & \infty      & \\
           n_{\iOm{0}}-1 & n_{\iOm{1}} & n_{\iOm{\infty}}
           & \upomega \\
           -n_{\iOm{0}}  & -n_{\iOm{1}} & -n_{\iOm{\infty}}+2
            &}
              \right\rbrace
              ~
       \mathrm{P}\left\lbrace
         \mqty{
           0         & 1         & \infty      & \\
           m_{\iOm{0}}  & m_{\iOm{1}}-1 & m_{\iOm{\infty}}+1
                                              & \upomega
     \\
           -m_{\iOm{0}} & -m_{\iOm{1}}   & -m_{\iOm{\infty}}+1
            &}
        \right\rbrace
              .
              \label{eq:X_solution><>}
\end{align}
 \item $n_{\iOm{0}}>m_{\iOm{0}}$, $n_{\iOm{1}}<m_{\iOm{1}}$ and $n_{\iOm{\infty}}<m_{\iOm{\infty}}$
   \begin{align}
       \mathrm{P}\left\lbrace
         \mqty{
           0        & 1         & \infty      & \\
           n_{\iOm{0}}-1 & n_{\iOm{1}} & n_{\iOm{\infty}}+1
           & \upomega \\
           -n_{\iOm{0}}  & -n_{\iOm{1}} & -n_{\iOm{\infty}}+1
            &}
              \right\rbrace
              ~
       \mathrm{P}\left\lbrace
         \mqty{
           0         & 1         & \infty      & \\
           m_{\iOm{0}}  & m_{\iOm{1}}-1 & m_{\iOm{\infty}}
           & \upomega \\
           -m_{\iOm{0}} & -m_{\iOm{1}}   & -m_{\iOm{\infty}}+2
            &}
        \right\rbrace
              .
              \label{eq:X_solution><<}
\end{align}
   %
 \item $n_{\iOm{0}}<m_{\iOm{0}}$, $n_{\iOm{1}}>m_{\iOm{1}}$ and $n_{\iOm{\infty}}>m_{\iOm{\infty}}$
   \begin{align}
       \mathrm{P}\left\lbrace
         \mqty{
           0        & 1         & \infty      & \\
           n_{\iOm{0}} & n_{\iOm{1}}-1 & n_{\iOm{\infty}}
           & \upomega \\
           -n_{\iOm{0}} & -n_{\iOm{1}}   & -n_{\iOm{\infty}}+2
            &}
              \right\rbrace
              ~
       \mathrm{P}\left\lbrace
         \mqty{
           0         & 1         & \infty      & \\
           m_{\iOm{0}}-1  & m_{\iOm{1}} & m_{\iOm{\infty}}+1
           & \upomega \\
           -m_{\iOm{0}}   & -m_{\iOm{1}}   & -m_{\iOm{\infty}}+1
            &}
        \right\rbrace
              .
              \label{eq:X_solution<>>}
\end{align}
 \item $n_{\iOm{0}}<m_{\iOm{0}}$, $n_{\iOm{1}}>m_{\iOm{1}}$ and $n_{\iOm{\infty}}<m_{\iOm{\infty}}$
   \begin{align}
       \mathrm{P}\left\lbrace
         \mqty{
           0        & 1         & \infty      & \\
           n_{\iOm{0}} & n_{\iOm{1}}-1 & n_{\iOm{\infty}}+1
           & \upomega \\
           -n_{\iOm{0}} & -n_{\iOm{1}}   & -n_{\iOm{\infty}}+1
            &}
              \right\rbrace
              ~
       \mathrm{P}\left\lbrace
         \mqty{
           0         & 1         & \infty      & \\
           m_{\iOm{0}}-1  & m_{\iOm{1}}   & m_{\iOm{\infty}}
           & \upomega \\
           -m_{\iOm{0}}   & -m_{\iOm{1}}   & -m_{\iOm{\infty}}+2
            &}
        \right\rbrace
              .
              \label{eq:X_solution<><}
\end{align}
 \item $n_{\iOm{0}}<m_{\iOm{0}}$, $n_{\iOm{1}}<m_{\iOm{1}}$
   \begin{align}
       \mathrm{P}\left\lbrace
         \mqty{
           0        & 1         & \infty      & \\
           n_{\iOm{0}} & n_{\iOm{1}}     & n_{\iOm{\infty}}
           & \upomega \\
           -n_{\iOm{0}} & -n_{\iOm{1}}   & -n_{\iOm{\infty}}+1
            &}
              \right\rbrace
              ~
       \mathrm{P}\left\lbrace
         \mqty{
           0         & 1         & \infty      & \\
           m_{\iOm{0}}-1  & m_{\iOm{1}}-1 & m_{\iOm{\infty}}+1
           & \upomega \\
           -m_{\iOm{0}} & -m_{\iOm{1}}   & -m_{\iOm{\infty}}+2
            &}
        \right\rbrace
              .
              \label{eq:X_solution<<}
\end{align}
  \end{enumerate}
We can summarize the parameters which follows from the previous list and enter
the solution in Table~\ref{table:coeffs_k}, where the obvious symmetry in the
exchange of $n$ and $m$ becomes manifest.

\begin{center}
  \captionof{table}{Integer shifts entering the hypergeometric
    parameters.}
  \label{table:coeffs_k}
  \begin{tabular}{
  c|c|c 
  |c c 
  |c c c 
  |c c c 
  }
  & &
& $\mathfrak{A}$ & $\mathfrak{B}$
& $\mathfrak{a}^{(L)}$ & $\mathfrak{b}^{(L)}$ & $\mathfrak{c}^{(L)}$
  & $\mathfrak{a}^{(R)}$ & $\mathfrak{b}^{(R)}$ & $\mathfrak{c}^{(R)}$
\\
  \hline
  $n_{\iOm{0}}>m_{\iOm{0}}$ &  $n_{\iOm{1}}>m_{\iOm{1}}$ & $n_{\iOm{\infty}}\lessgtr m_{\iOm{\infty}}$
  & -1 & -1
  & -1 & 0 & 0
  & 0 & +1 & +1
\\
  \hline
  $n_{\iOm{0}}>m_{\iOm{0}}$ &  $n_{\iOm{1}}<m_{\iOm{1}}$ & $n_{\iOm{\infty}}>m_{\iOm{\infty}}$
  & -1 & -1
  & -1 & +1 & 0
  & 0 & 0 & +1
  \\
  \hline
  $n_{\iOm{0}}>m_{\iOm{0}}$ &  $n_{\iOm{1}}<m_{\iOm{1}}$ & $n_{\iOm{\infty}}<m_{\iOm{\infty}}$
  & -1 & -1
  & 0 & 0 & 0
  & -1 & +1 & +1
\\
  \hline
  $n_{\iOm{0}}<m_{\iOm{0}}$ &  $n_{\iOm{1}}>m_{\iOm{1}}$ & $n_{\iOm{\infty}}>m_{\iOm{\infty}}$
  & -1 & -1
  & -1 & +1 & +1
  & 0 & 0 & 0
\\
  \hline
  $n_{\iOm{0}}<m_{\iOm{0}}$ &  $n_{\iOm{1}}>m_{\iOm{1}}$ & $n_{\iOm{\infty}}<m_{\iOm{\infty}}$
  & -1 & -1
  & 0 & 0 & +1
  & -1 & +1 & 0
\\
  \hline
  $n_{\iOm{0}}<m_{\iOm{0}}$ &  $n_{\iOm{1}}<m_{\iOm{1}}$ & $n_{\iOm{\infty}}\lessgtr m_{\iOm{\infty}}$
  & -1 & -1
  & 0 & +1 & +1
  & -1 & 0 & 0
\\
\end{tabular}
\end{center}

\subsection{The Basis of Solutions}
\label{sec:true_basis}

In the previous section we have produced one solution for each possible
ordering of the $n_{\iOm{t}}$ with respect to $m_{\iOm{t}}$.
This seems the end of the story but there are actually other solutions
and they are connected to the $\mathds{Z}_2$ in the isomorphism between
$\mathrm{SO}(4)$ and $\mathrm{SU}(2)\times \mathrm{SU}(2) /\mathds{Z}_2$.
Given any solution which is fixed by
$(\vb n_{\iOm{0}}, \vb n_{\iOm{1}}, \vb n_{\iOm{\infty}})\oplus
(\vb m_{\iOm{0}}, \vb m_{\iOm{1}}, \vb m_{\iOm{\infty}})$
we can replace any couple of $\vb n$ and $\vb m$ by
$\widehat{\vb n}$ and $\widehat{\vb m}$ and produce an apparently new solution.
For example we could consider
$(\widehat{\vb n}_{\iOm{0}}, \widehat{\vb n}_{\iOm{1}}, \vb n_{\iOm{\infty}})\oplus
(\vb m_{\iOm{0}}, \widehat{\vb m}_{\iOm{1}}, \widehat{\vb m}_{\iOm{\infty}})$.
The necessity of changing a couple is because the monodromies are
constrained by \eqref{eq:monodromy_relations}.
On the other hand the previous substitution would change the $\mathrm{SO}(4)$ in both
$\upomega=0$ and $\upomega=\infty$: it does not represent a new solution.
We are left therefore with three possibilities besides the original one:
\begin{align}
(\widehat{\vb n}_{\iOm{0}}, \widehat{\vb n}_{\iOm{1}}, \vb n_{\iOm{\infty}})&\oplus
  (\widehat{\vb m}_{\iOm{0}}, \widehat{\vb m}_{\iOm{1}}, \vb m_{\iOm{\infty}}),
  \nonumber\\
(\widehat{\vb n}_{\iOm{0}}, \vb n_{\iOm{1}}, \widehat{\vb n}_{\iOm{\infty}})&\oplus
(\vb m_{\iOm{0}}, \widehat{\vb m}_{\iOm{1}}, \widehat{\vb m}_{\iOm{\infty}}),
  \nonumber\\
  (\widehat{\vb n}_{\iOm{0}}, \vb n_{\iOm{1}}, \widehat{\vb n}_{\iOm{\infty}})&\oplus
 (\vb m_{\iOm{0}}, \widehat{\vb m}_{\iOm{1}}, \widehat{\vb m}_{\iOm{\infty}})
 .
\end{align}
Finally we want to gauge fix the $\mathds{Z}_2$ by letting
$\vb n_{\iOm{0}}^3 ,\vb m_{\iOm{0}}^3 >0$ as required by
\eqref{eq:maximal_torus_left} and \eqref{eq:maximal_torus_right}.
This eliminates the first two possibilities.
We are therefore left with two possible solutions
\begin{align}
  (\vb n_{\iOm{0}}, \vb n_{\iOm{1}}, \vb n_{\iOm{\infty}})&\oplus
(\vb m_{\iOm{0}}, \vb m_{\iOm{1}}, \vb m_{\iOm{\infty}}),
\nonumber\\
  (\vb n_{\iOm{0}}, \widehat{\vb n}_{\iOm{1}}, \widehat{\vb n}_{\iOm{\infty}})&\oplus
  (\vb m_{\iOm{0}}, \widehat{\vb m}_{\iOm{1}}, \widehat{\vb m}_{\iOm{\infty}}),
\end{align}
that is the original one and one which is obtained by acting with a
parity-like operator $P_2$ on the rotation parameters at $\upomega=1, \infty$
on both left and right sector at the same time. In order to accept it as a
further possible solution, we should now verify its independence with respect
to the first one.

Actually looking to Table~\ref{table:coeffs_k} we see that there are
only two different cases up to left-right symmetry.
The first case is
\begin{align}
  \Big\{
  (n_{\iOm{0}}>m_{\iOm{0}}, n_{\iOm{1}}>m_{\iOm{1}}, n_{\iOm{\infty}}> m_{\iOm{\infty}})
  ~,~
  (n_{\iOm{0}}>m_{\iOm{0}}, \hat n_{\iOm{1}}< \hat m_{\iOm{1}},
  \hat n_{\iOm{\infty}}< \hat m_{\iOm{\infty}})
  \Big\},
\end{align}
which is mapped to
\begin{align}
  \Big\{
  (n_{\iOm{0}}< m_{\iOm{0}}, n_{\iOm{1}}< m_{\iOm{1}}, n_{\iOm{\infty}}< m_{\iOm{\infty}})
  ~,~
  (n_{\iOm{0}}<m_{\iOm{0}}, \hat n_{\iOm{1}}>\hat m_{\iOm{1}},
  \hat n_{\iOm{\infty}}>\hat m_{\iOm{\infty}})
  \Big\},
\end{align}
by the left-right symmetry.
The second one is
\begin{align}
  \Big\{
  (n_{\iOm{0}}>m_{\iOm{0}}, n_{\iOm{1}}>m_{\iOm{1}}, n_{\iOm{\infty}}< m_{\iOm{\infty}})
~,~
  (n_{\iOm{0}}>m_{\iOm{0}}, \hat n_{\iOm{1}}<\hat m_{\iOm{1}}, \hat n_{\iOm{\infty}}>\hat m_{\iOm{\infty}})
  \Big\}
  ,
\end{align}
which is mapped to
\begin{align}
  \Big\{
  (n_{\iOm{0}}<m_{\iOm{0}}, n_{\iOm{1}}<m_{\iOm{1}}, n_{\iOm{\infty}}> m_{\iOm{\infty}})
~,~
  (n_{\iOm{0}}<m_{\iOm{0}}, \hat n_{\iOm{1}}>\hat m_{\iOm{1}}, \hat n_{\iOm{\infty}}<\hat m_{\iOm{\infty}})
\Big\}
  ,
\end{align}
by the left-right symmetry.

Let us now exam the two solutions in the two cases.
We first perform a generic computation which is common to the two
cases and then we explicitly specialize it.
Computing the hypergeometric parameters for the first
solution leads to:
\begin{align}
\left\{\begin{array}{l}
  a^{(L)}=n_{\iOm{0}}+n_{\iOm{1}}+n_{\iOm{\infty}}+\mathfrak{a}^{(L)}
\\
  b^{(L)}=n_{\iOm{0}}+n_{\iOm{1}}-n_{\iOm{\infty}}+\mathfrak{b}^{(L)}
\\
  c^{(L)}=2 n_{\iOm{0}}+\mathfrak{c}^{(L)}
       \end{array}
  \right.
  ,~~~~
\left\{\begin{array}{l}
 a^{(R)}=m_{\iOm{0}}+m_{\iOm{1}}+m_{\iOm{\infty}}+\mathfrak{a}^{(R)}
\\
  b^{(R)}=m_{\iOm{0}}+m_{\iOm{1}}-m_{\iOm{\infty}}+\mathfrak{b}^{(R)}
\\
  c^{(R)}=2 m_{\iOm{0}}+1+\mathfrak{c}^{(R)}
       \end{array}
  \right.,
\end{align}
where the values of the constants can be read from Table~\ref{table:coeffs_k}.
Then we compute the $K^{(L)}$ and $K^{(R)}$ factors using \eqref{eq:K_factor_value}.
Therefore the first solution is:
\begin{align}
    \partial_\upomega \chi_1 =&
  (-\upomega)^{n_{\iOm{0}}+m_{\iOm{0}}-1 }
  (1-\upomega)^{n_{\iOm{1}}+m_{\iOm{1}}-1 } \times
                              \nonumber\\
  & \times
  \mqty(
  F(a^{(L)}, b^{(L)}; c^{(L)}; \upomega) \\
  K^{(L)} (-\upomega)^{1-c^{(L)}}
  F(a^{(L)}+1-c^{(L)}, b^{(L)}+1-c^{(L)}; 2-c^{(L)}; \upomega)
  )
  \nonumber\\
  &\times
  \mqty(
  F(a^{(R)}, b^{(R)}; c^{(R)}; \upomega) \\
  K^{(R)} (-\upomega)^{1-c^{(R)}}
  F(a^{(R)}+1-c^{(R)}, b^{(R)}+1-c^{(R)}; 2-c^{(R)}; \upomega)
  )^T.
\end{align}

The parameters of the second solution read
\begin{align}
  &
    \left\{\begin{array}{ll}
\hat  a^{(L)}= n_{\iOm{0}}+\hat n_{\iOm{1}}+ \hat n_{\iOm{\infty}}
         +\hat {\mathfrak{a}}^{(L)}
         &= c^{(L)}-a^{(L)}
         +\mathfrak{a}^{(L)}-\mathfrak{c}^{(L)}+
         \hat {\mathfrak{a}}^{(L)}+1
\\
\hat   b^{(L)}=n_{\iOm{0}}+\hat n_{\iOm{1}}- \hat n_{\iOm{\infty}}+
             \hat {\mathfrak{b}}^{(L)}
         &= c^{(L)}-b^{(L)}
         +\mathfrak{b}^{(L)}-\mathfrak{c}^{(L)}+
         \hat {\mathfrak{b}}^{(L)}
\\
             \hat   c^{(L)}=2 n_{\iOm{0}} +\hat {\mathfrak{c}}^{(L)}
             &= c^{(L)} -\mathfrak{c}^{(L)}+\hat {\mathfrak{c}}^{(L)}
       \end{array}
  \right.,
  \nonumber\\
  &
    \left\{\begin{array}{ll}
\hat  a^{(R)}= m_{\iOm{0}}+\hat m_{\iOm{1}}+ \hat m_{\iOm{\infty}}
         +\hat {\mathfrak{a}}^{(R)}
         &= c^{(R)}-a^{(R)}
         +\mathfrak{a}^{(R)}-\mathfrak{c}^{(R)}+
         \hat {\mathfrak{a}}^{(R)}+1
\\
\hat   b^{(R)}=m_{\iOm{0}}+\hat m_{\iOm{1}}- \hat m_{\iOm{\infty}}+
             \hat {\mathfrak{b}}^{(R)}
         &= c^{(R)}-b^{(R)}
         +\mathfrak{b}^{(R)}-\mathfrak{c}^{(R)}+
         \hat {\mathfrak{b}}^{(R)}
\\
             \hat   c^{(R)}=2 m_{\iOm{0}} +\hat {\mathfrak{c}}^{(R)}
             &= c^{(R)} -\mathfrak{c}^{(R)}+\hat {\mathfrak{c}}^{(R)}
       \end{array}
               \right.
               .
\end{align}
We see that the two cases differ only for the constants and not for
the structure.

\subsubsection{Case 1}
\label{sec:case1}
We start with the case $n_{\iOm{0}}>m_{\iOm{0}}$, $n_{\iOm{1}}>m_{\iOm{1}}$ and
$n_{\iOm{\infty}}> m_{\iOm{\infty}}$
for which the second solution is
$n_{\iOm{0}}>m_{\iOm{0}}$, $\hat n_{\iOm{1}}< \hat m_{\iOm{1}}$ and
$\hat n_{\iOm{\infty}}< \hat m_{\iOm{\infty}}$
The parameters for the second are explicitly
\begin{align}
  &
    \left\{\begin{array}{l}
\hat  a^{(L)}= c^{(L)}-a^{(L)}
\\
\hat   b^{(L)}= c^{(L)}-b^{(L)}
\\
\hat   c^{(L)}= c^{(L)}
       \end{array}
  \right.
,~~~~
    \left\{\begin{array}{l}
\hat  a^{(R)}= c^{(R)}-a^{(R)}
\\
\hat   b^{(R)}= c^{(R)}-b^{(R)}+1
\\
\hat   c^{(R)}= c^{(R)} +1
       \end{array}
               \right.
               .
\end{align}
The $K$ factors are
\begin{equation}
  \hat K^{(L)}= K^{(L)},~~~~
  \hat K^{(R)}= \frac{K^{(R)}}{a^{(R)} (c^{(R)}-b^{(R)})}
  .
\end{equation}
Using Euler relation
\begin{equation}
  F(a,b;c; \upomega) =(1-\upomega)^{c-a-b} F(c-a,c-b;c; \upomega)
  ,
\end{equation}
we can finally write the second solution as
\begin{align}
    \partial_\upomega \chi_2 =&
  (-\upomega)^{n_{\iOm{0}}+m_{\iOm{0}}-1}
  (1-\upomega)^{n_{\iOm{1}}+m_{\iOm{1}}} \times
                              \nonumber\\
  & \times
  \mqty(
  F(a^{(L)}, b^{(L)}; c^{(L)}; \upomega) \\
  K^{(L)} (-\upomega)^{1-c^{(L)}}
  F(a^{(L)}+1-c^{(L)}, b^{(L)}+1-c^{(L)}; 2-c^{(L)}; \upomega)
  )
  \nonumber\\
  &\times
  \mqty(
  F(a^{(R)}+1, b^{(R)}; c^{(R)}+1; \upomega) \\
  \hat K^{(R)}
  (-\upomega)^{-c^{(R)}}
  F(a^{(R)}+1-c^{(R)}, b^{(R)}-c^{(R)}; 1-c^{(R)}; \upomega)
  )^T
  ,
\end{align}
in which the left basis is exactly equal to the first solution while
the right basis differs for $a^{(R)}\rightarrow a^{(R)}+1$ and
$c^{(R)}\rightarrow c^{(R)}+1$.

\subsubsection{Case 2}
\label{sec:case2}

Consider now the second case $n_{\iOm{0}}>m_{\iOm{0}}$, $n_{\iOm{1}}>m_{\iOm{1}}$ and
$n_{\iOm{\infty}}< m_{\iOm{\infty}}$.
For the second solution we have
$n_{\iOm{0}}> m_{\iOm{0}}$, $\hat n_{\iOm{1}}< \hat m_{\iOm{1}}$ and
$\hat n_{\iOm{\infty}}> \hat m_{\iOm{\infty}}$ and the parameters are explicitly
\begin{align}
  &
    \left\{\begin{array}{l}
\hat  a^{(L)}= c^{(L)}-a^{(L)}-1
\\
\hat   b^{(L)}= c^{(L)}-b^{(L)}+1
\\
\hat   c^{(L)}= c^{(L)}
       \end{array}
  \right.
,~~~~
    \left\{\begin{array}{l}
\hat  a^{(R)}= c^{(R)}-a^{(R)}
\\
\hat   b^{(R)}= c^{(R)}-b^{(R)}
\\
\hat   c^{(R)}= c^{(R)}
       \end{array}
               \right.
               .
\end{align}
The $K$ factors are
\begin{equation}
  \hat K^{(L)}= K^{(L)}\frac{(b^{(L)}-1)(c^{(L)}-a^{(L)}-1)}{a^{(L)}(c^{(L)}-b^{(L)})},~~~~
  \hat K^{(R)}= K^{(R)}
  .
\end{equation}
Using Euler relation we can finally write the second solution for the
second case as
\begin{align}
    \partial_\upomega \chi_2 =&
  (-\upomega)^{n_{\iOm{0}}+m_{\iOm{0}}-1}
  (1-\upomega)^{n_{\iOm{1}}+m_{\iOm{1}}} \times
                              \nonumber\\
  & \times
  \mqty(
  F(a^{(L)}+1, b^{(L)}-1; c^{(L)}; \upomega) \\
  \hat K^{(L)}
  (-\upomega)^{1-c^{(L)}}
  F(a^{(L)}+2-c^{(L)}, b^{(L)}-c^{(L)}; 2-c^{(L)}; \upomega)
  )
  \nonumber\\
  &\times
  \mqty(
  F(a^{(R)}, b^{(R)}; c^{(R)}; \upomega) \\
  K^{(R)} (-\upomega)^{1-c^{(R)}}
  F(a^{(R)}+1-c^{(R)}, b^{(R)}+1-c^{(R)}; 2-c^{(R)}; \upomega)
  )^T
  ,
\end{align}
in which the right basis is exactly equal to the first solution while
the left basis differs for $a^{(L)}\rightarrow a^{(L)}+1$ and
$b^{(L)}\rightarrow b^{(L)}-1$.

\subsection{The Solution}
  In the previous section we have shown that there are two independent
  solutions, therefore the general solution for
  $\partial_\upomega \chi$ obviously reads
  \begin{equation}
\partial_\upomega \chi= C_1 \partial_\upomega \chi_1 + C_2 \partial_\upomega \chi_2
\label{eq:general_solution}
.
\end{equation}
  Therefore the final solution depends now only on two complex
  constants, $C_1$ and $C_2$ which we can fix imposing the global conditions
  in \eqref{eq:discontinuity_bc}, i.e. the second equation for all
  $t$'s in the solution \eqref{eq:classical_solution}.
  Since the three target space intersection
  points always define a triangle on a 2-dimensional plane, we can
  impose the boundary conditions knowing two angles formed by the sides (i.e.
  the branes between two intersections) and the length of one of
  them.
  We already fixed the parameters of the rotations, then we need to
  compute the length of one of the sides.
  and consider, for instance, the length of the side
  $X(x_{\overline{t}+1},x_{\overline{t}+1}) - X(x_{\overline{t}-1}, x_{\overline{t}-1})$:
Explicitly we impose the four real equations in spinorial formalism
  \begin{equation}
    \int_0^1 \dd{\upomega} \partial_\upomega \mathcal{X}(\upomega)
    +
    \UL^{\dagger}(\vb{n}_{\iD{\overline{t}}})
    ~\int_0^1 \dd{\bar\upomega} \partial_\upomega \mathcal{X}(\bar\upomega)
    ~\UR(\vb{m}_{\iD{\overline{t}}})
  =
  f_{\iD{\overline{t}+1}\,(s)}-f_{\iD{\overline{t}-1}\,(s)}
  ,
\end{equation}
where we have used the mapping \eqref{eq:def_omega} to write the
integrals directly in $\upomega$ variables.
This equation has then enough degrees of freedom to fix completely
the two complex parameters $C_1$ and $C_2$,
thus completing the determination of the full solution in its general form.

  \section{Recovering the \texorpdfstring{$\mathrm{SU}(2)$}{SU(2)} and the
  Abelian Solution}

  Before analyzing further the result, we first show how this general
  procedure automatically includes the solution with both pure $\mathrm{SU}(2)$
  and Abelian rotations of the D-branes.
  The Abelian solution emerges from the general construction as a
  limit and replicates the known result for Abelian
  $\mathrm{SO}(2) \times \mathrm{SO}(2) \subset \mathrm{SO}(4)$
  rotations in the case of a factorized space
  $\mathds{R}^4 = \mathds{R}^2 \times \mathds{R}^2$.

  \subsection{Abelian Limit of the \texorpdfstring{$\mathrm{SU}(2)$}{SU(2)}
  Monodromies}

\begin{figure}[t]
    \centering
    \def\svgwidth{0.8\textwidth}
    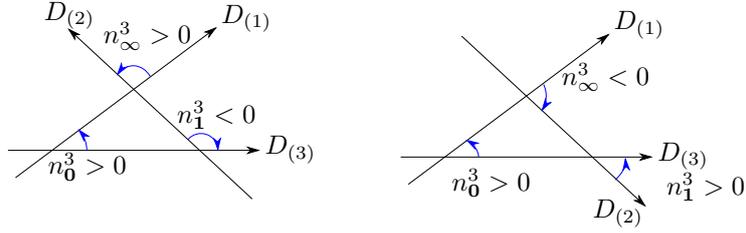
    \caption{The Abelian limit when the triangle has all acute angles.
      This corresponds to the cases  $n_{\iOm{0}} + n_{\iOm{\infty}}< \frac{1}{2}$ and
      $n_{\iOm{0}}< n_{\iOm{\infty}}$ which are exchanged under the parity $P_2$.}
    \label{fig:Abelian_angles_1}
  \end{figure}

\begin{figure}[t]
    \centering
    \def\svgwidth{0.8\textwidth}
    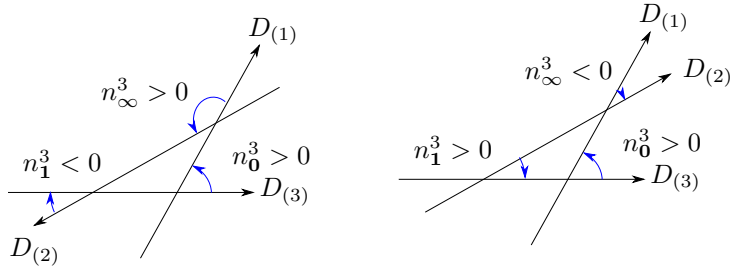
    \caption{The Abelian limit when the triangle has one obtuse angle.
      This corresponds to the cases $n_{\iOm{0}} + n_{\iOm{\infty}}> \frac{1}{2}$ and
      $n_{\iOm{0}}> n_{\iOm{\infty}}$ which are exchanged under the parity $P_2$.}
    \label{fig:Abelian_angles_2}
  \end{figure}

  We want now to compute the parameter $\vb{n}_{\iOm{1}}$ when we
  are given two Abelian rotation in $\upomega=0$ and $\upomega=\infty$
  using the standard expression for two $\mathrm{SU}(2)$ element multiplication
  given in \eqref{eq:product_in_SU2}.
  We can summarize the results in Table~\ref{table:Abelian_composition}.
  \begin{center}
\captionof{table}{Abelian limit of $\mathrm{SU}(2)$ monodromies}
  \label{table:Abelian_composition}
    \begin{tabular}{
  c  c  
  | c |c 
  | c c 
  | c 
}
$\vb{n}_{\iOm{0}}$  & $\vb{n}_{\iOm{\infty}}$
& &
& $n_{\iOm{1}}$ & $\vb{n}_{\iOm{1}}$
& $\sum_t \vb{n}_{\iD{t}}$
  \\
  \hline
$n_{\iOm{0}} \vb{k}$ & $n_{\iOm{\infty}} \vb{k}$
&  $n_{\iOm{0}} + n_{\iOm{\infty}}< \frac{1}{2}$  & $n_{\iOm{0}}\lessgtr n_{\iOm{\infty}}$
& $n_{\iOm{0}} + n_{\iOm{\infty}}$ & -$n_{\iOm{1}} \vb{k}$ & 0
\\
\hline
$n_{\iOm{0}} \vb{k}$ & $n_{\iOm{\infty}} \vb{k}$
&  $n_{\iOm{0}} + n_{\iOm{\infty}}> \frac{1}{2}$  & $n_{\iOm{0}}\lessgtr n_{\iOm{\infty}}$
& $1-(n_{\iOm{0}} + n_{\iOm{\infty}})$ & $+n_{\iOm{1}} \vb{k}$ & $\vb{k}$
\\
  \hline
$n_{\iOm{0}} \vb{k}$ & $-n_{\iOm{\infty}} \vb{k}$
&  $n_{\iOm{0}} + n_{\iOm{\infty}}\lessgtr \frac{1}{2}$  & $n_{\iOm{0}}> n_{\iOm{\infty}}$
& $n_{\iOm{0}} - n_{\iOm{\infty}}$ & $-n_{\iOm{1}} \vb{k}$ & 0
\\
\hline
$n_{\iOm{0}} \vb{k}$ & $-n_{\iOm{\infty}} \vb{k}$
&  $n_{\iOm{0}} + n_{\iOm{\infty}}\lessgtr \frac{1}{2}$  & $n_{\iOm{0}}< n_{\iOm{\infty}}$
& $-n_{\iOm{0}} + n_{\iOm{\infty}}$ & $+n_{\iOm{1}} \vb{k}$ & 0
\end{tabular}
\end{center}
Notice that under the parity $P_2$ the previous four cases are grouped
into two sets
$\{ n_{\iOm{1}}=n_{\iOm{0}} + n_{\iOm{\infty}}, \hat n_{\iOm{1}}=-n_{\iOm{0}} + \hat
n_{\iOm{\infty}} \}$
and
$\{ n_{\iOm{1}}=1-(n_{\iOm{0}} + n_{\iOm{\infty}}), \hat n_{\iOm{1}}=+n_{\iOm{0}}-\hat
n_{\iOm{\infty}} \}$.
This can be also seen geometrically since the first group corresponds to the
same geometry which is depicted in Figure~\ref{fig:Abelian_angles_1} while the
second in Figure~\ref{fig:Abelian_angles_2}.
  Arbitrarily fixing the orientation of $D_{\iD{3}}$ we can in fact obtain
  these geometrical interpretations and since $n^3_{\iOm{0}}>0$ we can fix the
  orientation of $D_{\iD{1}}$. The orientation of $D_{\iD{2}}$ is then fixed
  relatively to $D_{\iD{1}}$ by the sign of $n^3_{\iOm{\infty}}$. The sign of
  $n^3_{\iOm{1}}$ then follows.


\begin{figure}[t]
    \centering
    \def\svgwidth{0.6\textwidth}
    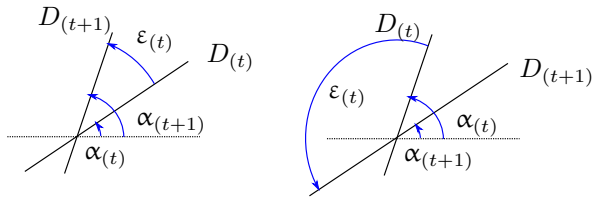
    \caption{The geometrical angles used in the usual geometrical
      approach to the Abelian configuration do not distinguish among
      the possible branes orientations. In fact we have $0\le
      \upalpha<1$ and $0<\upvarepsilon<1$.}
          \label{fig:usual_Abelian_angles}
  \end{figure}

The usual Abelian convention is more geometrical and visual therefore
it does not distinguish between the possible orientations of the
branes while this group approach does.
In fact comparing all possible brane orientations and the ensuing
group parameter $n^3$
with the usual angles used in the Abelian configuration depicted in
Figure~\ref{fig:usual_Abelian_angles}
we see that relation between the usual Abelian parameter $\epsilon_{\iD{t}}$
and the group one $n_{\iOm{t}}^3$ is given by
\begin{equation}
  \upvarepsilon_{\iOm{t}}=n_{\iOm{t}}^3 +\theta(-n^3_{\iOm{t}})
  \label{eq:Abelian_vs_n_simple_case}
  ,
\end{equation}
when all $m=0$.

\subsection{Abelian Limit of the Left Solutions}

Then we can compute the basis element for any entry of the
Table~\ref{table:coeffs_k} for any possible value of $n_1$ as given in
Table~\ref{table:Abelian_composition}.
Here we consider for simplicity the left sector of the solution:
everything can be stated in the same way for the right sector.

It turns out that either $K=0$ or $K=\infty$.
In the latter case we can absorb the infinite divergence in a constant
term in front of the solution and effectively use:
  \begin{eqnarray}
    \eval{D\;}_{K=0} & = & \mqty( 1 & \\ & 0 ),
    \\
    \eval{D\;}_{K=\infty} & = & \mqty( 0 & \\ & 1 )
                                                  .
  \end{eqnarray}
The result is then given in Table~\ref{table:Left_Abelian_solutions}. In this
table we have left some hypergeometrics in their symbolic form. However all of
them are elementary functions since either $a$ or $c-b$ is equal to $-1$.
\begin{table}
\captionof{table}
    { Abelian limit of the solutions
}
\label{table:Left_Abelian_solutions}
  \begin{tabular}{c| c| c  
    }
    $(\mathfrak{a}^{(L)},\mathfrak{b}^{(L)},\mathfrak{c}^{(L)})$ &
    $n_{\iOm{1}}$ &
                $\mathcal{B}^{(L)\, T}(z)$
    \\
    \hline
    $(- 1, 0, 0)$ &
    $n_{\iOm{0}} + n_{\iOm{\infty}}$ &
     $\left( \left(1-z\right)^{-2\,{ n_{\iOm{\infty}}}-2\,n_{\iOm{0}}+1} , 0 \right) $
    \\
    &
    $1-(n_{\iOm{0}} + n_{\iOm{\infty}})$ &
    $\left( 1 , 0 \right) $
    \\
    &
    $n_{\iOm{0}} - n_{\iOm{\infty}}$ &
    $\left( 0 , \left(-z\right)^{1-2\,n_{\iOm{0}}} \right) $
    \\
    &
    $-n_{\iOm{0}} + n_{\iOm{\infty}}$ &
    $\left( 1 , 0 \right) $
\\\hline
    $(- 1, 1, 0)$ &
    $n_{\iOm{0}} + n_{\iOm{\infty}}$ &
$\left( F\left(2\,{ n_{\iOm{\infty}}}+2\,n_{\iOm{0}}-1 , 2\,n_{\iOm{0}}+1 , 2\,n_{\iOm{0}} , z
 \right) , 0 \right) $
    \\
    &
    $1-(n_{\iOm{0}} + n_{\iOm{\infty}})$ &
    $\left( 1 , 0 \right) $
    \\
    &
    $n_{\iOm{0}} - n_{\iOm{\infty}}$ &
$\left( 0 , \left(-z\right)^{1-2\,n_{\iOm{0}}} \right) $
    \\
    &
    $-n_{\iOm{0}} + n_{\iOm{\infty}}$ &
$\left( 0 , \left(1-z\right)^{2\,n_{\iOm{0}}-2\,{ n_{\iOm{\infty}}}}\,\left(-z\right)
 ^{1-2\,n_{\iOm{0}}} \right) $
    \\
    \hline
    $(0, 0, 0)$ &
    $n_{\iOm{0}} + n_{\iOm{\infty}}$ &
$\left( \left(1-z\right)^{-2\,{ n_{\iOm{\infty}}}-2\,n_{\iOm{0}}} , 0 \right) $
    \\
    &
    $1-(n_{\iOm{0}} + n_{\iOm{\infty}})$ &
$\left( 0 , \left(1-z\right)^{2\,{ n_{\iOm{\infty}}}+2\,n_{\iOm{0}}-2}\,\left(-z
 \right)^{1-2\,n_{\iOm{0}}} \right) $
    \\
    &
    $n_{\iOm{0}} - n_{\iOm{\infty}}$ &
$\left( \left(1-z\right)^{2\,{ n_{\iOm{\infty}}}-2\,n_{\iOm{0}}} , 0 \right) $
    \\
    &
    $-n_{\iOm{0}} + n_{\iOm{\infty}}$ &
$\left( 1 , 0 \right) $
    \\
    \hline
    $(- 1, 1, 1)$ &
    $n_{\iOm{0}} + n_{\iOm{\infty}}$ &
$\left( \left(1-z\right)^{-2\,{ n_{\iOm{\infty}}}-2\,n_{\iOm{0}}+1} , 0 \right) $
    \\
    &
    $1-(n_{\iOm{0}} + n_{\iOm{\infty}})$ &
$\left( 1 , 0 \right) $
    \\
    &
    $n_{\iOm{0}} - n_{\iOm{\infty}}$ &
$\left( 0 , {F\left(-1 , 1-2\,{ n_{\iOm{\infty}}} , 1-2\,n_{\iOm{0}} , z\right)
 \left(-z\right)^{-2\,n_{\iOm{0}}}} \right) $
    \\
    &
    $-n_{\iOm{0}} + n_{\iOm{\infty}}$ &
$\left( 0 , {\left(1-z\right)^{-2\,{ n_{\iOm{\infty}}}+2\,n_{\iOm{0}}+1}
 \left(-z\right)^{-2\,n_{\iOm{0}}}} \right) $
    \\
    \hline
    $(0, 0, 1)$ &
    $n_{\iOm{0}} + n_{\iOm{\infty}}$ &
$\left( 0 , {\left(-z\right)^{-2\,n_{\iOm{0}}}} \right) $
    \\
    &
      $1-(n_{\iOm{0}} + n_{\iOm{\infty}})$ &
     $\left( 0 , {\left(1-z\right)^{2\,{ n_{\iOm{\infty}}}+2\,n_{\iOm{0}}-1}
     \left(-z\right)^{-2\,n_{\iOm{0}}}} \right) $
    \\
    &
    $n_{\iOm{0}} - n_{\iOm{\infty}}$ &
$\left( 0 , {\left(-z\right)^{-2\,n_{\iOm{0}}}} \right) $
    \\
    &
    $-n_{\iOm{0}} + n_{\iOm{\infty}}$ &
$\left( 1 , 0 \right) $
    \\
    \hline
    $(0, 1, 1)$ &
    $n_{\iOm{0}} + n_{\iOm{\infty}}$ &
$\left( \left(1-z\right)^{-2\,{ n_{\iOm{\infty}}}-2\,n_{\iOm{0}}} , 0 \right) $
    \\
    &
    $1-(n_{\iOm{0}} + n_{\iOm{\infty}})$ &
    $\left( 0 , {\left(1-z\right)^{2\,{ n_{\iOm{\infty}}}+2\,n_{\iOm{0}}-2}
    \left(-z\right)^{-2\,n_{\iOm{0}}}} \right) $
    \\
    &
    $n_{\iOm{0}} - n_{\iOm{\infty}}$ &
$\left( 0 , {\left(-z\right)^{-2\,n_{\iOm{0}}}} \right) $
    \\
    &
    $-n_{\iOm{0}} + n_{\iOm{\infty}}$ &
    $\left( 0 , {\left(1-z\right)^{2\,n_{\iOm{0}}-2\,{ n_{\iOm{\infty}}}}
      \left(-z\right)^{-2\,n_{\iOm{0}}}} \right) $
  \end{tabular}
\end{table}

  \subsection{The \texorpdfstring{$\mathrm{SU}_L(2)$}{SUL(2)} Limit}

We can recover the previously computed non Abelian $\mathrm{SU}(2)$ solution by
considering $m_{\iOm{t}}\sim 0$: this is case 1 of section \ref{sec:true_basis}.
The first thing we notice is that the left solution
$\mathcal{B}^{(L)}$ is always the same and matches the previous computation.
Despite so, the right sector seems to give different solutions when
different Abelian limits are taken.
Actually, examining all the possible solutions\footnote{
  We write possible because the $m_{\iOm{1}}=1-(m_{\iOm{0}} + m_{\iOm{\infty}})$
  case is not.
  },
we get that  all of them give the same answer in the limit
$m_{\iOm{t}}\rightarrow 0$, i.e. both $\mathcal{B}^{(R)}=(1,0)^T$ and $\mathcal{B}^{(R)}=(0,1)^T$.
The only difference is which solution is obtained from the case
$n_{\iOm{0}}>m_{\iOm{0}}$, $n_{\iOm{1}}>m_{\iOm{1}}$ and $n_{\iOm{\infty}}> m_{\iOm{\infty}}$
or from the $n_{\iOm{0}}>m_{\iOm{0}}$, $\hat n_{\iOm{1}}< \hat m_{\iOm{1}}$ and
$\hat n_{\iOm{\infty}}< \hat m_{\iOm{\infty}}$.
In any case we get a factorized solution of the form
$\mathcal{B}^{(L)} (C,C')^T$ which is what expected since the right
sector plays no role.

\subsection{Relating the Abelian Angles with the Group Parameters}
Using the explicit expression for the $\mathrm{SO}(4)$ and
$\mathrm{SU}(2) \times  \mathrm{SU}(2)$ it is easy to verify that
when the left and right $\mathrm{SU}(2)$ parameters are
$\vec n =n^3 \vec k$ and $\vec m =m^3 \vec k$
the rotation in plane $14$ is a $\mathrm{SO}(2)$ element
$\mqty( \cos(\theta) & \sin(\theta) \\ -\sin(\theta)&  \cos(\theta) )$
with angle $\theta=n^3-m^3$
and the one in plane $23$ is with angle $\theta=n^3+m^3$.

Comparing with the case with $m=0$ given in
\eqref{eq:Abelian_vs_n_simple_case} we can then guess that the general
relation between the group parameters and the usual Abelian angles is
given by
\begin{align}
  \upvarepsilon_{\iOm{t}}
&=n_{\iOm{t}}^3-m_{\iOm{t}}^3 +\theta(-(n^3_{\iOm{t}}-m_{\iOm{t}}^3)),
  \nonumber\\
  \upvarphi_{\iOm{t}}
&=n_{\iOm{t}}^3+m_{\iOm{t}}^3 +\theta(-(n^3_{\iOm{t}}+m_{\iOm{t}}^3)).
\label{eq:Abelian_vs_n_general_case}
\end{align}

  \subsection{Recovering the Abelian Result: an Example}

  In order to show how the Abelian limit works we consider the
following example.  We take case 1 as in section \ref{sec:case1} with
$n_{\iOm{1}}=1-(n_{\iOm{0}} + n_{\iOm{\infty}})$ and $m_{\iOm{1}}=-m_{\iOm{0}} +
m_{\iOm{\infty}}$.  which leads to two independent rational functions of
$\upomega_z$:
  \begin{align}
    \partial \mathcal{X}(\upomega_z)
   & =
    \mqty( i \partial \bar{ \mathcal{Z}}^{\bar 1}(\upomega_z)
    &  \partial {\mathcal{Z}}^{2}(\upomega_z)
    \\
    \partial \bar{ \mathcal{Z}}^{\bar 2}(\upomega_z)
    &
    i  \partial \mathcal{Z}^{1}(\upomega_z)
      )
    =
    \nonumber\\
    & =
    \mqty(
      i  \partial ( \mathcal{X}^1(\upomega_z) - i \mathcal{X}^4(\upomega_z) )
    &
          \partial ( \mathcal{X}^2(\upomega_z) + i \mathcal{X}^3(\upomega_z) )
    \\
        \partial ( \mathcal{X}^2(\upomega_z) - i \mathcal{X}^3(\upomega_z) )
    &
       i   \partial ( \mathcal{X}^1(\upomega_z) + i \mathcal{X}^4(\upomega_z) )
) =
  \nonumber\\
   & =
    \mqty( 0
    & C_1\,
    (-\upomega_z)^{\upvarepsilon_{\iOm{0}}-1} (1-\upomega_z)^{\upvarepsilon_{\iOm{1}}-1}
    \\
    0
    & C_2\,
    (-\upomega_z)^{-\upvarphi_{\iOm{0}}} (1-\upomega_z)^{-\upvarphi_{\iOm{1}}}
    ),
  \end{align}
  where   $C_1$, $C_2$ are constants as in
  \eqref{eq:general_solution}.
    This is the known result for the Abelian case, where we have
  two different $\mathrm{U}(1)$ sectors undergoing two different rotations
$
    \mathrm{U}_1(1) \times \mathrm{U}_2(1) \subset \mathrm{SU}_L(2) \times
    \mathrm{SU}_R(2)
$.
  In the previous expression we have used
\eqref{eq:Abelian_vs_n_general_case} to write the relation between the
usual Abelian angles and the group parameters as
  \begin{align}
    \upvarepsilon_{\iOm{0}} = n_{\iOm{0}} - m_{\iOm{0}}, &&
    \upvarepsilon_{\iOm{1}} = n_{\iOm{1}} - m_{\iOm{1}}, &&
    \upvarepsilon_{\iOm{\infty}} = n_{\iOm{\infty}} + m_{\iOm{\infty}},
\nonumber\\
    && \sum_t \upvarepsilon_{\iOm{t}} = 1 &&
    \label{eq:Abelian_rotation_first}
  \end{align}
  and
\begin{align}
    \upvarphi_{\iOm{0}} = n_{\iOm{0}} + m_{\iOm{0}}, &&
    \upvarphi_{\iOm{1}} = n_{\iOm{1}} + m_{\iOm{1}}, &&
    \upvarphi_{\iOm{\infty}} = n_{\iOm{\infty}} - m_{\iOm{\infty}},
\nonumber\\
   &&  \sum_t \upvarphi_{\iOm{t}} = 2, &&
    \label{eq:Abelian_rotation_second}
\end{align}
  in order to approach the usual notation in the literature.
  As usual
  $
    \partial \mathcal{Z}^1(\upomega_z)
    \ne
   [\partial \overline{\mathcal{Z}}^{\bar 1}(\upomega_z)]^*
  $.
  We can now build the Abelian solution to show the characteristic
analytical structure of the Abelian limit.  Explicitly we get
  \begin{align}
    \mqty( i \bar Z^{\bar 1}(u,\bar u)
    &   Z^{2}(u,\bar u)
    \\
    \bar {Z}^{\bar 2}(u,\bar u)
    &
      i  Z^{1}(u,\bar u)
      )
&=
\mqty(
    i \bar f^{\bar 1}_{\iD{\overline{t}-1}}
+i \int_0^{\bar \upomega_{\bar u}} d\upomega \partial
          {\mathcal{Z}}^{1}
&
      f^{2}_{\iD{\overline{t}-1}}
+
\int_0^{\upomega_u} d\upomega \partial  {\mathcal{Z}}^{ 2}
\\
     \bar f^{\bar 2}_{\iD{\overline{t}-1}}
    +
    \int_0^{\bar \upomega_{\bar u}} d\upomega \partial   {\mathcal{Z}}^{2}
    &
     i f^{1}_{\iD{\overline{t}-1}}
+
i \int_0^{\upomega_u} d\upomega \partial {\mathcal{Z}}^{1}
)
,
    \end{align}
where for simplicity we have chosen $R_{\iD{\overline{t}}}=\mathds{1}_4$ so
that $U_{\iD{\overline{t}}}$ in \eqref{eq:U_brane_t} is mapped
to the $\mathrm{SU}(2) \times \mathrm{SU}(2)$ element
$(i \upsigma_1, i \upsigma_1)$.
Notice however that $\vb{n}_{\iOm{t}}= n_{\iOm{t}}^3 \vb{k}$  implies
that ${v}^3_{\iD{{t}}}=0$ in
\eqref{eq:special_UL_brane_t}.
From the previous relations we see the usual holomorphicity
$\overline{Z}^{\bar 1}(\overline{u}) = \left( Z^1(u) \right)^*$
of the sector with $\sum_t \upvarepsilon_{\iOm{t}} = 1$
and 
$\overline{Z}^{\bar 2}(\overline{ u}) = \left( Z^2({u}) \right)^*$
of the sector with $\sum_t \upvarphi_{\iOm{t}} = 2$.

  \subsection{Abelian Limits}

  Following the example of the previous section
  it is possible to consider both cases given in Section~\ref{sec:case1} and
  Section~\ref{sec:case2} for all possible combinations of
  the expression of $n_{\iOm{1}}$ and $m_{\iOm{1}}$ for a total of $2 \cdot 4
  \cdot 4$ possible combinations.
  In all cases but 6 the solution in spinorial formalism is a
  $2\times 2$ matrix which has two non
  vanishing entries and hence two independent Abelian solutions.
  In the remaining 6 cases the matrix has only one non vanishing entry
  but the constraints on $n$ and $m$ are incompatible and therefore
  they should not be considered.
  The 6 inconsistent combinations are for case 1 when
  $\{n_{\iOm{1}}=n_{\iOm{0}} + n_{\iOm{\infty}},~ m_{\iOm{1}}=1-(m_{\iOm{0}} + m_{\iOm{\infty}})\}$
  and
  $\{n_{\iOm{1}}=1-(n_{\iOm{0}} + n_{\iOm{\infty}}),~ m_{\iOm{1}}=1-(m_{\iOm{0}} + m_{\iOm{\infty}})\}$
  and for case 2
  when $n_{\iOm{1}}=-n_{\iOm{0}} + n_{\iOm{\infty}}$.

  \section{The Physical Interpretation}

  In this section we would like to show some simple consequences of the explict
  classical solution for  the phenomenology of the branes at angles models. In
  particular we will focus on the value of the action which plays a fundamental
  role in the hierarchy of the Yukawa couplings.

  \subsection{Rewriting the Action}

  Once the solution to the boundary conditions has been found, it is possible
  to compute the classical action to show its contribution to the correlation
  functions of twist fields and Yukawa couplings. We use the equations of
  motion \eqref{eq:string_equation_of_motion} to simplify as much as possible
  the computation of the action \eqref{eq:string_action} and get:
  \begin{equation}
    4 \uppi \upalpha' \eval{S\;}_{\text{on-shell}}
    = i \sum\limits_{t=1}^3\,
    \sum_{m\in\{3, 4\}}
    g_{\iD{t},\, m}\,
    \int\limits_{x_t}^{x_{t-1}} \dd{x}
    \left( R_{\iD{t}} \right)_{m\,I}
    \eval{\left( X_L'(x) - X_R'(x) \right)^I}_{y=0^+},
    \label{eq:area_tmp}
  \end{equation}
  where $I = 1,\,2,\,3,\,4$ and $m=3,4$ are the transverse
directions in the well adapted frame with respect to the brane.
  Moreover, since the total
  number of D-branes is defined modulo $N_B = 3$, the interval defining
  $D_{\iD{1}}$ is split on two separate intervals, namely:
  \begin{equation*}
    \left[ x_1, x_3 \right] = \left[ x_1, +\infty \right) \cup \left( -\infty,
    x_3 \right],
  \end{equation*}
  as it is visually shown in Figure~\ref{fig:finite_cuts}.
  To proceed further we notice that for $x_t < x < x_{t-1}$ we have:
  \begin{equation*}
    X(x+iy,x-iy) = X^*(x+iy, x-iy) \Rightarrow X_L^*(x-iy) = X_R(x-iy) +
    Y_{\iD{t}},
  \end{equation*}
  where $Y_{\iD{t}}$ is a constant factor which cannot
  depend on the particular brane $D_{\iD{t}}$ and must be a real.
  From the continuity of $X_L(u)$ and $X_R(\overline{u})$ on the
  worldsheet intersection point we get
  \begin{equation*}
    \lim\limits_{x \to x_t^+} X(x, x) = \lim\limits_{x \to x_t^-} X(x, x)
  \end{equation*}
  which does not allow $Y_{\iD{t}}$ to depend on the brane while the
  reality of $X(u,\overline{u})$ implies that $\Im Y = 0$.
Then \eqref{eq:area_tmp} becomes:
  \begin{align}
    4 \uppi \upalpha' \eval{S}_{\text{on-shell}}
    &=
    -2 \sum\limits_{t=1}^3\,    \sum_{m\in\{3, 4\}}
    g_{\iD{t},\, m}\,
    \eval{\Im \left( R_{\iD{t}} \right)_{m \,I}
    X_L^I(x+i0^+)}^{x=x_{t-1}}_{x=x_t}
\nonumber\\
    &=
      -2 \sum\limits_{t=1}^3\,
    g^{(\perp)}_{\iD{t},\, I}\,
    \eval{\Im
      X_{ L}^I(x+i0^+)}^{x=x_{t-1}}_{x=x_t} \in \mathds{R}
      ,
    \label{eq:action_with_imaginary_part}
  \end{align}
  where $g^{(\perp)}_{\iD{t},\, I}
  = \sum_{m\in\{3, 4\}} \left( R_{\iD{t}}^{-1} \right)_{I \,m} g_{\iD{t},\, m}$
  is the transverse shift of $D_{\iD{t}}$ in global coordinates and
  because of this is perpendicular to $f_{\iD{t-1}}-f_{\iD{t}}$ which
  is tangent to $D_{\iD{t}}$, i.e.
\begin{equation}
  g^{(\perp)}_{\iD{t},\, I} (f_{\iD{t-1}}-f_{\iD{t}})^I=0
  .
\label{eq:g_perp_Delta_f}
\end{equation}

  \subsection{Holomorphic Case}
  In this case there exist global complex coordinates for which
  the string solution is holomorphic, i.e.
  \begin{equation}
    Z^i(u, \bar u)=Z^i_L(u)
    ,~~~~
    \bar Z^{\bar i}(u, \bar u)=\bar Z^{\bar i}(\bar u)= \left(
      Z^i_L(u) \right)^*
    ,
  \end{equation}
where $i=1$ in the Abelian case and $i=1,2$ in the $\mathrm{SU}(2)$ case.
In these cases we have $f^i_{\iD{t}}= Z^i_L(x_t+i 0^+)$ and because of
this
the previous equations \eqref{eq:g_perp_Delta_f} and
\eqref{eq:action_with_imaginary_part} become
\begin{align}
  &\Re\left(
    g^{(\perp)}_{\iD{t},\, i} (f_{\iD{t-1}}-f_{\iD{t}})^i
  \right)
    =0
    \nonumber\\
      4 \uppi \upalpha' \eval{S}_{\text{on-shell}}
    &=
      -2 \sum\limits_{t=1}^3\,
      \Im \left(
     g^{(\perp)}_{\iD{t},\, i}\,
      (f_{\iD{t-1}}-f_{\iD{t}})^i
      \right)
,
  \end{align}
where the last equation shows that the action can be expressed only
using the global data.

In the Abelian case where $i=1$ we can further simplify the action and
give a clear geometrical meaning.
We notice that given to complex numbers $a, b \in \mathds{C}$ such
that $\Re(a^* b)=0$ then $\Im(a^* b) = \pm \abs{a} \abs{b}$.
This can be seen either by direct computation or by using a
$\mathrm{U}(1)$ rotation to set $b$ equal to $\abs{b}$.
Since the action is positive then we can write
\begin{align}
         \eval{S}_{\text{on-shell}}
  &=
    \frac{1}{2 \uppi \upalpha'}
       \sum\limits_{t=1}^3\,
     \frac{1}{2} \abs{g^{(\perp)}_{\iD{t}}} \,
      \abs{f_{\iD{t-1}}-f_{\iD{t}}}
,
  \end{align}
where a factor $\frac{1}{2}$ comes from raising the
$g^{(\perp)}_{\iD{t}\, i= 1}$
complex index.
We now see that the right hand side is the sum of the areas of the
triangles having as base the interval between two intersection points
on a given brane $D_{\iD{t}}$ and as height the distance between the
brane and the origin as shown in Figure~\ref{fig:branes_at_angles}.

For the $\mathrm{SU}(2)$ case we can use a $\mathrm{SU}(2)$ rotation
to bring  $(f_{\iD{t-1}}-f_{\iD{t}})^i$ to the form
$\norm{f_{\iD{t-1}}-f_{\iD{t}}} \delta^i_1$, then each term of the action can
be interpreted again as an area of a triangle where the distance between the
interaction points is the base. Also in this case a kind of flatness is playing
a role to give the value of the action.

  \subsection{The General Non Abelian Case}

  \begin{figure}[t]
    \centering
    \def\svgwidth{0.6\textwidth}
    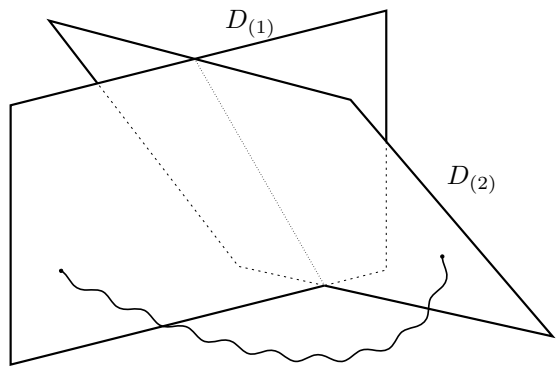
    \caption{This is  a pictorical 3-dimensional representation of two D2-branes
      intersecting in the Euclidean space $\mathds{R}^3$ along a line (in
      $\mathds{R}^4$ the intersection is a point since the co-dimension of each
      brane is 2): since it is no longer constrained on a bi-dimensional plane,
      the string must be deformed in order to stretch between two consecutive
      branes. Its action will therefore be larger than the planar area.}
    \label{fig:brane3d}
  \end{figure}

In the general case there does not seem to be any possible way of
computing the action \eqref{eq:action_with_imaginary_part} in term of
the global data.
It does not seem to be any kind of flatness involved and probably the
action is bigger than in the holomorphic case since the string is no
longer confined to a plane and, given the nature of the rotation, its
worldsheet has to bend in order to be attached to the brane as
pictorially shown in Figure~\ref{fig:brane3d} in the case of a
3-dimensional space.
   The general case we considered then differs from the known
factorized case by an additional contribution in the on-shell action
which can be intuitively understood as a small ``bump'' of the string
worldsheet in proximity of the boundary.

  The physical consequence is an exponential
  suppression of the contribution of the classical action to the correlators of
  twist fields and to the Yukawa coupling with respect to the
  holomorphic case.

\vskip 1cm
\noindent {\large {\bf Acknowledgments}}

This work is partially supported by the Compagnia di San Paolo
contract ``MAST: Modern Applications of String Theory''
TO-Call3-2012-0088 and by the MIUR PRIN Contract 2015MP2CX4 
``Non-perturbative Aspects Of Gauge Theories And Strings"

  \footnotesize
  \bibliographystyle{utphys}
  \addcontentsline{toc}{section}{References}
  \bibliography{non-abelian-twists}
  \normalsize

  \clearpage
  \pagenumbering{roman}
  \numberwithin{figure}{section}
  \numberwithin{table}{section}

  \appendix

  \section{The Isomorphism in Details}\label{sec:isomorphism}

  In this appendix we discuss our conventions for $\mathrm{SU}(2)$ and show the details
  on the constructions of the isomorphism between $\mathrm{SO}(4)$ and a class
  of equivalence of $\mathrm{SU}(2) \times \mathrm{SU}(2)$.

  \subsection{\texorpdfstring{$\mathrm{SU}(2)$}{SU(2)} Conventions}
  We choose to parameterize any $\mathrm{SU}(2)$ matrix with a vector $\vb{n}
  \in \mathds{R}^3$ such that:
  \begin{equation}
    U(\vb{n}) = \cos(2\uppi n) \mathds{1}_2 + i \frac{\vb{n} \cdot
    \vb{\upsigma}}{n} \sin(2\uppi n),
  \end{equation}
  where $n = \norm{\vb{n}}$ and $0 \le n \le \frac{1}{2}$ with the
  identification of all $\vb{n}$ when $n=\frac{1}{2}$ since in this case $
  U(\vb{n})= -\mathds{1}_2$. The parametrization is  such that:
  \begin{eqnarray}
    \left( U(\vb{n}) \right)^* = & \upsigma^2 U(\vb{n}) \upsigma^2 & =
    U(\widetilde{\vb{n}}),
    \\
    \left( U(\vb{n}) \right)^{\dagger} = & \left( U(\widetilde{\vb{n}})
    \right)^T & = U(-\vb{n}),
    \\
    -U(\vb{n}) = & U(\widehat{\vb{n}}) &
    \label{eq:U_props}
  \end{eqnarray}
  where $\widetilde{\vb{n}} = \left( -n^1, +n^2, -n^3 \right)$ and
  $\widehat{\vb{n}} = - \left(\frac{1}{2} -n \right) \vb{n} / n$.

  The product of two elements is given by
  $U(\vb{n}\circ \vb{m})=U(\vb{n}) U(\vb{m})$ or more explicitly by:
  \begin{equation}
    \small
    \begin{split}
      \cos(2\uppi \norm{\vb{n}\circ \vb{m}}) & = \cos(2\uppi n)
      \cos(2\uppi m) - \sin(2\uppi n) \sin(2\uppi m) \frac{\vb{n}}{n}
      \cdot\frac{\vb{m}}{m},
      \\
      \sin(2\uppi \norm{\vb{n}\circ \vb{m}}) \frac{\vb{n}\circ
      \vb{m}} {\norm{\vb{n}\circ \vb{m}}} & = \cos(2\uppi n)
      \sin(2\uppi m) \frac{\vb{m}}{m} + \sin(2\uppi n) \cos(2\uppi m)
      \frac{\vb{n}}{n}.
    \end{split}
    \label{eq:product_in_SU2}
  \end{equation}

  \subsection{The Isomorphism}

  Let $I = 1,\, 2,\, 3,\, 4$ as in the main text and define:
  \begin{eqnarray*}
    \uptau_I & = & \left( i \mathds{1}_2, \vb{\upsigma} \right),
  \end{eqnarray*}
  where $\vb{\upsigma} = \left( \upsigma^1, \upsigma^2, \upsigma^3 \right)$ are
  the usual Pauli matrices. It is then easy to show that:
  \begin{equation}
    \begin{split}
      \left( \uptau_I \right)^{\dagger} & = \upeta_{I J} {\uptau}^I,
      \\
      \left( \uptau^I \right)^{*} & = - \upsigma_2 \uptau_I \upsigma_2,
    \end{split}
    \label{eq:tau_props}
  \end{equation}
  where $\upeta_{IJ} = \mathrm{diag}(-1,1,1,1)$. We have the following useful
  relations:
  \begin{eqnarray*}
    \tr(\uptau_I) & = & 2 i \updelta_{I1},
    \\
    \tr(\uptau_I \uptau_J) & = & 2 \upeta_{IJ},
    \\
    \tr(\uptau_I \left( \uptau_J \right)^{\dagger}) & = & 2 \updelta_{IJ}.
  \end{eqnarray*}

  Now consider a vector in this spinor representation:
  \begin{equation*}
    X_{(s)} = X^I \uptau_I.
  \end{equation*}
  We can recover its components using the previous properties:
  \begin{equation*}
    X^I = \frac{1}{2} \updelta^{IJ} \tr(X_{(s)} \left( \uptau_J
    \right)^{\dagger}) = \frac{1}{2} \upeta^{IJ} \tr(X_{(s)} \uptau_J).
  \end{equation*}
  If the vector $X^I$ is real, using the properties in \eqref{eq:tau_props},
  then we have
  \begin{equation}
    \begin{split}
      X_{(s)}^\dagger = & X^I \upeta_{I J} \uptau^J = \frac{1}{2}\tr(X_{(s)}
      \uptau_I) \uptau^I,
      \\
      X_{(s)}^* = & - \upsigma_2 X_{(s)}\upsigma_2.
    \end{split}
    \label{eq:X_dagger}
  \end{equation}

  A rotation in the spinor representation is defined as:
  \begin{equation}
    X'_{(s)} = \UL(\vb{n}) X_{(s)} \UR^{\dagger}(\vb{m})
  \end{equation}
  and it is equivalent to:
  \begin{equation}
    \left( X' \right)^I = R^I_{\,J} X^J
  \end{equation}
  through
  \begin{equation}
    R_{IJ} = \frac{1}{2} \tr(\left( \uptau_I \right)^{\dagger} \UL(\vb{n})
    \uptau_J \UR^{\dagger}(\vb{m})).
  \end{equation}
  $R$ is indeed the matrix we are looking for since
  \begin{equation*}
    \tr(X'_{(s)} X^{'\dagger}_{(s)}) = \tr(X_{(s)} X^{\dagger}_{(s)})
    \Rightarrow R_{I\,K} R^*_{J \,K} = \updelta_{I \,J}.
  \end{equation*}

  It is then necessary to show that $R$ is a real matrix. From the second
  equation in \eqref{eq:tau_props} and the first equation in \eqref{eq:U_props}
  we get:
  \begin{equation*}
    R_{N M }
    = \frac{1}{2} \upeta_{N I} \upeta_{M J}
    \tr(\uptau_I ^{\dagger} \UR\uptau_J \UL^{\dagger})
    = \frac{1}{2}
    \tr(\uptau_N  \UR\uptau_M ^\dagger \UL^{\dagger})
    = ( R_{N M} )^*.
  \end{equation*}
  The property $R\in \mathrm{SO}(4, \mathds{R})$ can also be shown by a direct
  computation of the determinant using the parametrization of the
  $\mathrm{SU}(2)$ matrices: we find
  \begin{equation*}
    \det(R) = 1.
  \end{equation*}
  Moreover the explicit  choice of the basis $\uptau$ ensures $R$ to be a real
  matrix.

  Since $\left\lbrace \UL, \UR \right\rbrace$ and $\left\lbrace -\UL, -\UR
  \right\rbrace$ generate the same $\mathrm{SO}(4)$ matrix then the correct
  isomorphism takes the form:
  \begin{equation*}
    \mathrm{SO}(4) \cong \frac{\mathrm{SU}(2) \times
    \mathrm{SU}(2)}{\mathds{Z}_2}.
  \end{equation*}

  \section{The Parameters of the Hypergeometric Function}
  \label{sec:parameters}

  \subsection{Consistency Conditions for \texorpdfstring{$\mathrm{U}(2)$}{U(2)}
  and \texorpdfstring{$\mathrm{U}(1,1)$}{U(1,1)} Monodromies}
  In the main text we have set
  \begin{equation}
          D M_{\iOm{\infty}} \left( D \right)^{-1} = e^{-2\uppi i
            \updelta_{\iOm{\infty}}} \mathcal{L}(\vb{n}_{\iOm{\infty}})
          ,
\end{equation}
where $\mathcal{L}(\vb{n}_{\iOm{\infty}})$ is a $\mathrm{SU}(2)$ matrix.
This is a somewhat strong statement which may imply and implies some
consistency conditions.
The previous equation implies
\begin{equation}
  [D M_{\iOm{\infty}} \left( D \right)^{-1} ]^\dagger
  =
    [D M_{\iOm{\infty}} \left( D \right)^{-1} ]^{-1}
,
\end{equation}
which can be rewritten as
\begin{equation}
  \widetilde{M}_{\iOm{\infty}}^{-1}   ~{\mathcal C}^\dagger D^\dagger D {\mathcal C}
  =
  {\mathcal C}^\dagger D^\dagger D {\mathcal C}
  ~\widetilde{M}_{\iOm{\infty}}^{-1}
  .
  \end{equation}
Since $\widetilde{M}_{\iOm{\infty}}$ is a generic diagonal matrix the
previous equation implies that the off-diagonal elements of
$ {\mathcal C}^\dagger D^\dagger D {\mathcal C}$ must vanish.
This means that
\begin{align}
  |K|^{-2}
  &=
    - \frac{
    {\mathcal C}_{21} {\mathcal C}^*_{22}
    }{
{\mathcal C}_{11} {\mathcal C}^*_{12}
    } =
    \nonumber\\
    &=-
    \frac{
    \Upgamma^*(a) \Upgamma(b) \Upgamma(c-a) \Upgamma^*(c-b)
    }{
    \Upgamma(1-a) \Upgamma^*(1-b) \Upgamma^*(1-c+a) \Upgamma(1-c+b)
    } =
    \nonumber\\
    &=
    -\frac{1}{\uppi^4}
    \abs{\Upgamma(a) \Upgamma(b) \Upgamma(c-a) \Upgamma(c-b)}^2 \times
    \nonumber\\
    &\times
    ~\sin(\uppi a) \sin^*(\uppi (c-a))
    ~(\sin(\uppi b) \sin^*(\uppi (c-b)))^*
    .
\end{align}
For real $a, b$ and $c$ this means that
\begin{equation}
\sin(\uppi a) \sin(\uppi (c-a))
~\sin(\uppi b) \sin(\uppi (c-b))<0
\label{eq:constraint_from_K^2}
.
\end{equation}
The previous equation is invariant under integer shift of any of the
three parameters therefore we can limit to consider what happens to
the fractional parts $0\le \{a\}, \{b\}, \{c\}<1 $.
Finally we get that the previous equation, i.e. the original
position of having $U(2)$ monodromies requires
\begin{align}
  \qq{either}
  0\le \{b\}< \{c\}< \{a\}<1
  \qq{or}
  0\le \{a\}< \{c\}< \{b\}<1
  .
  \label{eq:K_consistency_condition}
  \end{align}

  Should we require the monodromies be in $U(1,1)$, as required by
  moving rotated branes, then we would get
  \begin{align}
  |K|^{-2}
  &=
    + \frac{
    {\mathcal C}_{21} {\mathcal C}^*_{22}
    }{
{\mathcal C}_{11} {\mathcal C}^*_{12}
    }
    ,
  \end{align}
  which would imply
\begin{align}
  \qq{either}
  0\le \{c\}< \{a\},\{b\}<1
  \qq{or}
  0\le \{a\}, \{b\}< \{c\}<1
  .
  \end{align}

  \subsection{Fixing the Parameters}

  In this appendix we show in a detailed way how to compute the parameters of
  the basis of hypergeometric functions we used in the main text. The relation
  between such parameters and the $\mathrm{SU}(2)$ matrices are computed
  requiring that the monodromies induced by the choice of the parameters equal
  the monodromies of the rotations of the D-branes.

  The monodromy in $\upomega_{\overline{t}-1} = 0$ is simpler to compute since
  we chose $\mathcal{L}(\vb{n}_{\iOm{0}})$ and
  $\mathcal{R}(\widetilde{\vb{m}}_{\iOm{0}})$ to be diagonal. We impose:
  \begin{eqnarray*}
    \mqty( 1 & \\ & e^{-2\uppi i c^{(L)}} ) & = & e^{-2\uppi i
    \updelta_{\iOm{0}}^{(L)}} \mqty( e^{2\uppi i n_{\iOm{0}}} & \\ & e^{-2\uppi i
    n_{\iOm{0}}} ),
    \\
    \mqty( 1 & \\ & e^{-2\uppi i c^{(R)}} ) & = & e^{-2\uppi i
    \updelta_{\iOm{0}}^{(R)}} \mqty( e^{-2\uppi i m_{\iOm{0}}} & \\ & e^{2\uppi i
    m_{\iOm{0}}} ),
  \end{eqnarray*}
  where $n^3_{\iOm{0}}=n_{\iOm{0}}$ and $m^3_{\iOm{0}}=m_{\iOm{0}}$  with $0\le  n_{\iOm{0}},
  m_{\iOm{0}} <1 $ with the conventions of \eqref{eq:maximal_torus_left} and
  \eqref{eq:maximal_torus_right}. In the left  sector we therefore find:
  \begin{equation}
    \begin{split}
      \updelta_{\iOm{0}}^{(L)} & = n_{\iOm{0}}+
      k_{\updelta^{(L)}_{\iOm{0}}}
      , \qq{where} k_{\updelta^{(L)}_{\iOm{0}}}\in\mathds{Z},
      \\
      c^{(L)} & = 2 n_{\iOm{0}}+ k_c, \qq{where} k_c\in\mathds{Z}.
    \label{eq:cL}
    \end{split}
  \end{equation}
  Since $e^{i 4 \uppi  \updelta_{\iOm{0}}^{(L)}}$ is the determinant of the right
  hand side the range of definition of $\updelta_{\iOm{0}}^{(L)}$ is $\upalpha \le
  \updelta_{\iOm{0}}^{(L)} \le  \upalpha +\frac{1}{2}$ since $0\le
  n_{\iOm{0}}< \frac{1}{2}$  we can simply take
  $\upalpha=0$ and set
  \begin{equation*}
    \updelta_{\iOm{0}}^{(L)}  = n_{\iOm{0}}.
  \end{equation*}
  Analogous results hold in the right sector. From the third equation in
  \eqref{eq:parameters_equality_zero} and from the first equation in
  \eqref{eq:cL} we find:
  \begin{equation*}
    -A + n_{\iOm{0}} + m_{\iOm{0}} \in \mathds{Z}.
  \end{equation*}

  We now need to fix the $6$ parameters $a^{(L)}, b^{(L)},
  \updelta^{(L)}_{\iOm{\infty}}$, $B$ and $\abs{K^{(L)}}, \mathcal{T}^{(L)}$. Our
  strategy is first to find $3$ equations to determine $a^{(L)},
  b^{(L)},\updelta^{(L)}_{\iOm{\infty}}$ and then fix the remaining ones. Clearly
  everything holds true also for the right sector. All these equations follow
  from imposing the requests \eqref{eq:parameters_equality_infty}. The first
  two equations for $a^{(L)},b^{(L)},\updelta^{(L)}_{\iOm{\infty}}$ follow easily by
  considering the trace of \eqref{eq:parameters_equality_infty}:
  \begin{equation*}
    e^{i\uppi ( a^{(L)} + b^{(L)} )} \cos(\uppi( a^{(L)} - b^{(L)} ) ) =
    e^{-2\uppi i \updelta^{(L)}_{\infty}} \cos(2\uppi n_{\iOm{\infty}}),
  \end{equation*}
  which translates into:
  \begin{equation*}
    \begin{split}
      \updelta^{(L)}_{\iOm{\infty}} & = -\frac{1}{2}(a^{(L)} +
      b^{(L)})
      + \frac{1}{2}
      k_{\updelta^{(L)}_{\iOm{\infty}}}, \qq{where} k_{\updelta_{\iOm{\infty}}} \in
      \mathds{Z},
      \\
      a^{(L)} - b^{(L)} & = 2 (-1)^{p^{(L)}} n_{\iOm{\infty}} +
      k_{\updelta^{(L)}_{\iOm{\infty}}} (-1)^{q^{(L)}} + 2 k'_{a b}, \qq{where} k'_{a
      b}\in \mathds{Z},
    \end{split}
  \end{equation*}
  with $p^{(L)},q^{(L)} \in \left\lbrace 0, 1 \right\rbrace$. A change of value
  of $p^{(L)}$ corresponds to an exchange between $a$ and $b$: since the
  hypergeometric function is symmetric in $a$ and $b$ we can fix $p^{(L)}=0$.
  Redefining $k'$ we can always set $q^{(L)}=0$, hence we can write
  \begin{equation}
    a^{(L)} - b^{(L)}  = 2 n_{\iOm{\infty}} + k_{\updelta^{(L)}_{\iOm{\infty}}} + 2 k_{a
    b}, \qq{where} k_{a b}\in \mathds{Z}.
    \label{eq:aL-bL}
  \end{equation}
  A discussion of the possible values of $k_{\updelta^{(L)}_{\iOm{\infty}}} $ is
  analogous to what done for the monodromy around $\upomega_{\overline{t}-1} =
  0$
  but with an important difference: $ \frac{1}{2}(a^{(L)} + b^{(L)})$
 may a priori take values in an interval of width $1$.
 Because of this
 since also in this case we have $\upalpha \le \updelta_{\iOm{\infty}}^{(L)} \le
 \upalpha +\frac{1}{2}$ with $\upalpha$ \emph{a priori} arbitrary
 we cannot choose $k_{\updelta^{(L)}_{\iOm{\infty}}} =0$ but we have to
 consider $k_{\updelta^{(L)}_{\infty}} =0,1$.

  We then find a third relation by considering:
  \begin{equation*}
    \Im\left( e^{+2\uppi i \updelta_{\iOm{\infty}}^{(L)}} D^{(L)}
    M_{\iOm{\infty}}^{(L)} \left( D^{(L)} \right)^{-1} \right)_{11} = \Im\left(
    \mathcal{L}(n_{\iOm{\infty}}) \right)_{11}.
  \end{equation*}
  With the help of
  \begin{equation*}
    \det \mathcal{C} = \frac{\sin(\uppi c^{(L)}) }{\sin(\uppi
    (a^{(L)}-b^{(L)}))},
  \end{equation*}
  and the second equation in \eqref{eq:cL} and \eqref{eq:aL-bL}, it leads to:
  \begin{equation*}
    \cos(\uppi( a^{(L)} + b^{(L)} - c^{(L)} )) =
    (-1)^{k_c+k_{\updelta^{(L)}_{\iOm{\infty}}} } \cos(2\uppi \mathcal{A}^{(L)}),
  \end{equation*}
  where
  \begin{equation}
    \cos(2\uppi \mathcal{A}^{(L)})
    =
    \cos(2\uppi n_{\iOm{0}}) \cos(2\uppi
    n_{\iOm{\infty}}) - \sin(2\uppi n_{\iOm{0}}) \sin(2\uppi n_{\iOm{\infty}})
    \frac{n_{\iOm{\infty}}^3}{n_{\iOm{\infty}}}.
\label{eq:cos_n1}
  \end{equation}
  The rotation parameter in the third interaction point
  $\upomega_{\overline{t}+1} = 1$ is connected with the previous expression as
  \begin{equation*}
    \cos(2\uppi \mathcal{A}^{(L)}) = \cos(2\uppi {n}_{\iOm{1}}).
  \end{equation*}
  Then we can write:
  \begin{eqnarray*}
    a^{(L)} + b^{(L)} - c^{(L)} &=& 2 (-1)^{f^{(L)}} n_{\iOm{1}} +k_c+
    k_{\updelta^{(L)}_{\iOm{\infty}}} + 2 k_{abc}, \qq{where} k_{abc}\in \mathds{Z},
  \end{eqnarray*}
  with $f^{(L)} \in \left\lbrace 0, 1 \right\rbrace$.

  We then fix the $B$  parameter in the third equation of
  \eqref{eq:parameters_equality_infty} requiring:
  \begin{equation*}
    A + B - n_{\iOm{0}} - m_{\iOm{0}}
    - (-1)^{f^{(L)}} n_{\iOm{1}} - (-1)^{f^{(R)}} m_{\iOm{1}}
    \in \mathds{Z}.
  \end{equation*}

  We can summarize the results so far as
  \begin{eqnarray*}
    a & = & n_{\iOm{0}} + (-1)^{f^{(L)}} n_{\iOm{1}} + n_{\iOm{\infty}} + m_a,
    \\
    b & = & n_{\iOm{0}} + (-1)^{f^{(L)}} n_{\iOm{1}} - n_{\iOm{\infty}} + m_b,
    \\
    c & = & 2 n_{\iOm{0}} + m_c,
    \\
    \updelta_{\iOm{0}}^{(L)}  & = & n_{\iOm{0}},
    \\
    \updelta_{\iOm{\infty}}^{(L)} & = & -n_{\iOm{0}} - (-1)^{f^{(L)}} n_{\iOm{1}}
                                    +m_c+ 2 m_\updelta
    \\
    A & = & n_{\iOm{0}} + m_{\iOm{0}} + m_A,
    \\
    B & = & (-1)^{f^{(L)}} n_{\iOm{1}} + (-1)^{f^{(R)}} m_{\iOm{1}} + m_B,
  \end{eqnarray*}
  where all the factors $m$ are integers.

  Finally we determine $K^{(L)}$. To this purpose we consider:
  \begin{equation}
    \left( D^{(L)} M_{\iOm{\infty}} \left( D^{(L)} \right)^{-1} \right)_{21}  =
    e^{-2\uppi i \updelta_{\iOm{\infty}}^{(L)}} \left( \mathcal{L}(n_{\iOm{\infty}})
    \right)_{21},
    \label{eq:fixing_K_21}
  \end{equation}
  and get:
  \begin{eqnarray}
    K^{(L)} & = &
    -\frac{1}{2 \uppi^2}
                  \Upgamma(1-a^{(L)}) \Upgamma(1-b^{(L)})
    \Upgamma(a^{(L)}+1-c^{(L)}) \Upgamma(b^{(L)}+1-c^{(L)}) \times
    \nonumber\\
    & \times &
    \sin(\uppi c) \sin(\uppi (a-b) ) \frac{n^1_{\iOm{\infty}} + i
              n^2_{\iOm{\infty}}}{n_{\iOm{\infty}}}
              \nonumber\\
    &=&
    -\frac{(-1)^{m_a+m_b+m_c} }{2 \uppi^2}
                  \Upgamma(1-a^{(L)}) \Upgamma(1-b^{(L)})
    \Upgamma(a^{(L)}+1-c^{(L)}) \Upgamma(b^{(L)}+1-c^{(L)}) \times
    \nonumber\\
    & \times &
    \sin(2 \uppi n_{\iOm{0}}) \sin(2 \uppi n_{\iOm{\infty}}) \frac{n^1_{\iOm{\infty}} + i
              n^2_{\iOm{\infty}}}{n_{\iOm{\infty}}}
              .
\label{eq:app_B_K21}
  \end{eqnarray}

  \subsection{Checking the Consistency of the Solution}
  Given the previous solution we can now check the consistency condition
  \eqref{eq:K_consistency_condition} with the help  of
  \eqref{eq:product_in_SU2}.
  Another way of performing this check is to compute $K^{(L)}$  from
    \begin{equation}
    \left( D^{(L)} M_{\iOm{\infty}} \left( D^{(L)} \right)^{-1} \right)_{12}  =
    e^{-2\uppi i \updelta_{\iOm{\infty}}^{(L)}} \left( \mathcal{L}(n_{\iOm{\infty}})
    \right)_{12},
  \end{equation}
  instead of \eqref{eq:fixing_K_21}.
  The result is
    \begin{eqnarray}
    \frac{1}{K^{(L)}} & = &
    \frac{(-1)^{m_a+m_b+m_c} }{2 \uppi^2}
                  \Upgamma(a^{(L)}) \Upgamma(b^{(L)})
    \Upgamma(-a^{(L)}+c^{(L)}) \Upgamma(-b^{(L)}+c^{(L)}) \times
    \nonumber\\
    & \times &
              \sin(2 \uppi n_{\iOm{0}}) \sin(2 \uppi n_{\iOm{\infty}})
              \frac{n^1_{\iOm{\infty}} -i n^2_{\iOm{\infty}}}{n_{\iOm{\infty}}}
              .
\label{eq:app_B_K12}
  \end{eqnarray}
  This expression and \eqref{eq:app_B_K21} are compatible only if
  \begin{align}
    \frac{(n^1_{\iOm{\infty}})^2 + (n^2_{\iOm{\infty}})^2}{n^2_{\iOm{\infty}}}
    &=
     -4 \frac{
\sin(\uppi a) \sin(\uppi(c-a))\sin(\uppi b) \sin(\uppi(c-b))
      }{
      \sin^2(\uppi c) \sin^2(\uppi(a-b))
      }
      .
      \label{eq: n12+n22}
    \end{align}
Notice that this equation may be true only if the constraint found
before and expressed in \eqref{eq:constraint_from_K^2} is true.
To proof it we rewrite \eqref{eq:cos_n1} as
  \begin{align}
    \frac{(n^3_{\iOm{\infty}})^2}{n^2_{\iOm{\infty}}}
    &=
      \frac{
      (
      \cos(\uppi (a-b)) \cos(\uppi c)- \cos(\uppi(a+b-c))
      )^2
      }{
      \sin^2(\uppi c) \sin^2(\uppi(a-b))
      }
      ,
\end{align}
and then we verify that the sum of the right hand side of this
equation and the  the right hand side of \eqref{eq: n12+n22} is equal
to $1$.

\end{document}

%% file: img/branes-at-angles.pdf_tex
\begingroup%
  \makeatletter%
  \providecommand\color[2][]{%
    \errmessage{(Inkscape) Color is used for the text in Inkscape, but the package 'color.sty' is not loaded}%
    \renewcommand\color[2][]{}%
  }%
  \providecommand\transparent[1]{%
    \errmessage{(Inkscape) Transparency is used (non-zero) for the text in Inkscape, but the package 'transparent.sty' is not loaded}%
    \renewcommand\transparent[1]{}%
  }%
  \providecommand\rotatebox[2]{#2}%
  \newcommand*\fsize{\dimexpr\f@size pt\relax}%
  \newcommand*\lineheight[1]{\fontsize{\fsize}{#1\fsize}\selectfont}%
  \ifx\svgwidth\undefined%
    \setlength{\unitlength}{257.26047483bp}%
    \ifx\svgscale\undefined%
      \relax%
    \else%
      \setlength{\unitlength}{\unitlength * \real{\svgscale}}%
    \fi%
  \else%
    \setlength{\unitlength}{\svgwidth}%
  \fi%
  \global\let\svgwidth\undefined%
  \global\let\svgscale\undefined%
  \makeatother%
  \begin{picture}(1,0.76663542)%
    \lineheight{1}%
    \setlength\tabcolsep{0pt}%
    \put(0,0){\includegraphics[width=\unitlength,page=1]{branes-at-angles.pdf}}%
    \put(0.84263173,0.35076381){\color[rgb]{0,0,0}\makebox(0,0)[lt]{\lineheight{0}\smash{\begin{tabular}[t]{l}$X^1$\end{tabular}}}}%
    \put(0.37837622,0.73404809){\color[rgb]{0,0,0}\makebox(0,0)[lt]{\lineheight{0}\smash{\begin{tabular}[t]{l}$X^2$\end{tabular}}}}%
    \put(0,0){\includegraphics[width=\unitlength,page=2]{branes-at-angles.pdf}}%
    \put(0.31219038,0.15896038){\color[rgb]{0,0,0}\makebox(0,0)[lt]{\lineheight{0}\smash{\begin{tabular}[t]{l}$D_{\iD{1}}$\end{tabular}}}}%
    \put(0.17346322,0.46159691){\color[rgb]{0,0,0}\makebox(0,0)[lt]{\lineheight{0.80000001}\smash{\begin{tabular}[t]{l}$D_{\iD{2}}$\end{tabular}}}}%
    \put(0.58494162,0.40717731){\color[rgb]{0,0,0}\makebox(0,0)[lt]{\lineheight{0}\smash{\begin{tabular}[t]{l}$D_{\iD{3}}$\end{tabular}}}}%
    \put(0.14858524,0.22079414){\color[rgb]{0,0,0}\makebox(0,0)[lt]{\lineheight{0}\smash{\begin{tabular}[t]{l}$f_{\iD{1}}$\end{tabular}}}}%
    \put(0.4597515,0.6317968){\color[rgb]{0,0,0}\makebox(0,0)[lt]{\lineheight{0}\smash{\begin{tabular}[t]{l}$f_{\iD{2}}$\end{tabular}}}}%
    \put(0.66654893,0.07252184){\color[rgb]{0,0,0}\makebox(0,0)[lt]{\lineheight{0}\smash{\begin{tabular}[t]{l}$f_{\iD{3}}$\end{tabular}}}}%
    \put(0,0){\includegraphics[width=\unitlength,page=3]{branes-at-angles.pdf}}%
    \put(0.35272736,0.26002516){\color[rgb]{0,0,0}\makebox(0,0)[lt]{\lineheight{0}\smash{\begin{tabular}[t]{l}$g_{\iD{1}}$\end{tabular}}}}%
    \put(0.30210448,0.36941682){\color[rgb]{0,0,0}\makebox(0,0)[lt]{\lineheight{0}\smash{\begin{tabular}[t]{l}$g_{\iD{2}}$\end{tabular}}}}%
    \put(0.4886858,0.38440996){\color[rgb]{0,0,0}\makebox(0,0)[lt]{\lineheight{0}\smash{\begin{tabular}[t]{l}$g_{\iD{3}}$\end{tabular}}}}%
    \put(0,0){\includegraphics[width=\unitlength,page=4]{branes-at-angles.pdf}}%
    \put(0.63987354,0.17175434){\color[rgb]{0,0,0}\makebox(0,0)[lt]{\lineheight{0}\smash{\begin{tabular}[t]{l}$\pi \upalpha_{\iD{1}}$\end{tabular}}}}%
    \put(0,0){\includegraphics[width=\unitlength,page=5]{branes-at-angles.pdf}}%
    \put(0.16328442,0.2932239){\color[rgb]{0,0,0}\makebox(0,0)[lt]{\lineheight{0}\smash{\begin{tabular}[t]{l}$\pi \upalpha_{\iD{2}}$\end{tabular}}}}%
    \put(0.5284708,0.71025944){\color[rgb]{0,0,0}\makebox(0,0)[lt]{\lineheight{0}\smash{\begin{tabular}[t]{l}$\pi \upalpha_{\iD{3}}$\end{tabular}}}}%
    \put(0,0){\includegraphics[width=\unitlength,page=6]{branes-at-angles.pdf}}%
  \end{picture}%
\endgroup%

%% file: img/branes-conformal-gauge.pdf_tex
\begingroup%
  \makeatletter%
  \providecommand\color[2][]{%
    \errmessage{(Inkscape) Color is used for the text in Inkscape, but the package 'color.sty' is not loaded}%
    \renewcommand\color[2][]{}%
  }%
  \providecommand\transparent[1]{%
    \errmessage{(Inkscape) Transparency is used (non-zero) for the text in Inkscape, but the package 'transparent.sty' is not loaded}%
    \renewcommand\transparent[1]{}%
  }%
  \providecommand\rotatebox[2]{#2}%
  \newcommand*\fsize{\dimexpr\f@size pt\relax}%
  \newcommand*\lineheight[1]{\fontsize{\fsize}{#1\fsize}\selectfont}%
  \ifx\svgwidth\undefined%
    \setlength{\unitlength}{182.16883725bp}%
    \ifx\svgscale\undefined%
      \relax%
    \else%
      \setlength{\unitlength}{\unitlength * \real{\svgscale}}%
    \fi%
  \else%
    \setlength{\unitlength}{\svgwidth}%
  \fi%
  \global\let\svgwidth\undefined%
  \global\let\svgscale\undefined%
  \makeatother%
  \begin{picture}(1,0.3477229)%
    \lineheight{1}%
    \setlength\tabcolsep{0pt}%
    \put(0,0){\includegraphics[width=\unitlength,page=1]{branes-conformal-gauge.pdf}}%
    \put(0.21224687,0.32948417){\color[rgb]{0,0,0}\makebox(0,0)[lt]{\lineheight{0}\smash{\begin{tabular}[t]{l}$\Im u$\end{tabular}}}}%
    \put(0.89529592,0.01355223){\color[rgb]{0,0,0}\makebox(0,0)[lt]{\lineheight{0}\smash{\begin{tabular}[t]{l}$\Re u$\end{tabular}}}}%
    \put(0.84412925,0.1504503){\color[rgb]{0,0,0}\makebox(0,0)[lt]{\lineheight{0}\smash{\begin{tabular}[t]{l}$D_{\iD{1}}$\end{tabular}}}}%
    \put(0,0){\includegraphics[width=\unitlength,page=2]{branes-conformal-gauge.pdf}}%
    \put(0.66688889,0.14740003){\color[rgb]{0,0,0}\makebox(0,0)[lt]{\lineheight{0}\smash{\begin{tabular}[t]{l}$D_{\iD{2}}$\end{tabular}}}}%
    \put(0.46080069,0.15014473){\color[rgb]{0,0,0}\makebox(0,0)[lt]{\lineheight{0}\smash{\begin{tabular}[t]{l}$D_{\iD{3}}$\end{tabular}}}}%
    \put(0.00643408,0.14865471){\color[rgb]{0,0,0}\makebox(0,0)[lt]{\lineheight{0}\smash{\begin{tabular}[t]{l}$D_{\iD{1}}$\end{tabular}}}}%
    \put(0.04167137,0.02037686){\color[rgb]{0,0,0}\makebox(0,0)[lt]{\lineheight{0}\smash{\begin{tabular}[t]{l}$x_{N_B}$\end{tabular}}}}%
    \put(0.33970692,0.02063565){\color[rgb]{0,0,0}\makebox(0,0)[lt]{\lineheight{0}\smash{\begin{tabular}[t]{l}$x_{3}$\end{tabular}}}}%
    \put(0.58523209,0.02141983){\color[rgb]{0,0,0}\makebox(0,0)[lt]{\lineheight{0}\smash{\begin{tabular}[t]{l}$x_{2}$\end{tabular}}}}%
    \put(0.76089317,0.02141987){\color[rgb]{0,0,0}\makebox(0,0)[lt]{\lineheight{0}\smash{\begin{tabular}[t]{l}$x_{1}$\end{tabular}}}}%
    \put(0.16704824,0.02119642){\color[rgb]{0,0,0}\makebox(0,0)[lt]{\lineheight{0}\smash{\begin{tabular}[t]{l}$\cdots$\end{tabular}}}}%
    \put(0.16780425,0.15052048){\color[rgb]{0,0,0}\makebox(0,0)[lt]{\lineheight{0}\smash{\begin{tabular}[t]{l}$\cdots$\end{tabular}}}}%
  \end{picture}%
\endgroup%

%% file: img/finite-cuts.pdf_tex
\begingroup%
  \makeatletter%
  \providecommand\color[2][]{%
    \errmessage{(Inkscape) Color is used for the text in Inkscape, but the package 'color.sty' is not loaded}%
    \renewcommand\color[2][]{}%
  }%
  \providecommand\transparent[1]{%
    \errmessage{(Inkscape) Transparency is used (non-zero) for the text in Inkscape, but the package 'transparent.sty' is not loaded}%
    \renewcommand\transparent[1]{}%
  }%
  \providecommand\rotatebox[2]{#2}%
  \ifx\svgwidth\undefined%
    \setlength{\unitlength}{183.27074667bp}%
    \ifx\svgscale\undefined%
      \relax%
    \else%
      \setlength{\unitlength}{\unitlength * \real{\svgscale}}%
    \fi%
  \else%
    \setlength{\unitlength}{\svgwidth}%
  \fi%
  \global\let\svgwidth\undefined%
  \global\let\svgscale\undefined%
  \makeatother%
  \begin{picture}(1,0.39441195)%
    \put(0,0){\includegraphics[width=\unitlength,page=1]{finite-cuts.pdf}}%
    \put(0.23732465,0.35689043){\color[rgb]{0,0,0}\makebox(0,0)[lb]{\smash{$\Im u$}}}%
    \put(0.91626688,0.11270004){\color[rgb]{0,0,0}\makebox(0,0)[lb]{\smash{$\Re u$}}}%
    \put(0,0){\includegraphics[width=\unitlength,page=2]{finite-cuts.pdf}}%
    \put(0.06777473,0.04964162){\color[rgb]{0,0,0}\makebox(0,0)[lb]{\smash{$x_{4}$}}}%
    \put(0.31163684,0.04989886){\color[rgb]{0,0,0}\makebox(0,0)[lb]{\smash{$x_{3}$}}}%
    \put(0.52076477,0.05067832){\color[rgb]{0,0,0}\makebox(0,0)[lb]{\smash{$x_{2}$}}}%
    \put(0.8263235,0.05067835){\color[rgb]{0,0,0}\makebox(0,0)[lb]{\smash{$x_{1}$}}}%
    \put(0,0){\includegraphics[width=\unitlength,page=3]{finite-cuts.pdf}}%
  \end{picture}%
\endgroup%

%% file: img/hypergeometric-cuts.pdf_tex
\begingroup%
  \makeatletter%
  \providecommand\color[2][]{%
    \errmessage{(Inkscape) Color is used for the text in Inkscape, but the package 'color.sty' is not loaded}%
    \renewcommand\color[2][]{}%
  }%
  \providecommand\transparent[1]{%
    \errmessage{(Inkscape) Transparency is used (non-zero) for the text in Inkscape, but the package 'transparent.sty' is not loaded}%
    \renewcommand\transparent[1]{}%
  }%
  \providecommand\rotatebox[2]{#2}%
  \ifx\svgwidth\undefined%
    \setlength{\unitlength}{155.41420884bp}%
    \ifx\svgscale\undefined%
      \relax%
    \else%
      \setlength{\unitlength}{\unitlength * \real{\svgscale}}%
    \fi%
  \else%
    \setlength{\unitlength}{\svgwidth}%
  \fi%
  \global\let\svgwidth\undefined%
  \global\let\svgscale\undefined%
  \makeatother%
  \begin{picture}(1,0.46510659)%
    \put(0,0){\includegraphics[width=\unitlength,page=1]{hypergeometric-cuts.pdf}}%
    \put(0.10046991,0.42085969){\color[rgb]{0,0,0}\makebox(0,0)[lb]{\smash{$\Im \upomega_z$}}}%
    \put(0.89940959,0.17833561){\color[rgb]{0,0,0}\makebox(0,0)[lb]{\smash{$\Re \upomega_z$}}}%
    \put(0,0){\includegraphics[width=\unitlength,page=2]{hypergeometric-cuts.pdf}}%
    \put(0.0847975,0.06487667){\color[rgb]{0,0,0}\makebox(0,0)[lb]{\smash{$0$}}}%
    \put(0.30932208,0.05721781){\color[rgb]{0,0,0}\makebox(0,0)[lb]{\smash{$1$}}}%
    \put(0.94257337,0.06205993){\color[rgb]{0,0,0}\makebox(0,0)[lb]{\smash{$\infty$}}}%
    \put(0,0){\includegraphics[width=\unitlength,page=3]{hypergeometric-cuts.pdf}}%
  \end{picture}%
\endgroup%

%% file: img/Abelian_angles_case1.pdf_tex
\begingroup%
  \makeatletter%
  \providecommand\color[2][]{%
    \errmessage{(Inkscape) Color is used for the text in Inkscape, but the package 'color.sty' is not loaded}%
    \renewcommand\color[2][]{}%
  }%
  \providecommand\transparent[1]{%
    \errmessage{(Inkscape) Transparency is used (non-zero) for the text in Inkscape, but the package 'transparent.sty' is not loaded}%
    \renewcommand\transparent[1]{}%
  }%
  \providecommand\rotatebox[2]{#2}%
  \newcommand*\fsize{\dimexpr\f@size pt\relax}%
  \newcommand*\lineheight[1]{\fontsize{\fsize}{#1\fsize}\selectfont}%
  \ifx\svgwidth\undefined%
    \setlength{\unitlength}{816.64086535bp}%
    \ifx\svgscale\undefined%
      \relax%
    \else%
      \setlength{\unitlength}{\unitlength * \real{\svgscale}}%
    \fi%
  \else%
    \setlength{\unitlength}{\svgwidth}%
  \fi%
  \global\let\svgwidth\undefined%
  \global\let\svgscale\undefined%
  \makeatother%
  \begin{picture}(1,0.2830841)%
    \lineheight{1}%
    \setlength\tabcolsep{0pt}%
    \put(0,0){\includegraphics[width=\unitlength,page=1]{Abelian_angles_case1.pdf}}%
    \put(0.34933769,0.09038652){\color[rgb]{0,0,0}\makebox(0,0)[lt]{\lineheight{1.25000012}\smash{\begin{tabular}[t]{l}$D_{\iD{3}}$\end{tabular}}}}%
    \put(0.28964196,0.26947381){\color[rgb]{0,0,0}\makebox(0,0)[lt]{\lineheight{1.25000012}\smash{\begin{tabular}[t]{l}$D_{\iD{1}}$\end{tabular}}}}%
    \put(0.04942886,0.27343344){\color[rgb]{0,0,0}\makebox(0,0)[lt]{\lineheight{1.25000012}\smash{\begin{tabular}[t]{l}$D_{\iD{2}}$\end{tabular}}}}%
    \put(0.05461488,0.06269528){\color[rgb]{0,0,0}\makebox(0,0)[lt]{\lineheight{1.25}\smash{\begin{tabular}[t]{l}$n^3_{\iOm{0}}>0$\end{tabular}}}}%
    \put(0,0){\includegraphics[width=\unitlength,page=2]{Abelian_angles_case1.pdf}}%
    \put(0.12994782,0.24109199){\color[rgb]{0,0,0}\makebox(0,0)[lt]{\lineheight{1.25}\smash{\begin{tabular}[t]{l}$n^3_{\iOm{\infty}}>0$\end{tabular}}}}%
    \put(0,0){\includegraphics[width=\unitlength,page=3]{Abelian_angles_case1.pdf}}%
    \put(0.22994621,0.13301883){\color[rgb]{0,0,0}\makebox(0,0)[lt]{\lineheight{1.25}\smash{\begin{tabular}[t]{l}$n^3_{\iOm{1}}<0$\end{tabular}}}}%
    \put(0,0){\includegraphics[width=\unitlength,page=4]{Abelian_angles_case1.pdf}}%
    \put(0.88200762,0.08120258){\color[rgb]{0,0,0}\makebox(0,0)[lt]{\lineheight{1.25000012}\smash{\begin{tabular}[t]{l}$D_{\iD{3}}$\end{tabular}}}}%
    \put(0.82231189,0.26028982){\color[rgb]{0,0,0}\makebox(0,0)[lt]{\lineheight{1.25000012}\smash{\begin{tabular}[t]{l}$D_{\iD{1}}$\end{tabular}}}}%
    \put(0.79350357,0.00323504){\color[rgb]{0,0,0}\makebox(0,0)[lt]{\lineheight{1.25000012}\smash{\begin{tabular}[t]{l}$D_{\iD{2}}$\end{tabular}}}}%
    \put(0.60253017,0.04036875){\color[rgb]{0,0,0}\makebox(0,0)[lt]{\lineheight{1.25}\smash{\begin{tabular}[t]{l}$n^3_{\iOm{0}}>0$\end{tabular}}}}%
    \put(0,0){\includegraphics[width=\unitlength,page=5]{Abelian_angles_case1.pdf}}%
    \put(0.75139217,0.18475055){\color[rgb]{0,0,0}\makebox(0,0)[lt]{\lineheight{1.25}\smash{\begin{tabular}[t]{l}$n^3_{\iOm{\infty}}<0$\end{tabular}}}}%
    \put(0,0){\includegraphics[width=\unitlength,page=6]{Abelian_angles_case1.pdf}}%
    \put(0.89351593,0.03499128){\color[rgb]{0,0,0}\makebox(0,0)[lt]{\lineheight{1.25}\smash{\begin{tabular}[t]{l}$n^3_{\iOm{1}}>0$\end{tabular}}}}%
  \end{picture}%
\endgroup%

%% file: img/Abelian_angles_case2.pdf_tex
\begingroup%
  \makeatletter%
  \providecommand\color[2][]{%
    \errmessage{(Inkscape) Color is used for the text in Inkscape, but the package 'color.sty' is not loaded}%
    \renewcommand\color[2][]{}%
  }%
  \providecommand\transparent[1]{%
    \errmessage{(Inkscape) Transparency is used (non-zero) for the text in Inkscape, but the package 'transparent.sty' is not loaded}%
    \renewcommand\transparent[1]{}%
  }%
  \providecommand\rotatebox[2]{#2}%
  \newcommand*\fsize{\dimexpr\f@size pt\relax}%
  \newcommand*\lineheight[1]{\fontsize{\fsize}{#1\fsize}\selectfont}%
  \ifx\svgwidth\undefined%
    \setlength{\unitlength}{834.89808786bp}%
    \ifx\svgscale\undefined%
      \relax%
    \else%
      \setlength{\unitlength}{\unitlength * \real{\svgscale}}%
    \fi%
  \else%
    \setlength{\unitlength}{\svgwidth}%
  \fi%
  \global\let\svgwidth\undefined%
  \global\let\svgscale\undefined%
  \makeatother%
  \begin{picture}(1,0.33221064)%
    \lineheight{1}%
    \setlength\tabcolsep{0pt}%
    \put(0,0){\includegraphics[width=\unitlength,page=1]{Abelian_angles_case2.pdf}}%
    \put(0.34169851,0.08592619){\color[rgb]{0,0,0}\makebox(0,0)[lt]{\lineheight{1.25000012}\smash{\begin{tabular}[t]{l}$D_{\iD{3}}$\end{tabular}}}}%
    \put(0.3259873,0.30480477){\color[rgb]{0,0,0}\makebox(0,0)[lt]{\lineheight{1.25000012}\smash{\begin{tabular}[t]{l}$D_{\iD{1}}$\end{tabular}}}}%
    \put(0.00256503,0.00373823){\color[rgb]{0,0,0}\makebox(0,0)[lt]{\lineheight{1.25000012}\smash{\begin{tabular}[t]{l}$D_{\iD{2}}$\end{tabular}}}}%
    \put(0.30435313,0.12928674){\color[rgb]{0,0,0}\makebox(0,0)[lt]{\lineheight{1.25}\smash{\begin{tabular}[t]{l}$n^3_{\iOm{0}}>0$\end{tabular}}}}%
    \put(0,0){\includegraphics[width=\unitlength,page=2]{Abelian_angles_case2.pdf}}%
    \put(0.12647628,0.21669111){\color[rgb]{0,0,0}\makebox(0,0)[lt]{\lineheight{1.25}\smash{\begin{tabular}[t]{l}$n^3_{\iOm{\infty}}>0$\end{tabular}}}}%
    \put(0,0){\includegraphics[width=\unitlength,page=3]{Abelian_angles_case2.pdf}}%
    \put(0.02064846,0.11940986){\color[rgb]{0,0,0}\makebox(0,0)[lt]{\lineheight{1.25}\smash{\begin{tabular}[t]{l}$n^3_{\iOm{1}}<0$\end{tabular}}}}%
    \put(0,0){\includegraphics[width=\unitlength,page=4]{Abelian_angles_case2.pdf}}%
    \put(0.87170329,0.10389247){\color[rgb]{0,0,0}\makebox(0,0)[lt]{\lineheight{1.25000012}\smash{\begin{tabular}[t]{l}$D_{\iD{3}}$\end{tabular}}}}%
    \put(0.8559921,0.32277102){\color[rgb]{0,0,0}\makebox(0,0)[lt]{\lineheight{1.25000012}\smash{\begin{tabular}[t]{l}$D_{\iD{1}}$\end{tabular}}}}%
    \put(0.91548456,0.24837237){\color[rgb]{0,0,0}\makebox(0,0)[lt]{\lineheight{1.25000012}\smash{\begin{tabular}[t]{l}$D_{\iD{2}}$\end{tabular}}}}%
    \put(0.83435796,0.14725297){\color[rgb]{0,0,0}\makebox(0,0)[lt]{\lineheight{1.25}\smash{\begin{tabular}[t]{l}$n^3_{\iOm{0}}>0$\end{tabular}}}}%
    \put(0,0){\includegraphics[width=\unitlength,page=5]{Abelian_angles_case2.pdf}}%
    \put(0.7015946,0.24566063){\color[rgb]{0,0,0}\makebox(0,0)[lt]{\lineheight{1.25}\smash{\begin{tabular}[t]{l}$n^3_{\iOm{\infty}}<0$\end{tabular}}}}%
    \put(0,0){\includegraphics[width=\unitlength,page=6]{Abelian_angles_case2.pdf}}%
    \put(0.55065327,0.13737614){\color[rgb]{0,0,0}\makebox(0,0)[lt]{\lineheight{1.25}\smash{\begin{tabular}[t]{l}$n^3_{\iOm{1}}>0$\end{tabular}}}}%
  \end{picture}%
\endgroup%

%% file: img/Usual_Abelian_Angles.pdf_tex
\begingroup%
  \makeatletter%
  \providecommand\color[2][]{%
    \errmessage{(Inkscape) Color is used for the text in Inkscape, but the package 'color.sty' is not loaded}%
    \renewcommand\color[2][]{}%
  }%
  \providecommand\transparent[1]{%
    \errmessage{(Inkscape) Transparency is used (non-zero) for the text in Inkscape, but the package 'transparent.sty' is not loaded}%
    \renewcommand\transparent[1]{}%
  }%
  \providecommand\rotatebox[2]{#2}%
  \newcommand*\fsize{\dimexpr\f@size pt\relax}%
  \newcommand*\lineheight[1]{\fontsize{\fsize}{#1\fsize}\selectfont}%
  \ifx\svgwidth\undefined%
    \setlength{\unitlength}{355.22695349bp}%
    \ifx\svgscale\undefined%
      \relax%
    \else%
      \setlength{\unitlength}{\unitlength * \real{\svgscale}}%
    \fi%
  \else%
    \setlength{\unitlength}{\svgwidth}%
  \fi%
  \global\let\svgwidth\undefined%
  \global\let\svgscale\undefined%
  \makeatother%
  \begin{picture}(1,0.32270433)%
    \lineheight{1}%
    \setlength\tabcolsep{0pt}%
    \put(0,0){\includegraphics[width=\unitlength,page=1]{Usual_Abelian_Angles.pdf}}%
    \put(0.05611951,0.31335108){\color[rgb]{0,0,0}\makebox(0,0)[lt]{\lineheight{1.25}\smash{\begin{tabular}[t]{l}$D_{(t+1)}$\end{tabular}}}}%
    \put(0.36134758,0.24589212){\color[rgb]{0,0,0}\makebox(0,0)[lt]{\lineheight{1.25}\smash{\begin{tabular}[t]{l}$D_{(t)}$\end{tabular}}}}%
    \put(0.23304382,0.13058122){\color[rgb]{0,0,0}\makebox(0,0)[lt]{\lineheight{1.25}\smash{\begin{tabular}[t]{l}$\upalpha_{(t+1)}$\end{tabular}}}}%
    \put(0.23239422,0.27983575){\color[rgb]{0,0,0}\makebox(0,0)[lt]{\lineheight{1.25}\smash{\begin{tabular}[t]{l}$\upvarepsilon_{(t)}$\end{tabular}}}}%
    \put(0.13884617,0.06854071){\color[rgb]{0,0,0}\makebox(0,0)[lt]{\lineheight{1.25}\smash{\begin{tabular}[t]{l}$\upalpha_{(t)}$\end{tabular}}}}%
    \put(0,0){\includegraphics[width=\unitlength,page=2]{Usual_Abelian_Angles.pdf}}%
    \put(0.66455532,0.30052786){\color[rgb]{0,0,0}\makebox(0,0)[lt]{\lineheight{1.25}\smash{\begin{tabular}[t]{l}$D_{(t)}$\end{tabular}}}}%
    \put(0.81110206,0.12696335){\color[rgb]{0,0,0}\makebox(0,0)[lt]{\lineheight{1.25}\smash{\begin{tabular}[t]{l}$\upalpha_{(t)}$\end{tabular}}}}%
    \put(0.58031959,0.18048261){\color[rgb]{0,0,0}\makebox(0,0)[lt]{\lineheight{1.25}\smash{\begin{tabular}[t]{l}$\upvarepsilon_{(t)}$\end{tabular}}}}%
    \put(0.71690441,0.06492283){\color[rgb]{0,0,0}\makebox(0,0)[lt]{\lineheight{1.25}\smash{\begin{tabular}[t]{l}$\upalpha_{(t+1)}$\end{tabular}}}}%
    \put(0.92734545,0.21828234){\color[rgb]{0,0,0}\makebox(0,0)[lt]{\lineheight{1.25}\smash{\begin{tabular}[t]{l}$D_{(t+1)}$\end{tabular}}}}%
  \end{picture}%
\endgroup%

%% file: img/brane3d.pdf_tex
\begingroup%
  \makeatletter%
  \providecommand\color[2][]{%
    \errmessage{(Inkscape) Color is used for the text in Inkscape, but the package 'color.sty' is not loaded}%
    \renewcommand\color[2][]{}%
  }%
  \providecommand\transparent[1]{%
    \errmessage{(Inkscape) Transparency is used (non-zero) for the text in Inkscape, but the package 'transparent.sty' is not loaded}%
    \renewcommand\transparent[1]{}%
  }%
  \providecommand\rotatebox[2]{#2}%
  \ifx\svgwidth\undefined%
    \setlength{\unitlength}{808.2369106bp}%
    \ifx\svgscale\undefined%
      \relax%
    \else%
      \setlength{\unitlength}{\unitlength * \real{\svgscale}}%
    \fi%
  \else%
    \setlength{\unitlength}{\svgwidth}%
  \fi%
  \global\let\svgwidth\undefined%
  \global\let\svgscale\undefined%
  \makeatother%
  \begin{picture}(1,0.66311263)%
    \put(0,0){\includegraphics[width=\unitlength,page=1]{brane3d.pdf}}%
    \put(0.39674996,0.62174419){\color[rgb]{0,0,0}\makebox(0,0)[lb]{\smash{$D_{(1)}$}}}%
    \put(0.79765484,0.33808391){\color[rgb]{0,0,0}\makebox(0,0)[lb]{\smash{$D_{(2)}$}}}%
    \put(0,0){\includegraphics[width=\unitlength,page=2]{brane3d.pdf}}%
  \end{picture}%
\endgroup%

%% file: non-abelian-twists.bbl
\providecommand{\href}[2]{#2}\begingroup\raggedright\begin{thebibliography}{10}

\bibitem{Chamoun:2003pf}
N.~Chamoun, S.~Khalil, and E.~Lashin, ``{Fermion masses and mixing in
  intersecting branes scenarios},''
  \href{http://dx.doi.org/10.1103/PhysRevD.69.095011}{{\em Phys. Rev.}
  {\bfseries D69} (2004) 095011},
\href{http://arxiv.org/abs/hep-ph/0309169}{{\ttfamily arXiv:hep-ph/0309169
  [hep-ph]}}.

\bibitem{Cremades:2003qj}
D.~Cremades, L.~E. Ibanez, and F.~Marchesano, ``{Yukawa couplings in
  intersecting D-brane models},''
  \href{http://dx.doi.org/10.1088/1126-6708/2003/07/038}{{\em JHEP} {\bfseries
  07} (2003) 038},
\href{http://arxiv.org/abs/hep-th/0302105}{{\ttfamily arXiv:hep-th/0302105
  [hep-th]}}.

\bibitem{Cvetic:2009mt}
M.~Cvetic, I.~Garcia-Etxebarria, and R.~Richter, ``{Branes and instantons
  intersecting at angles},''
  \href{http://dx.doi.org/10.1007/JHEP01(2010)005}{{\em JHEP} {\bfseries 01}
  (2010) 005},
\href{http://arxiv.org/abs/0905.1694}{{\ttfamily arXiv:0905.1694 [hep-th]}}.

\bibitem{Abel:2006yk}
S.~A. Abel and M.~D. Goodsell, ``{Realistic Yukawa Couplings through Instantons
  in Intersecting Brane Worlds},''
  \href{http://dx.doi.org/10.1088/1126-6708/2007/10/034}{{\em JHEP} {\bfseries
  10} (2007) 034},
\href{http://arxiv.org/abs/hep-th/0612110}{{\ttfamily arXiv:hep-th/0612110
  [hep-th]}}.

\bibitem{Chen:2008rx}
C.-M. Chen, T.~Li, V.~E. Mayes, and D.~V. Nanopoulos, ``{Yukawa Corrections
  from Four-Point Functions in Intersecting D6-Brane Models},''
  \href{http://dx.doi.org/10.1103/PhysRevD.78.105015}{{\em Phys. Rev.}
  {\bfseries D78} (2008) 105015},
\href{http://arxiv.org/abs/0807.4216}{{\ttfamily arXiv:0807.4216 [hep-th]}}.

\bibitem{Abel:2004ue}
S.~A. Abel and B.~W. Schofield, ``{One-loop Yukawas on intersecting branes},''
  \href{http://dx.doi.org/10.1088/1126-6708/2005/06/072}{{\em JHEP} {\bfseries
  06} (2005) 072},
\href{http://arxiv.org/abs/hep-th/0412206}{{\ttfamily arXiv:hep-th/0412206
  [hep-th]}}.

\bibitem{Abel:2003fk}
S.~A. Abel, M.~Masip, and J.~Santiago, ``{Flavor changing neutral currents in
  intersecting brane models},''
  \href{http://dx.doi.org/10.1088/1126-6708/2003/04/057}{{\em JHEP} {\bfseries
  04} (2003) 057},
\href{http://arxiv.org/abs/hep-ph/0303087}{{\ttfamily arXiv:hep-ph/0303087
  [hep-ph]}}.

\bibitem{Angelantonj:2000hi}
C.~Angelantonj, I.~Antoniadis, E.~Dudas, and A.~Sagnotti, ``{Type I strings on
  magnetized orbifolds and brane transmutation},''
  \href{http://dx.doi.org/10.1016/S0370-2693(00)00907-2}{{\em Phys. Lett.}
  {\bfseries B489} (2000) 223--232},
\href{http://arxiv.org/abs/hep-th/0007090}{{\ttfamily arXiv:hep-th/0007090
  [hep-th]}}.

\bibitem{Bertolini:2005qh}
M.~Bertolini, M.~Billo, A.~Lerda, J.~F. Morales, and R.~Russo, ``{Brane world
  effective actions for D-branes with fluxes},''
  \href{http://dx.doi.org/10.1016/j.nuclphysb.2006.02.044}{{\em Nucl. Phys.}
  {\bfseries B743} (2006) 1--40},
\href{http://arxiv.org/abs/hep-th/0512067}{{\ttfamily arXiv:hep-th/0512067
  [hep-th]}}.

\bibitem{Bianchi:2005yz}
M.~Bianchi and E.~Trevigne, ``{The Open story of the magnetic fluxes},''
  \href{http://dx.doi.org/10.1088/1126-6708/2005/08/034}{{\em JHEP} {\bfseries
  08} (2005) 034},
\href{http://arxiv.org/abs/hep-th/0502147}{{\ttfamily arXiv:hep-th/0502147
  [hep-th]}}.

\bibitem{Pesando:2009tt}
I.~Pesando, ``{Open and Closed String Vertices for branes with magnetic field
  and T-duality},'' \href{http://dx.doi.org/10.1007/JHEP02(2010)064}{{\em JHEP}
  {\bfseries 02} (2010) 064},
\href{http://arxiv.org/abs/0910.2576}{{\ttfamily arXiv:0910.2576 [hep-th]}}.

\bibitem{Forste:2018kub}
S.~Förste and C.~Liyanage, ``{Yukawa couplings from magnetized D-brane models
  on non-factorisable tori},''
\href{http://arxiv.org/abs/1802.05136}{{\ttfamily arXiv:1802.05136 [hep-th]}}.

\bibitem{Kiritsis:1993jk}
E.~Kiritsis and C.~Kounnas, ``{String propagation in gravitational wave
  backgrounds},'' \href{http://dx.doi.org/10.1016/0370-2693(94)90655-6}{{\em
  Phys. Lett.} {\bfseries B320} (1994) 264--272},
  \href{http://arxiv.org/abs/hep-th/9310202}{{\ttfamily arXiv:hep-th/9310202
  [hep-th]}}.
[Addendum: Phys. Lett.B325,536(1994)].

\bibitem{DAppollonio:2003zow}
G.~D'Appollonio and E.~Kiritsis, ``{String interactions in gravitational wave
  backgrounds},'' \href{http://dx.doi.org/10.1016/j.nuclphysb.2003.09.020}{{\em
  Nucl. Phys.} {\bfseries B674} (2003) 80--170},
\href{http://arxiv.org/abs/hep-th/0305081}{{\ttfamily arXiv:hep-th/0305081
  [hep-th]}}.

\bibitem{Berkooz:2004yy}
M.~Berkooz, B.~Durin, B.~Pioline, and D.~Reichmann, ``{Closed strings in Misner
  space: Stringy fuzziness with a twist},''
  \href{http://dx.doi.org/10.1088/1475-7516/2004/10/002}{{\em JCAP} {\bfseries
  0410} (2004) 002},
\href{http://arxiv.org/abs/hep-th/0407216}{{\ttfamily arXiv:hep-th/0407216
  [hep-th]}}.

\bibitem{DAppollonio:2004ppq}
G.~D'Appollonio and E.~Kiritsis, ``{D-branes and BCFT in Hpp-wave
  backgrounds},'' \href{http://dx.doi.org/10.1016/j.nuclphysb.2005.01.020}{{\em
  Nucl. Phys.} {\bfseries B712} (2005) 433--512},
\href{http://arxiv.org/abs/hep-th/0410269}{{\ttfamily arXiv:hep-th/0410269
  [hep-th]}}.

\bibitem{Gava:1997jt}
E.~Gava, K.~S. Narain, and M.~H. Sarmadi, ``{On the bound states of p-branes
  and (p+2)-branes},''
  \href{http://dx.doi.org/10.1016/S0550-3213(97)00508-7}{{\em Nucl. Phys.}
  {\bfseries B504} (1997) 214--238},
\href{http://arxiv.org/abs/hep-th/9704006}{{\ttfamily arXiv:hep-th/9704006
  [hep-th]}}.

\bibitem{Duo:2007he}
D.~Duo, R.~Russo, and S.~Sciuto, ``{New twist field couplings from the
  partition function for multiply wrapped D-branes},''
  \href{http://dx.doi.org/10.1088/1126-6708/2007/12/042}{{\em JHEP} {\bfseries
  12} (2007) 042},
\href{http://arxiv.org/abs/0709.1805}{{\ttfamily arXiv:0709.1805 [hep-th]}}.

\bibitem{David:2000um}
J.~R. David, ``{Tachyon condensation in the D0 / D4 system},''
  \href{http://dx.doi.org/10.1088/1126-6708/2000/10/004}{{\em JHEP} {\bfseries
  10} (2000) 004},
\href{http://arxiv.org/abs/hep-th/0007235}{{\ttfamily arXiv:hep-th/0007235
  [hep-th]}}.

\bibitem{David:2000yn}
J.~R. David, ``{Tachyon condensation using the disc partition function},''
  \href{http://dx.doi.org/10.1088/1126-6708/2001/07/009}{{\em JHEP} {\bfseries
  07} (2001) 009},
\href{http://arxiv.org/abs/hep-th/0012089}{{\ttfamily arXiv:hep-th/0012089
  [hep-th]}}.

\bibitem{David:2001vm}
J.~R. David, M.~Gutperle, M.~Headrick, and S.~Minwalla, ``{Closed string
  tachyon condensation on twisted circles},''
  \href{http://dx.doi.org/10.1088/1126-6708/2002/02/041}{{\em JHEP} {\bfseries
  02} (2002) 041},
\href{http://arxiv.org/abs/hep-th/0111212}{{\ttfamily arXiv:hep-th/0111212
  [hep-th]}}.

\bibitem{Hashimoto:2003xz}
K.~Hashimoto and S.~Nagaoka, ``{Recombination of intersecting D-branes by local
  tachyon condensation},''
  \href{http://dx.doi.org/10.1088/1126-6708/2003/06/034}{{\em JHEP} {\bfseries
  06} (2003) 034},
\href{http://arxiv.org/abs/hep-th/0303204}{{\ttfamily arXiv:hep-th/0303204
  [hep-th]}}.

\bibitem{Burwick:1990tu}
T.~T. Burwick, R.~K. Kaiser, and H.~F. Muller, ``{General Yukawa couplings of
  strings on Z(N) orbifolds},''
\href{http://dx.doi.org/10.1016/0550-3213(91)90491-F}{{\em Nucl. Phys.}
  {\bfseries B355} (1991) 689--711}.

\bibitem{Stieberger:1992bj}
S.~Stieberger, D.~Jungnickel, J.~Lauer, and M.~Spalinski, ``{Yukawa couplings
  for bosonic Z(N) orbifolds: Their moduli and twisted sector dependence},''
  \href{http://dx.doi.org/10.1142/S0217732392002457}{{\em Mod. Phys. Lett.}
  {\bfseries A7} (1992) 3059--3070},
\href{http://arxiv.org/abs/hep-th/9204037}{{\ttfamily arXiv:hep-th/9204037
  [hep-th]}}.

\bibitem{Erler:1992gt}
J.~Erler, D.~Jungnickel, M.~Spalinski, and S.~Stieberger, ``{Higher twisted
  sector couplings of Z(N) orbifolds},''
  \href{http://dx.doi.org/10.1016/0550-3213(93)90348-S}{{\em Nucl. Phys.}
  {\bfseries B397} (1993) 379--416},
\href{http://arxiv.org/abs/hep-th/9207049}{{\ttfamily arXiv:hep-th/9207049
  [hep-th]}}.

\bibitem{Anastasopoulos:2011gn}
P.~Anastasopoulos, M.~Bianchi, and R.~Richter, ``{On closed-string twist-field
  correlators and their open-string descendants},''
\href{http://arxiv.org/abs/1110.5359}{{\ttfamily arXiv:1110.5359 [hep-th]}}.

\bibitem{Anastasopoulos:2011hj}
P.~Anastasopoulos, M.~Bianchi, and R.~Richter, ``{Light stringy states},''
  \href{http://dx.doi.org/10.1007/JHEP03(2012)068}{{\em JHEP} {\bfseries 03}
  (2012) 068},
\href{http://arxiv.org/abs/1110.5424}{{\ttfamily arXiv:1110.5424 [hep-th]}}.

\bibitem{Anastasopoulos:2013sta}
P.~Anastasopoulos, M.~D. Goodsell, and R.~Richter, ``{Three- and Four-point
  correlators of excited bosonic twist fields},''
  \href{http://dx.doi.org/10.1007/JHEP10(2013)182}{{\em JHEP} {\bfseries 10}
  (2013) 182},
\href{http://arxiv.org/abs/1305.7166}{{\ttfamily arXiv:1305.7166 [hep-th]}}.

\bibitem{Sciuto:1969vz}
S.~Sciuto, ``{The general vertex function in dual resonance models},''
\href{http://dx.doi.org/10.1007/BF02755622}{{\em Lett. Nuovo Cim.} {\bfseries
  2S1} (1969) 411--418}.

\bibitem{DellaSelva:1970bj}
A.~Della~Selva and S.~Saito, ``{A simple expression for the sciuto
  three-reggeon vertex-generating duality},''
  \href{http://dx.doi.org/10.1007/BF02755329}{{\em Lett. Nuovo Cim.} {\bfseries
  4S1} (1970) 689--692}.
[Lett. Nuovo Cim.4,689(1970)].

\bibitem{Pesando:2014owa}
I.~Pesando, ``{Correlators of arbitrary untwisted operators and excited twist
  operators for $N$ branes at angles},''
  \href{http://dx.doi.org/10.1016/j.nuclphysb.2014.06.010}{{\em Nucl. Phys.}
  {\bfseries B886} (2014) 243--287},
\href{http://arxiv.org/abs/1401.6797}{{\ttfamily arXiv:1401.6797 [hep-th]}}.

\bibitem{Pesando:2012cx}
I.~Pesando, ``{Green functions and twist correlators for $N$ branes at
  angles},'' \href{http://dx.doi.org/10.1016/j.nuclphysb.2012.08.016}{{\em
  Nucl. Phys.} {\bfseries B866} (2013) 87--123},
\href{http://arxiv.org/abs/1206.1431}{{\ttfamily arXiv:1206.1431 [hep-th]}}.

\bibitem{Pesando:2011ce}
I.~Pesando, ``{The generating function of amplitudes with $N$ twisted and L
  untwisted states},'' \href{http://dx.doi.org/10.1142/S0217751X15501213}{{\em
  Int. J. Mod. Phys.} {\bfseries A30} no.~21, (2015) 1550121},
\href{http://arxiv.org/abs/1107.5525}{{\ttfamily arXiv:1107.5525 [hep-th]}}.

\bibitem{Pesando:2003ww}
I.~Pesando, ``{Multibranes boundary states with open string interactions},''
  \href{http://dx.doi.org/10.1016/j.nuclphysb.2007.10.002}{{\em Nucl. Phys.}
  {\bfseries B793} (2008) 211--245},
\href{http://arxiv.org/abs/hep-th/0310027}{{\ttfamily arXiv:hep-th/0310027
  [hep-th]}}.

\bibitem{DiVecchia:2007dh}
P.~Di~Vecchia, A.~Liccardo, R.~Marotta, I.~Pesando, and F.~Pezzella, ``{Wrapped
  magnetized branes: two alternative descriptions?},''
  \href{http://dx.doi.org/10.1088/1126-6708/2007/11/100}{{\em JHEP} {\bfseries
  11} (2007) 100},
\href{http://arxiv.org/abs/0709.4149}{{\ttfamily arXiv:0709.4149 [hep-th]}}.

\bibitem{Pesando:2011yd}
I.~Pesando, ``{Strings in an arbitrary constant magnetic field with arbitrary
  constant metric and stringy form factors},''
  \href{http://dx.doi.org/10.1007/JHEP06(2011)138}{{\em JHEP} {\bfseries 06}
  (2011) 138},
\href{http://arxiv.org/abs/1101.5898}{{\ttfamily arXiv:1101.5898 [hep-th]}}.

\bibitem{DiVecchia:2011mf}
P.~Di~Vecchia, R.~Marotta, I.~Pesando, and F.~Pezzella, ``{Open strings in the
  system D5/D9},'' \href{http://dx.doi.org/10.1088/1751-8113/44/24/245401}{{\em
  J. Phys.} {\bfseries A44} (2011) 245401},
\href{http://arxiv.org/abs/1101.0120}{{\ttfamily arXiv:1101.0120 [hep-th]}}.

\bibitem{Pesando:2013qda}
I.~Pesando, ``{Light cone quantization and interactions of a new closed bosonic
  string inspired to D1 string},''
  \href{http://dx.doi.org/10.1016/j.nuclphysb.2013.07.022}{{\em Nucl. Phys.}
  {\bfseries B876} (2013) 1--15},
\href{http://arxiv.org/abs/1305.2710}{{\ttfamily arXiv:1305.2710 [hep-th]}}.

\bibitem{Pesando:2014sca}
I.~Pesando, ``{Canonical quantization of a string describing $N$ branes at
  angles},'' \href{http://dx.doi.org/10.1016/j.nuclphysb.2014.10.005}{{\em
  Nucl. Phys.} {\bfseries B889} (2014) 120--155},
\href{http://arxiv.org/abs/1407.4627}{{\ttfamily arXiv:1407.4627 [hep-th]}}.

\bibitem{Inoue:1987ak}
K.~Inoue, M.~Sakamoto, and H.~Takano, ``{NONABELIAN ORBIFOLDS},''
\href{http://dx.doi.org/10.1143/PTP.78.908}{{\em Prog. Theor. Phys.} {\bfseries
  78} (1987) 908}.

\bibitem{Inoue:1990ci}
K.~Inoue and S.~Nima, ``{String interactions on nonAbelian orbifold},''
\href{http://dx.doi.org/10.1143/PTP.84.702}{{\em Prog. Theor. Phys.} {\bfseries
  84} (1990) 702--727}.

\bibitem{Gato:1990mx}
B.~Gato, ``{Vertex operators, nonAbelian orbifolds and the Riemann-Hilbert
  problem},''
\href{http://dx.doi.org/10.1016/0550-3213(90)90485-V}{{\em Nucl. Phys.}
  {\bfseries B334} (1990) 414--430}.

\bibitem{Frampton:2000mq}
P.~H. Frampton and T.~W. Kephart, ``{Classification of conformality models
  based on nonAbelian orbifolds},''
  \href{http://dx.doi.org/10.1103/PhysRevD.64.086007}{{\em Phys. Rev.}
  {\bfseries D64} (2001) 086007},
\href{http://arxiv.org/abs/hep-th/0011186}{{\ttfamily arXiv:hep-th/0011186
  [hep-th]}}.

\bibitem{Pesando:2015fpj}
I.~Pesando, ``{Towards a fully stringy computation of Yukawa couplings on non
  factorized tori and non abelian twist correlators (I): the classical solution
  and action},'' \href{http://dx.doi.org/10.1016/j.nuclphysb.2016.06.013}{{\em
  Nucl. Phys.} {\bfseries B910} (2016) 618--664},
\href{http://arxiv.org/abs/1512.07920}{{\ttfamily arXiv:1512.07920 [hep-th]}}.

\bibitem{NIST:DLMF}
``{NIST Digital Library of Mathematical Functions}.'' Http://dlmf.nist.gov/,
  release 1.0.19 of 2018-06-22.
\newblock \url{http://dlmf.nist.gov/}. F.~W.~J. Olver, A.~B. {Olde Daalhuis},
  D.~W. Lozier, B.~I. Schneider, R.~F. Boisvert, C.~W. Clark, B.~R. Miller and
  B.~V. Saunders, eds.

\end{thebibliography}\endgroup
